\documentclass{aa}
\usepackage{graphicx}
\usepackage{txfonts}
\usepackage[
        colorlinks=true,
        linkcolor=black,
        citecolor=black,
        filecolor=black,
        urlcolor=black,
        bookmarks=true,
        bookmarksopen=true,
        bookmarksopenlevel=3,
        plainpages=false,
        pdfpagelabels=true,
]{hyperref}
\usepackage[ugly]{units}
\usepackage{wasysym}
\usepackage[english=american]{csquotes}
\usepackage{longtable}
\usepackage{booktabs}
\usepackage{subfigure}
\usepackage{natbib}
\usepackage[multiple,symbol]{footmisc}
\bibpunct{(}{)}{;}{a}{}{,}

\begin{document} 

\title{Transmission spectroscopy of the hot Jupiter TrES-3\,b: Disproof of an overly large Rayleigh-like feature\thanks{Based on (1) data obtained with the STELLA robotic telescopes in Tenerife, an AIP facility jointly operated by AIP and IAC, (2) observations collected at the German-Spanish Astronomical Center, Calar Alto, jointly operated by the Max-Planck-Institut für Astronomie Heidelberg and the Instituto de Astrofísica de Andalucía (CSIC) and (3) observations made with the Italian Telescopio Nazionale Galileo (TNG) operated on the island of La Palma by the Fundación Galileo Galilei of the INAF (Istituto Nazionale di Astrofisica) at the Spanish Observatorio del Roque de los Muchachos of the Instituto de Astrofisica de Canarias.}\fnmsep\thanks{Newly observed photometric data from Sects.~\ref{sec:new-transits} and \ref{sec:new-activity} are only available in electronic form at the CDS via anonymous ftp to cdsarc.u-strasbg.fr (130.79.128.5) or via \url{http://cdsweb.u-strasbg.fr/cgi-bin/qcat?J/A+A/}} }
\titlerunning{Transmission spectroscopy of the hot Jupiter TrES-3b}
\author{F.\,Mackebrandt \inst{1,2}\fnmsep\thanks{\email{mackebrandt@mps.mpg.de}}
                \and M.\,Mallonn \inst{1}
                \and J.\,M. Ohlert \inst{3,4}
                \and T.\,Granzer \inst{1}
                \and S. Lalitha \inst{5}
                \and A.\,Garc\'{i}a Mu\~{n}oz \inst{6}
                \and N.\,P.\,Gibson \inst{7}
                \and J.\,W.\,Lee \inst{8,9}
                \and A.\,Sozzetti \inst{10}
                \and J.\,D.\,Turner \inst{11}
                \and M.\,Va\v{n}ko \inst{12}
                \and K.\,G.\,Strassmeier \inst{1}
                }
\institute{
        Leibniz-Institut für Astrophysik Potsdam (AIP), An der Sternwarte 16, 14482 Potsdam, Germany
        \and Max-Planck-Institut für Sonnensystemforschung, Justus-von-Liebig-Weg 3, 37077 G\"ottingen, Germany
        \and Michael Adrian Observatorium, Astronomie Stiftung Trebur, 65468 Trebur, Germany
        \and University of Applied Sciences, Technische Hochschule Mittelhessen,
        61169 Friedberg, Germany
        \and Tata Institute of Fundamental Research, Homi Bhabha Road, Mumbai 400005, India 0000-0001-8102-3033
        \and Zentrum für Astronomie und Astrophysik, Technische Universit\"at Berlin, Hardenbergstraße 36, 10623 Berlin, Germany
        \and Astrophysics Research Centre, School of Mathematics and Physics, Queens University Belfast, Belfast BT7 1NN, UK
        \and Korea Astronomy and Space Science Institute, Daejeon, 34055, Korea
        \and Astronomy and Space Science Major, Korea University of Science and Technology, Daejeon 34113, Korea
        \and INAF - Osservatorio Astrofisico di Torino, via Osservatorio 20, 10025 Pino Torinese, Italy
        \and Department of Astronomy, University of Virginia, Charlottesville, VA 22904, USA
        \and Astronomical Institute, Slovak Academy of Sciences, 059 60 Tatransk\'a Lomnica, Slovakia
        }
\date{Received January 27, 2017; accepted September 04, 2017}
% \abstract{}{}{}{}{} 
% 5 {} token are mandatory
\abstract
% context heading (optional)
% {} leave it empty if necessary  
{Transit events of extrasolar planets offer the opportunity to study the composition of their atmospheres. Previous work on transmission spectroscopy of the close-in gas giant TrES-3\,b revealed an increase in absorption towards blue wavelengths of very large amplitude in terms of atmospheric pressure scale heights, too large to be explained by Rayleigh-scattering in the planetary atmosphere.}
% aims heading (mandatory)
{We present a follow-up study of the optical transmission spectrum of the hot Jupiter TrES-3\,b to investigate the strong increase in opacity towards short wavelengths found by a previous study. Furthermore, we aim to estimate the effect of stellar spots on the transmission spectrum.}
% methods heading (mandatory)
{This work uses previously published long slit spectroscopy transit data of the Gran Telescopio Canarias (GTC) and published broad band observations as well as new observations in different bands from the near-UV to the near-IR, for a homogeneous transit light curve analysis. Additionally, a long-term photometric monitoring of the TrES-3 host star was performed.}
% results heading (mandatory)
{Our newly analysed GTC spectroscopic transit observations show a slope of much lower amplitude than previous studies. We conclude from our results the previously reported increasing signal towards short wavelengths is not intrinsic to the TrES-3 system. Furthermore, the broad band spectrum favours a flat spectrum. Long-term photometric monitoring rules out a significant modification of the transmission spectrum by unocculted star spots.}
% conclusions heading (optional), leave it empty if necessary 
{}

\maketitle
% flatex input: [intro.tex]
\section{Introduction}
Transiting exoplanets present a unique opportunity to probe the upper regions of their atmospheres. During transit, the stellar light passes through the annulus of the planet’s atmosphere and is absorbed until an optical depth of about unity. Wavelength-dependent variations in the atmospheres optical depth cause variations in that transit depth. This yields information about atoms, molecules, and condensates in the outer (optically thin) region of the planetary atmosphere.

The magnitude of the wavelength dependence of the transit depth can be approximated by the area of the planetary atmosphere compared to the area of the star. Therefore, the observability of the transmission signal is maximised for planets with a large planet-to-star radius ratio and a large atmospheric pressure scale height. The majority of atmospheric characterisations have been accomplished for hot Jupiter exoplanets \citep{sing_2016}. However, observations of transmission spectra of Neptune-sized planets \citep[e.g.][]{fraine_2014} and super-Earths \citep[e.g.][]{kreidberg_2014} have also been published.

The vertical extent of the atmospheric annulus probed by transmission spectroscopy at optical wavelengths is about five to ten scale heights for gas giants \citep{seager_2009,burrows_2014,madhusudhan_2014}. This signal size is predicted for atomic and molecular absorption in cloud/haze-free atmosphere models as well as for Rayleigh-scattering in hazy atmospheres \citep{fortney_2010,miller-ricci_2010,wakeford_2015}. The vast majority of detected transmission signals is in agreement with this amplitude or lower \citep[e.g.][]{sing_2016,sedaghati_2016,kreidberg_2015,fraine_2014,nikolov_2016,gibson_2017}. Stronger amplitudes measured in the UV can be understood as Ly$ \alpha $ absorption in the planetary escaping exosphere \citep{ehrenreich_2015,vidal-madjar_2003}. However,  measured transmission signals of more than ten scale heights amplitude at optical wavelengths have been proposed \citep{southworth_2015,mancini_2014,turner_2016,parviainen_2016}. The physical interpretation of these results is challenging, because according to our current understanding these signals cannot entirely be caused by the planetary atmosphere. Does stellar activity influence the measurements \citep{oshagh_2013}, or does contamination from a nearby star play a role \citep{southworth_2016}? Are the values of the planetary scale height underestimated, which would point to a severely wrong estimation of planetary parameters, or are the uncertainties in the transmission spectroscopy measurements underestimated? Could there be missing physics in the modelling of exoplanet transmission spectra?

Ground-based transmission spectroscopy is often conducted with transit observations in low-resolution \citep[e.g.][]{sing_2012,lendl_2016,mallonn_2016a} or high-resolution spectroscopy \citep[e.g.][]{snellen_2008,wood_2011}. However, very broad spectral features in the planetary spectrum can also be detected by spectrophotometry using multiple broad band filters \citep{mancini_2013,nascimbeni_2015,kirk_2016,mallonn_2015a}. Examples of such spectral features are an increase in opacity towards blue wavelengths caused by scattering \citep{nascimbeni_2015} or an increase in opacity towards central optical wavelengths caused by TiO absorption \citep{evans_2016}. The broad filters allow for sub-millimag photon noise in the transit light curves with medium-sized telescopes on the ground \citep{southworth_2017,mallonn_2016a,kirk_2016} and even metre-sized telescopes can contribute significantly to a characterisation of the planet atmosphere when multiple transit light curves of individual filters are analysed together \citep{mallonn_2015b,dragomir_2015}. 

The exoplanet TrES-3\,b was discovered by \citet{odonovan_2007}. The planet was identified by two different transit surveys -- the Trans-atlantic Exoplanet Survey (TrES) and the Hungarian Automated telescope Network (HATNet). The planet is a massive Hot Jupiter of $ \unit[1.899]{M_{\jupiter}} $ with a short orbital period of $ \unit[1.3]{d} $ \citep{southworth_2011}. Attention must be paid to the high impact parameter $ b = 0.84 $ \citep{southworth_2011} which indicates an almost grazing orbit.

TrES-3\,b is only moderately favourable for transmission spectroscopy because of its rather small scale height compared to other targets observationally investigated. The spectral features of about five scale heights amplitude or less predicted by atmospheric models translate to a variation in transit depth smaller than the precision routinely achieved nowadays in ground-based transmission spectroscopy studies.

\citet{parviainen_2016} (hereafter P16) obtained the first transmission spectrum of the hot Jupiter TrES-3\,b and reported an overly large Rayleigh-like feature from $ \unit[650]{nm} $ on towards shorter wavelengths. To our knowledge, it is the strongest signal in units of scale height among all hot Jupiter transmission spectra and so far cannot be theoretically explained (P16). However, P16 analysed only a single transit measurement.

In this work we target the reliability and repeatability of this outstanding transmission signal. Therefore, we reanalyse the Gran Telescopio Canarias (GTC) transit observation used by P16 and homogeneously analyse 36 published and 15 newly observed broad band transit observations in nine band passes from the near-UV until the near-IR to build a broad band spectrophotometry transmission spectrum for comparison. Our aim is to verify whether the signal is intrinsic to the TrES-3 system or rather caused by artefacts in the data or aspects of the data analysis, rather than a detailed atmospheric characterisation.

We performed a photometric long-term monitoring of the host star TrES-3 to search for rotational modulation potentially caused by star spots \citep{strassmeier_2009}. Unocculted star spots on the visible hemisphere of the star influence the transit photometry and therefore the derived transit parameters \citep{pont_2013}. This effect is wavelength dependent and can potentially introduce a slope in the transmission spectrum similar to a spectral scattering signature produced by Rayleigh or Mie scattering by condensate particles \citep{oshagh_2014}. The monitoring program allows us to estimate an upper limit of the spot filling factor and the modification of the measured transit depth.

This paper is structured as follows: Section 2 and Section 3 describe the observations and the data reduction. In Section 4, the analysis is presented. The results are reported in Section 5, followed by the Discussion and Conclusions in Section 6.

% flatex input: [observations.tex]
\section{Observations}
The published broad band transit light curve observations are mostly observed in bands of long optical wavelengths. However, the Rayleigh-like feature is pronounced at short and intermediate optical wavelengths. We extended the small sample of published observations in Johnson \textit{U}, \textit{B}, and \textit{V} with 15 new transit light curve observations. 

Our new observations along the ones we collected from previous works are summarised in Table~\ref{tab:photo-obs}. Our analysis is based on more than 50 transit observations in nine different bands.
% flatex input: [tab-observations.tex]
\begin{table}[htp]
\caption{Overview of the analysed transit observations of TrES-3\,b. Columns are the photometric band, source reference (if applicable), date of observation, cadence, number of data points $ N_{data} $ and dispersion of the residuals as root mean square (rms) of the observations after subtracting a transit model. The last column indicates the observation number.} 
\label{tab:photo-obs}
\centering
\begin{tabular}{crcrrrr}
        \toprule
        Band & Ref & Date of obs. & cadence & $N_{data}$ & rms & \# \\
         &  & & $ \unit{/s} $ & & $ \unit{/mmag} $ & \\
        \midrule
\textit{U}   & 1 & 2016-04-25 & 40 & 253 & 2.73 &   1\\
& 2 & 2011-10-14 & 67 & 151 & 5.03 &   2\\
&   2   & 2011-11-04 & 67 & 115 & 3.59 &   3\\
&   2   & 2012-03-25 & 69 & 151 & 3.20 &   4\\
\textit{B}   & 3 & 2016-05-02 & 45 & 121 & 2.19 &   5\\
& 4 & 2007-04-08 & 70 & 126 & 1.53 &   6\\
& 2 & 2012-04-11 & 47 & 314 & 2.41 &   7\\
& 5 & 2016-04-12 & 102 & 91 & 3.01 &   8\\
&   5   & 2016-04-25 & 102 & 77 & 1.65 &   9\\
&   5   & 2016-05-19 & 102 & 91 & 2.39 &   10\\
&   5   & 2016-06-01 & 102 & 75 & 1.64 &   11\\
&   5   & 2016-06-05 & 102 & 90 & 1.49 &   12\\
&   5   & 2016-06-18 & 102 & 92 & 2.09 &   13\\
\textit{V}   & 6 & 2007-04-24 & 82 & 139 & 1.72 &   14\\
& 7 & 2016-03-09 & 69 & 172 & 2.12 &   15\\
&   7   & 2016-04-29 & 69 & 146 & 1.94 &   16\\
&   7   & 2016-05-02 & 69 & 147 & 2.57 &   17\\
&   7   & 2016-05-19 & 69 & 181 & 1.96 &   18\\
& 2 & 2009-07-04 & 30 & 270 & 2.81 &   19\\
& 8 & 2016-04-19 & 90 & 130 & 4.14 &   20\\
\textit{RISE}   & 9 & 2008-03-08 & 8 & 1350 & 2.92 &   21\\
&   9   & 2008-05-28 & 8 & 1350 & 3.39 &   22\\
&   9   & 2008-06-14 & 8 & 1350 & 2.18 &   23\\
&   9   & 2008-07-01 & 8 & 1350 & 2.48 &   24\\
&   9   & 2008-07-22 & 8 & 1350 & 3.27 &   25\\
\textit{r’}   & 10 & 2009-05-14 & 45 & 238 & 1.16 &   26\\
&   10   & 2010-05-16 & 45 & 248 & 1.23 &   27\\
&   10   & 2010-10-12 & 45 & 246 & 1.09 &   28\\
&   10   & 2011-03-24 & 45 & 371 & 0.74 &   29\\
&   10   & 2011-06-21 & 45 & 306 & 1.43 &   30\\
&   10   & 2011-08-24 & 45 & 168 & 1.34 &   31\\
& 6 & 2008-03-27 & 44 & 291 & 1.48 &   32\\
\textit{R}   & 11 & 2010-05-25 & 100 & 99 & 3.45 &   33\\
&   11   & 2010-06-11 & 100 & 80 & 2.95 &   34\\
&   11   & 2010-06-15 & 100 & 80 & 3.70 &   35\\
&   11   & 2010-06-28 & 100 & 77 & 3.69 &   36\\
& 12 & 2010-06-15 & 100 & 158 & 1.70 &   37\\
& 2 & 2009-06-22 & 21 & 581 & 1.80 &   38\\
& 13 & 2009-08-01 & 45 & 345 & 4.95 &   39\\
&   13   & 2010-04-27 & 35 & 180 & 4.17 &   40\\
&   13   & 2010-06-30 & 40 & 238 & 3.68 &   41\\
\textit{792} & 14 & 2009-08-10 & 140 & 71 & 0.68 &   42\\
\textit{I}   & 6 & 2008-03-09 & 44 & 253 & 2.01 &   43\\
&   6   & 2008-03-27 & 44 & 246 & 1.75 &   44\\
&   6   & 2008-04-12 & 63 & 166 & 2.14 &   45\\
& 7 & 2016-07-19 & 69 & 169 & 1.34 &   46\\
&   7   & 2016-08-25 & 69 & 171 & 1.33 &   47\\
\textit{z'}   & 4 & 2007-03-26 & 150 & 150 & 1.41 &   48\\
& 6 & 2007-03-25 & 104 & 134 & 1.24 &   49\\
& 5 & 2016-06-28 & 72 & 119 & 2.84 &   50\\
&   5   & 2016-07-15 & 102 & 125 & 1.85 &   51\\

        \bottomrule
\end{tabular}

\tablebib{
        (1)~TNG; (2)~\citet{turner_2013}; (3) CAHA; (4) \citet{odonovan_2007}; (5) STELLA; (6) \citet{sozzetti_2009}; (7) T1T; (8) VBO; (9) \citet{gibson_2009}; (10) \citet{kundurthy_2013}; (11) \citet{jiang_2013}; (12) \citet{lee_2011}; (13) \citet{vanko_2013}; (14) \citet{colon_2010}
}
\end{table}
% flatex input end: [tab-observations.tex]

\subsection{Published transit observations}
\subsubsection{Spectroscopy}
P16 published the first transmission spectrum of TrES-3\,b, observed  with the OSIRIS spectrograph as part of the \enquote{GTC exoplanet transit spectroscopy survey} \citep[see also][]{murgas_2014,palle_2016}. The observation covers a spectral range from $ \unit[530]{nm} $ to $ \unit[950]{nm} $ and was performed on July 8, 2014. 

\subsubsection{Photometry}
We collected 36 transit light curves from the literature obtained in Johnson/Cousin and Sloan filters from the \textit{U} band to the \textit{z'} band. The data were originally published by \citet{odonovan_2007,gibson_2009,sozzetti_2009,colon_2010,lee_2011,sada_2012,vanko_2013,jiang_2013,turner_2013} and \citet{kundurthy_2013}. The light curves by \citet{gibson_2009} were obtained with the RISE instrument on the Liverpool Telescope with a transmission from about $ \unit[500]{nm} $ to $ \unit[700]{nm} $. They are hereafter labelled as \enquote{\textit{RISE}}.

A transit of TrES-3\,b was also observed with the GTC by \citet{colon_2010}. They used the OSIRIS instrument in a tunable filter imaging mode. Two quasi simultaneous high-precision light curves were obtained in bandpasses centred at $ \unit[790.2]{nm} $ and $ \unit[794.4]{nm} $ each with a width of $ \unit[2]{nm} $. Due to the spectral affinity of the light curves, we decided to merge them together. This light curve is hereafter labelled \enquote{\textit{792}}.

\subsection{New transit observations} \label{sec:new-transits}

\subsubsection{Calar Alto Observatory}
One observation was performed with the 2.2 m telescope at the Calar Alto Observatory (CAHA) with the instrument CAFOS in the \textit{B} band on May 02, 2016. The weather conditions were stated as a photometric night with good transparency and a mean seeing of about $ \unit[1.2]{\arcsec} $. The telescope was defocused to an object PSF of about $ \unit[2.5]{\arcsec} $ with an integration time of $ \unit[45]{s} $ for each image.

\subsubsection{STELLar Activity}
STELLar Activity (STELLA) is a robotic observatory with two $ \unit[1.2]{m} $ fully automatic telescopes located at the Teide Observatory on the island of Tenerife with an Echelle spectrograph (SES) and a wide field imager WiFSIP \citep{strassmeier_2004}. It is owned and operated by the Leibniz-Institut für Astrophysik (AIP). Exoplanet transit photometry with WiFSIP was already performed for GJ 1214b, HAT-P-32b and HAT-P-12b \citep{mallonn_2015b,seeliger_2014,teske_2013}. We observed six \textit{B} band transits of TrES-3\,b with WiFSIP (April 12, 25; May 19; June 01, 05, 18, 2016), as well as three transits in the \textit{z'} band (June 28; July 15, 2016).

\subsubsection{Telescopio Nazionale Galileo}
One observation was performed at the $\unit[3.58]{m} $ Telescopio Nazionale Galileo (TNG) in the \textit{U} band on April 25, 2016. A $ 2\times2 $ binning mode was used to reduce the read-out-time. The telescope was defocused to spread PSF of the object to about $ \unit[2.5]{\arcsec} $. Each image was integrated with $ \unit[40]{s} $ exposures.

\subsubsection{Trebur-1m telescope}
The Trebur-$\unit[1]{m}$ telescope (T1T) is a $ \unit[1.2]{m} $ telescope operated by a foundation as a part of the Michael Adrian Observatory. It is located in Trebur, near Frankfurt am Main, Germany. TrES-3b has already been observed using T1T \citep{vanko_2013}. We obtained four new transit observations by the T1T in the \textit{V} band (March 09; April 29; May 02 and 19, 2016) and two observations in the \textit{I} band (July 19; August 25, 2016).

\subsubsection{Vainu Bappu Observatory}
The Vainu Bappu Observatory (VBO) located in India is owned and operated by the Indian Institute of Astrophysics. It hosts several meter-sized telescopes. We received one observation performed with the $ \unit[1.3]{m} $ JCB-Telescope in the \textit{V} band on April 19, 2016.

\subsection{Photometric long-term monitoring} \label{sec:new-activity}
We monitored TrES-3 with STELLA/WiFSIP in 2016 over the course of four months from March 7 to July 10, 2016. Because of the very low $ v \sin i $ value of $ \unit[1.5]{km\,s^{-1}} $ \citep{sozzetti_2009} we expect a stellar rotation period of $ > \! \unit[10]{days} $ and aimed for a photometric data point approximately every third day. We observed in two filters, Johnson \textit{B} and Johnson \textit{V}, in blocks of three exposures each. The exposure times were 90 and 60 seconds, respectively. However, due to a failure in the observation setting, only single exposures per filter were obtained in the first month of the campaign. During the time span of about 120 days of the campaign, we obtained data on 43 nights.

% flatex input end: [observations.tex]

% flatex input: [reduction.tex]
\section{Data reduction}
\subsection{Transit data}

\subsubsection{Spectroscopy}
The GTC/OSIRIS data were reduced by routines written in ESO-Midas. The bias value was extracted from the overscan regions per frame and subtracted from the flat, arc lamp, and science frames. We flatfielded the science frames with a master flat created by averaging 100 normalised individual flat frames observed in the morning after the observations. We fitted a polynomial of second order along the centroid of the spectrum in dispersion direction to estimate and correct for the misalignment regarding the pixel rows.
The background flux at the stellar spectrum was estimated by a linear interpolation per pixel column between sky stripes on both sides of the spectrum. The one-dimensional (1D) spectra were extracted as the simple flux sum within a certain aperture. We tested which sky stripe and aperture width minimises the scatter in the spectrophotometric light curves. We found wider apertures than P16, namely a 52 pixel-wide aperture for the reference object and a width of 80 pixels for the exoplanet host star. Since the full width at half maximum (FWHM) of the spectral spatial profile varies during the observations, we also tested a FWHM-dependent aperture with different scaling factors. However, this varying aperture width resulted in a higher noise level in the spectrophotometric light curves than the fixed aperture width. The optimal sky stripes had a width of 80 pixels each. We measured a mild drift in dispersion direction relative to the first exposure by fitting a Gaussian profile to the telluric O$_{2}$ Fraunhofer A line per spectra and approximated it with a second-order polynomial over time. Prior to the wavelength calibration, we subtracted this value that accounted for approximately one pixel at maximum. Then, we used the combined HgAr, Xe, and Ne arc lamp frames taken with the $ \unit[1]{''} $ long slit for wavelength calibration. 

\subsubsection{Photometry}
The reduction of the transit data obtained on STELLA, TNG, CAHA, VBO and T1T was performed homogeneously with a customised ESO-MIDAS pipeline, which calls the photometry software SExtractor \citep{bertin_1996}. A bias value was extracted from the overscan regions and subtracted, and a normalised masterflat was applied to the science images by division. A cosmic ray correction was performed within SExtractor. Aperture photometry was carried out with circular apertures (MAG\_APER in SExtractor) and automatically adjusted elliptical apertures (MAG\_AUTO in SExtractor). The algorithm repeats the light curve extraction many times with different aperture sizes to determine the aperture that minimises the scatter in the differential target light curve. In the majority of our observations the circular aperture with fixed radius (MAG\_APER) yielded a more stable photometry than the flexible elliptical aperture (MAG\_AUTO).

We performed differential photometry using the flux sum of multiple stars as a reference light curve. The pipeline looks for a combination of comparison stars which minimises the point-to-point scatter (root mean square, rms) of the target light curve. The comparison stars are weighted according to their individual light curve quality in terms of rms \citep{broeg_2005}. The rms of the target light curve is computed not only on the out-of-transit data but also on the residuals of the entire time series after subtracting a transit fit including a first-order polynomial in time for detrending.

\subsection{Long-term photometry}
\begin{figure}[t]
        % flatex input: [./images/activity/photo.tex]
% GNUPLOT: LaTeX picture with Postscript
\begingroup
  \makeatletter
  \providecommand\color[2][]{%
    \GenericError{(gnuplot) \space\space\space\@spaces}{%
      Package color not loaded in conjunction with
      terminal option `colourtext'%
    }{See the gnuplot documentation for explanation.%
    }{Either use 'blacktext' in gnuplot or load the package
      color.sty in LaTeX.}%
    \renewcommand\color[2][]{}%
  }%
  \providecommand\includegraphics[2][]{%
    \GenericError{(gnuplot) \space\space\space\@spaces}{%
      Package graphicx or graphics not loaded%
    }{See the gnuplot documentation for explanation.%
    }{The gnuplot epslatex terminal needs graphicx.sty or graphics.sty.}%
    \renewcommand\includegraphics[2][]{}%
  }%
  \providecommand\rotatebox[2]{#2}%
  \@ifundefined{ifGPcolor}{%
    \newif\ifGPcolor
    \GPcolortrue
  }{}%
  \@ifundefined{ifGPblacktext}{%
    \newif\ifGPblacktext
    \GPblacktexttrue
  }{}%
  % define a \g@addto@macro without @ in the name:
  \let\gplgaddtomacro\g@addto@macro
  % define empty templates for all commands taking text:
  \gdef\gplbacktext{}%
  \gdef\gplfronttext{}%
  \makeatother
  \ifGPblacktext
    % no textcolor at all
    \def\colorrgb#1{}%
    \def\colorgray#1{}%
  \else
    % gray or color?
    \ifGPcolor
      \def\colorrgb#1{\color[rgb]{#1}}%
      \def\colorgray#1{\color[gray]{#1}}%
      \expandafter\def\csname LTw\endcsname{\color{white}}%
      \expandafter\def\csname LTb\endcsname{\color{black}}%
      \expandafter\def\csname LTa\endcsname{\color{black}}%
      \expandafter\def\csname LT0\endcsname{\color[rgb]{1,0,0}}%
      \expandafter\def\csname LT1\endcsname{\color[rgb]{0,1,0}}%
      \expandafter\def\csname LT2\endcsname{\color[rgb]{0,0,1}}%
      \expandafter\def\csname LT3\endcsname{\color[rgb]{1,0,1}}%
      \expandafter\def\csname LT4\endcsname{\color[rgb]{0,1,1}}%
      \expandafter\def\csname LT5\endcsname{\color[rgb]{1,1,0}}%
      \expandafter\def\csname LT6\endcsname{\color[rgb]{0,0,0}}%
      \expandafter\def\csname LT7\endcsname{\color[rgb]{1,0.3,0}}%
      \expandafter\def\csname LT8\endcsname{\color[rgb]{0.5,0.5,0.5}}%
    \else
      % gray
      \def\colorrgb#1{\color{black}}%
      \def\colorgray#1{\color[gray]{#1}}%
      \expandafter\def\csname LTw\endcsname{\color{white}}%
      \expandafter\def\csname LTb\endcsname{\color{black}}%
      \expandafter\def\csname LTa\endcsname{\color{black}}%
      \expandafter\def\csname LT0\endcsname{\color{black}}%
      \expandafter\def\csname LT1\endcsname{\color{black}}%
      \expandafter\def\csname LT2\endcsname{\color{black}}%
      \expandafter\def\csname LT3\endcsname{\color{black}}%
      \expandafter\def\csname LT4\endcsname{\color{black}}%
      \expandafter\def\csname LT5\endcsname{\color{black}}%
      \expandafter\def\csname LT6\endcsname{\color{black}}%
      \expandafter\def\csname LT7\endcsname{\color{black}}%
      \expandafter\def\csname LT8\endcsname{\color{black}}%
    \fi
  \fi
    \setlength{\unitlength}{0.0500bp}%
    \ifx\gptboxheight\undefined%
      \newlength{\gptboxheight}%
      \newlength{\gptboxwidth}%
      \newsavebox{\gptboxtext}%
    \fi%
    \setlength{\fboxrule}{0.5pt}%
    \setlength{\fboxsep}{1pt}%
\begin{picture}(4988.00,3458.00)%
    \gplgaddtomacro\gplbacktext{%
      \csname LTb\endcsname%
      \put(915,2152){\makebox(0,0)[r]{\strut{}-0.015}}%
      \put(915,2334){\makebox(0,0)[r]{\strut{}-0.010}}%
      \put(915,2515){\makebox(0,0)[r]{\strut{}-0.005}}%
      \put(915,2697){\makebox(0,0)[r]{\strut{}0.000}}%
      \put(915,2878){\makebox(0,0)[r]{\strut{}0.005}}%
      \put(915,3059){\makebox(0,0)[r]{\strut{}0.010}}%
      \put(915,3241){\makebox(0,0)[r]{\strut{}0.015}}%
      \put(1047,1751){\makebox(0,0){\strut{}}}%
      \put(1533,1751){\makebox(0,0){\strut{}}}%
      \put(2020,1751){\makebox(0,0){\strut{}}}%
      \put(2506,1751){\makebox(0,0){\strut{}}}%
      \put(2992,1751){\makebox(0,0){\strut{}}}%
      \put(3478,1751){\makebox(0,0){\strut{}}}%
      \put(3965,1751){\makebox(0,0){\strut{}}}%
      \put(4451,1751){\makebox(0,0){\strut{}}}%
      \put(4937,1751){\makebox(0,0){\strut{}}}%
    }%
    \gplgaddtomacro\gplfronttext{%
      \csname LTb\endcsname%
      \put(145,2696){\rotatebox{-270}{\makebox(0,0){\strut{}$\Delta m_B$}}}%
    }%
    \gplgaddtomacro\gplbacktext{%
      \csname LTb\endcsname%
      \put(915,670){\makebox(0,0)[r]{\strut{}-0.015}}%
      \put(915,856){\makebox(0,0)[r]{\strut{}-0.010}}%
      \put(915,1041){\makebox(0,0)[r]{\strut{}-0.005}}%
      \put(915,1227){\makebox(0,0)[r]{\strut{}0.000}}%
      \put(915,1413){\makebox(0,0)[r]{\strut{}0.005}}%
      \put(915,1599){\makebox(0,0)[r]{\strut{}0.010}}%
      \put(915,1784){\makebox(0,0)[r]{\strut{}0.015}}%
      \put(1047,264){\makebox(0,0){\strut{}$440$}}%
      \put(1533,264){\makebox(0,0){\strut{}$460$}}%
      \put(2020,264){\makebox(0,0){\strut{}$480$}}%
      \put(2506,264){\makebox(0,0){\strut{}$500$}}%
      \put(2992,264){\makebox(0,0){\strut{}$520$}}%
      \put(3478,264){\makebox(0,0){\strut{}$540$}}%
      \put(3965,264){\makebox(0,0){\strut{}$560$}}%
      \put(4451,264){\makebox(0,0){\strut{}$580$}}%
      \put(4937,264){\makebox(0,0){\strut{}$600$}}%
    }%
    \gplgaddtomacro\gplfronttext{%
      \csname LTb\endcsname%
      \put(145,1227){\rotatebox{-270}{\makebox(0,0){\strut{}$\Delta m_V$}}}%
      \put(2992,-66){\makebox(0,0){\strut{}JD - 2457000}}%
    }%
    \gplbacktext
    \put(0,0){\includegraphics{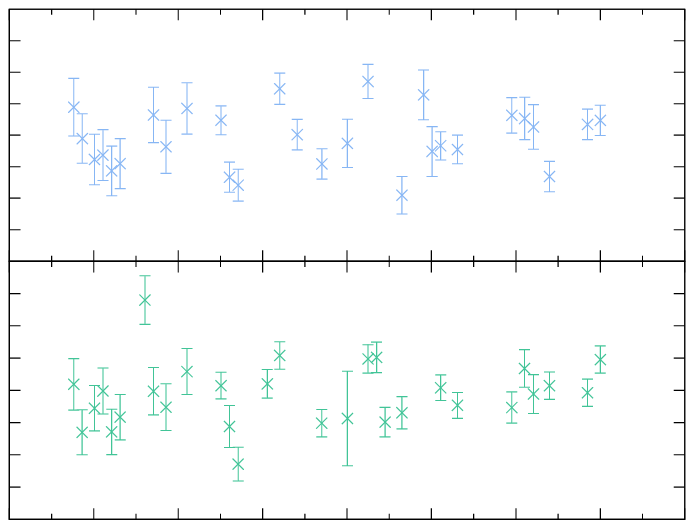}}%
    \gplfronttext
  \end{picture}%
\endgroup

% flatex input end: [./images/activity/photo.tex]
 
        \caption{Monitoring light curve of the host star TrES-3 using STELLA/WiFSIP in Johnson B (upper panel) and V (bottom panel). Vertical error-bars indicate the     $ \unit[1]{\sigma} $ uncertainty.}
        \label{fig:active-photo}
        \vspace{10pt}
%\end{figure}
%\begin{figure}[t]
        % flatex input: [./images/activity/period.tex]
% GNUPLOT: LaTeX picture with Postscript
\begingroup
  \makeatletter
  \providecommand\color[2][]{%
    \GenericError{(gnuplot) \space\space\space\@spaces}{%
      Package color not loaded in conjunction with
      terminal option `colourtext'%
    }{See the gnuplot documentation for explanation.%
    }{Either use 'blacktext' in gnuplot or load the package
      color.sty in LaTeX.}%
    \renewcommand\color[2][]{}%
  }%
  \providecommand\includegraphics[2][]{%
    \GenericError{(gnuplot) \space\space\space\@spaces}{%
      Package graphicx or graphics not loaded%
    }{See the gnuplot documentation for explanation.%
    }{The gnuplot epslatex terminal needs graphicx.sty or graphics.sty.}%
    \renewcommand\includegraphics[2][]{}%
  }%
  \providecommand\rotatebox[2]{#2}%
  \@ifundefined{ifGPcolor}{%
    \newif\ifGPcolor
    \GPcolortrue
  }{}%
  \@ifundefined{ifGPblacktext}{%
    \newif\ifGPblacktext
    \GPblacktexttrue
  }{}%
  % define a \g@addto@macro without @ in the name:
  \let\gplgaddtomacro\g@addto@macro
  % define empty templates for all commands taking text:
  \gdef\gplbacktext{}%
  \gdef\gplfronttext{}%
  \makeatother
  \ifGPblacktext
    % no textcolor at all
    \def\colorrgb#1{}%
    \def\colorgray#1{}%
  \else
    % gray or color?
    \ifGPcolor
      \def\colorrgb#1{\color[rgb]{#1}}%
      \def\colorgray#1{\color[gray]{#1}}%
      \expandafter\def\csname LTw\endcsname{\color{white}}%
      \expandafter\def\csname LTb\endcsname{\color{black}}%
      \expandafter\def\csname LTa\endcsname{\color{black}}%
      \expandafter\def\csname LT0\endcsname{\color[rgb]{1,0,0}}%
      \expandafter\def\csname LT1\endcsname{\color[rgb]{0,1,0}}%
      \expandafter\def\csname LT2\endcsname{\color[rgb]{0,0,1}}%
      \expandafter\def\csname LT3\endcsname{\color[rgb]{1,0,1}}%
      \expandafter\def\csname LT4\endcsname{\color[rgb]{0,1,1}}%
      \expandafter\def\csname LT5\endcsname{\color[rgb]{1,1,0}}%
      \expandafter\def\csname LT6\endcsname{\color[rgb]{0,0,0}}%
      \expandafter\def\csname LT7\endcsname{\color[rgb]{1,0.3,0}}%
      \expandafter\def\csname LT8\endcsname{\color[rgb]{0.5,0.5,0.5}}%
    \else
      % gray
      \def\colorrgb#1{\color{black}}%
      \def\colorgray#1{\color[gray]{#1}}%
      \expandafter\def\csname LTw\endcsname{\color{white}}%
      \expandafter\def\csname LTb\endcsname{\color{black}}%
      \expandafter\def\csname LTa\endcsname{\color{black}}%
      \expandafter\def\csname LT0\endcsname{\color{black}}%
      \expandafter\def\csname LT1\endcsname{\color{black}}%
      \expandafter\def\csname LT2\endcsname{\color{black}}%
      \expandafter\def\csname LT3\endcsname{\color{black}}%
      \expandafter\def\csname LT4\endcsname{\color{black}}%
      \expandafter\def\csname LT5\endcsname{\color{black}}%
      \expandafter\def\csname LT6\endcsname{\color{black}}%
      \expandafter\def\csname LT7\endcsname{\color{black}}%
      \expandafter\def\csname LT8\endcsname{\color{black}}%
    \fi
  \fi
    \setlength{\unitlength}{0.0500bp}%
    \ifx\gptboxheight\undefined%
      \newlength{\gptboxheight}%
      \newlength{\gptboxwidth}%
      \newsavebox{\gptboxtext}%
    \fi%
    \setlength{\fboxrule}{0.5pt}%
    \setlength{\fboxsep}{1pt}%
\begin{picture}(4988.00,3458.00)%
    \gplgaddtomacro\gplbacktext{%
      \csname LTb\endcsname%
      \put(516,1971){\makebox(0,0)[r]{\strut{}$0$}}%
      \put(516,2261){\makebox(0,0)[r]{\strut{}$2$}}%
      \put(516,2551){\makebox(0,0)[r]{\strut{}$4$}}%
      \put(516,2842){\makebox(0,0)[r]{\strut{}$6$}}%
      \put(516,3132){\makebox(0,0)[r]{\strut{}$8$}}%
      \put(648,1751){\makebox(0,0){\strut{}}}%
      \put(1077,1751){\makebox(0,0){\strut{}}}%
      \put(1506,1751){\makebox(0,0){\strut{}}}%
      \put(1935,1751){\makebox(0,0){\strut{}}}%
      \put(2364,1751){\makebox(0,0){\strut{}}}%
      \put(2793,1751){\makebox(0,0){\strut{}}}%
      \put(3221,1751){\makebox(0,0){\strut{}}}%
      \put(3650,1751){\makebox(0,0){\strut{}}}%
      \put(4079,1751){\makebox(0,0){\strut{}}}%
      \put(4508,1751){\makebox(0,0){\strut{}}}%
      \put(4937,1751){\makebox(0,0){\strut{}}}%
      \put(4079,3204){\makebox(0,0)[l]{\strut{}B band}}%
      \put(2578,2829){\makebox(0,0)[l]{\strut{}FAP = 0.1}}%
      \put(2578,3174){\makebox(0,0)[l]{\strut{}FAP = 0.01}}%
    }%
    \gplgaddtomacro\gplfronttext{%
      \csname LTb\endcsname%
      \put(169,2696){\rotatebox{-270}{\makebox(0,0){\strut{}$\unit[power]{/a.u.}$}}}%
    }%
    \gplgaddtomacro\gplbacktext{%
      \csname LTb\endcsname%
      \put(516,484){\makebox(0,0)[r]{\strut{}$0$}}%
      \put(516,781){\makebox(0,0)[r]{\strut{}$2$}}%
      \put(516,1078){\makebox(0,0)[r]{\strut{}$4$}}%
      \put(516,1376){\makebox(0,0)[r]{\strut{}$6$}}%
      \put(516,1673){\makebox(0,0)[r]{\strut{}$8$}}%
      \put(648,264){\makebox(0,0){\strut{}$0$}}%
      \put(1077,264){\makebox(0,0){\strut{}$0.1$}}%
      \put(1506,264){\makebox(0,0){\strut{}$0.2$}}%
      \put(1935,264){\makebox(0,0){\strut{}$0.3$}}%
      \put(2364,264){\makebox(0,0){\strut{}$0.4$}}%
      \put(2793,264){\makebox(0,0){\strut{}$0.5$}}%
      \put(3221,264){\makebox(0,0){\strut{}$0.6$}}%
      \put(3650,264){\makebox(0,0){\strut{}$0.7$}}%
      \put(4079,264){\makebox(0,0){\strut{}$0.8$}}%
      \put(4508,264){\makebox(0,0){\strut{}$0.9$}}%
      \put(4937,264){\makebox(0,0){\strut{}$1$}}%
      \put(4079,1747){\makebox(0,0)[l]{\strut{}V band}}%
    }%
    \gplgaddtomacro\gplfronttext{%
      \csname LTb\endcsname%
      \put(169,1227){\rotatebox{-270}{\makebox(0,0){\strut{}$\unit[power]{/a.u.}$}}}%
      \put(2792,-66){\makebox(0,0){\strut{}$\unit[f]{/day^{-1}}$}}%
    }%
    \gplbacktext
    \put(0,0){\includegraphics{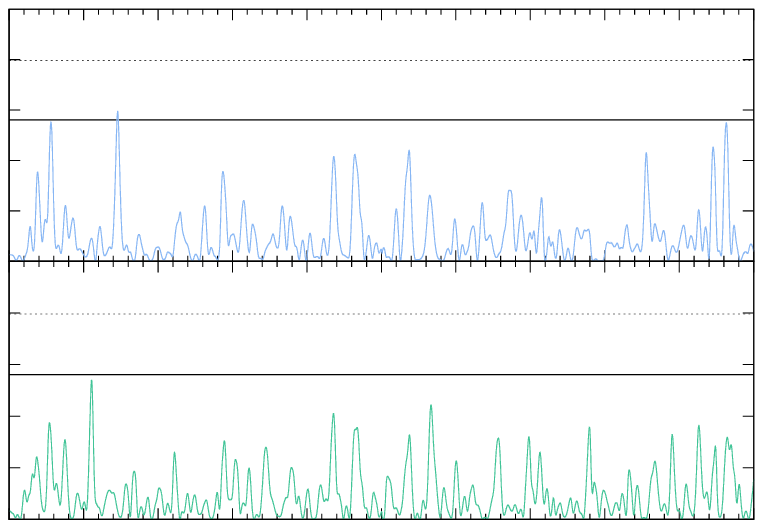}}%
    \gplfronttext
  \end{picture}%
\endgroup

% flatex input end: [./images/activity/period.tex]

%\begin{figure}[t]
        \caption{Lomb-Scargle periodogram of the monitoring light curves in B and V. The horizontal lines indicate the false alarm probability (FAP) of 0.1 and 0.01.}
        \label{fig:active-period}
\end{figure}

The bias- and flatfield correction was done with the STELLA data-reduction pipeline (Granzer, in preparation). We conducted aperture photometry with SExtractor applying the MAG\_AUTO option to automatically adjust elliptical apertures. This option provides the flexibility to account for the varying observing conditions from night to night. We performed differential photometry using the flux sum of multiple stars as a reference light curve. The choice of the comparison stars did not significantly affect the Lomb-Scargle periodograms in the following section. Out of our monitoring data from 43 days, several data points had to be discarded because of instrumental problems or low quality caused by a nearby bright moon. Therefore, we could finally make use of 28 data points in Johnson \textit{B} and 29 data points in Johnson \textit{V} after averaging the three exposures per night (Figure~\ref{fig:active-photo}). The rms in both light curves is about $ \unit[4.8]{mmag} $, which is slightly higher than the typical value of about $ 2.5 \text{ to } \unit[3]{mmag} $ achieved in similar monitoring campaigns for other exoplanet host stars with the same instrument \citep{mallonn_2016a,mallonn_2015a}. 

% flatex input end: [reduction.tex]

% flatex input: [analysis.tex]
\section{Analysis} \label{sec:analysis}
For the analysis of the light curves we used the publicly available software JKTEBOP in version 34 \citep{southworth_2004}\footnote{\url{http://www.astro.keele.ac.uk/jkt/codes/jktebop.html}}. This code is written in FORTRAN 77 and was originally used to fit models to the light curves of detached eclipsing binary stars but it is also an excellent code for analysing transiting extrasolar systems \citep{southworth_2008}. A Levenberg-Marquardt optimisation algorithm is used to find the best-fitting model. The transit fit parameters consist of the scaled planetary and stellar radius $ r_P = R_P/a $ and $ r_* = R_* / a $, the orbital inclination $ i $, the transit midtime $ T_0 $, the orbital Period $ P $ and the host star limb darkening coefficients, as well as an Nth-order polynomial detrending of the light curve as a function of time, with $ N \in \left\{ 1,2 \right\} $. The time stamps from all light curves analysed in this work were transferred to BJD$_{\rm TDB}$ following the recommendation of \citet{eastman_2010}.

\subsection{Orbital parameters}
In our analysis of the wavelength dependence of the planet-star radius ratio $ R_p/R_* $ we fixed the parameter-scaled stellar radius $r_* = R_*/a$, orbital inclination $ i $, orbital period $ P, $ and orbital eccentricity $ e $ to common values because they are not wavelength dependent. We fixed the eccentricity to zero \citep{sozzetti_2009}. The orbital period is kept to $ P = \unit[1.30618608]{days} $ \citep{christiansen_2011}. To find values for $ r_* $ and $ i $ we averaged the literature values found by \citet{christiansen_2011,jiang_2013,odonovan_2007,sozzetti_2009} and \citet{vanko_2013} with a weighted mean. We discarded the values by \citet{turner_2013} because of their large uncertainties compared to other studies. The individual values and averages are listed in Table~\ref{tab:param-ave}.

We used these averages as initial values for a white light curve fit of the GTC/OSIRIS spectra. The best-fit transit parameters agree very well to the averaged values and we used the latter one for our subsequent analysis. Using the average ensures suppression of observational systematic errors caused by using many different observations from divers telescopes.
% flatex input: [tab-param-ave.tex]
\begin{table}[t]
\caption{Literature values for stellar radius and inclination used for the weighted mean and the result of the white light curve analysis.} 
\label{tab:param-ave}
\centering
\begin{tabular}{cccc}
        \toprule
        Ref. & $ r_* $ & $ \unit[i]{/^{\circ}} $ & Date of Obs.\\
        \midrule
        1 & $ 0.1664 \pm 0.0204 $ &$ 81.99 \pm 0.30 $ & 2008-03-06 \\
        2 &  &$ 81.73 \pm 0.13 $ &  \\
        3 & $ 0.1685 \pm 0.0013 $ &$ 81.81 \pm 0.15 $ & 2010-05-25 \\
        3 & $ 0.1679 \pm 0.0015 $ &$ 81.75 \pm 0.15 $ & 2010-06-11 \\
        3 & $ 0.1684 \pm 0.0015 $ &$ 81.83 \pm 0.15 $ & 2010-06-15 \\
        3 & $ 0.1670 \pm 0.0015 $ &$ 81.95 \pm 0.15 $ & 2010-06-19 \\
        3 & $ 0.1691 \pm 0.0015 $ &$ 81.89 \pm 0.15 $ & 2010-06-28 \\
        4 & $ 0.1675 \pm 0.0009 $ &$ 81.95 \pm 0.06 $ & 2009-05-14 \\
        5 & $ 0.165 \pm 0.003 $ &$ 82.15 \pm 0.21 $ & 2007-03-26 \\
        6 & $ 0.1687 \pm 0.0016 $ &$ 81.85 \pm 0.16 $ & 2007-04-24 \\
        7 & $ 0.1682 \pm 0.0032 $ &$ 81.86 \pm 0.28 $ & 2010-09-06 \\
        7 & $ 0.1696 \pm 0.0027 $ &$ 81.76 \pm 0.15 $ & 2010-09-06 \\   
        \midrule
        average & $ 0.1671 \pm 0.0013 $ & $ 81.89 \pm 0.12 $ & \\
        white lc & $ 0.167 \pm 0.002 $ & $ 81.88 \pm 0.18 $ & \\
        \bottomrule
\end{tabular}

\tablebib{
        (1)~\citet{christiansen_2011}; (2)~\citet{gibson_2009}; (3)~\citet{jiang_2013}; (4)~\citet{kundurthy_2013}; (5)~\citet{odonovan_2007}; (6)~\citet{sozzetti_2009}; (7)~\citet{vanko_2013}
}
\end{table}
% flatex input end: [tab-param-ave.tex]

\subsection{The stellar limb darkening law}
The stellar limb darkening (LD) is a second-order effect modifying the transit shape but it is important for measuring the radius of the planet accurately.
In the case of an almost grazing transit, as for TrES-3\,b, the transit chord probes the stellar limb, but not the limb to centre brightness variations. In this case, the transit light curve is not suited for fitting the LD coefficients (LDCs) \citep{mueller_2013}. Instead, we rely on theoretical calculations of the LDCs. In this work, we use the LDCs published by \citet{claret_2013} and the non-linear
four-parameter law 
introduced by \citet{claret_2000}, which provides the most accurate description of the LD compared to other LD laws \citep{sing_2010}. Fixing the LDCs leads to biased values for the radius ratio of less than 0.5 per cent in the case of TrES-3b \citep{espinoza_2015}. This is smaller to the uncertainties we achieve in this work. To obtain the LDCs we run a linear interpolation from the tabulated values for the host star parameters $ T_{eff} = \unit[5650]{K} $ and $ \log g = 4.568 $ and average each coefficient over the wavelength range of the band. The obtained LDCs are listed in Table~\ref{tab:ldc}. 

% flatex input: [tab-ldc.tex]
\begin{table}[t]
\caption{Non-linear limb darkening coefficients after \citet{claret_2000} for the broad band and GTC/OSIRIS light curves used in this work.} 
\label{tab:ldc}
\centering
\begin{tabular}{ccccc}
        \toprule
        Band & $ c_1 $ & $ c_2 $ & $ c_3 $ & $ c_4 $\\
        \midrule
        \textit{U} & 1.1981 & -2.2854 & 3.1390 & -1.0637 \\
        \textit{B} & 0.6263 & -0.7628 & 1.8656 & -0.8232 \\
        \textit{V} & 0.6943 & -0.5840 & 1.3565 & -0.6166 \\
        \textit{RISE} & 0.8347 & -0.7900 & 1.3819 & -0.5936 \\
        \textit{r'} & 0.7700 & -0.6140 & 1.1863 & -0.5260 \\
        \textit{R} & 0.7910 & -0.6540 & 1.1616 & -0.5009 \\
        \textit{792} & 0.8633 & -0.7500 & 1.0327 & -0.4200 \\
        \textit{I} & 0.7865 & -0.5898 & 0.9052 & -0.3824 \\
        \textit{z'} & 0.7621 & -0.5362 & 0.7808 & -0.3306 \\
        \midrule
        $ \unit[5525]{\AA} $ & 0.8209 & -0.8307 & 1.5294 & -0.6622 \\
        $ \unit[5775]{\AA} $ & 0.8173 & -0.7658 & 1.4133 & -0.6158 \\
        $ \unit[6025]{\AA} $ & 0.8200 & -0.7317 & 1.3399 & -0.5903 \\
        $ \unit[6275]{\AA} $ & 0.8616 & -0.8024 & 1.3246 & -0.5644 \\
        $ \unit[6525]{\AA} $ & 0.8485 & -0.7003 & 1.1426 & -0.4900 \\
        $ \unit[6775]{\AA} $ & 0.8581 & -0.7636 & 1.2104 & -0.5127 \\
        $ \unit[7025]{\AA} $ & 0.8568 & -0.7586 & 1.1734 & -0.4929 \\
        $ \unit[7275]{\AA} $ & 0.8666 & -0.7795 & 1.1452 & -0.4717 \\
        $ \unit[7525]{\AA} $ & 0.8567 & -0.7518 & 1.0977 & -0.4544 \\
        $ \unit[7775]{\AA} $ & 0.8554 & -0.7359 & 1.0502 & -0.4336 \\
        $ \unit[8025]{\AA} $ & 0.8434 & -0.7115 & 1.0104 & -0.4172 \\
        $ \unit[8275]{\AA} $ & 0.8436 & -0.7208 & 0.9910 & -0.4042 \\
        $ \unit[8525]{\AA} $ & 0.8502 & -0.7392 & 0.9671 & -0.3865 \\
        $ \unit[8775]{\AA} $ & 0.8271 & -0.6750 & 0.8900 & -0.3604 \\
        $ \unit[9025]{\AA} $ & 0.8305 & -0.6903 & 0.8990 & -0.3636 \\
        $ \unit[9275]{\AA} $ & 0.8102 & -0.6633 & 0.8832 & -0.3610 \\
        \bottomrule
\end{tabular}
\end{table}
% flatex input end: [tab-ldc.tex]

\subsection{Third light contamination}
Light contamination by unresolved background stars or stellar companions can mimic slopes in the final transmission spectrum of the exoplanet. \citet{ngo_2015} found no companion for TrES-3 within 10 arcsec using the diffraction-limited direct imaging method. 

Due to the slit alignment during the GTC observation a faint star (2MASS 17520839+3732378, $ J = 15.23 $, $ K = 14.42 $) at a distance of about 18 arcsec is projected into the aperture of TrES-3. This star is redder than TrES-3 (2MASS: $ J-K = 0.81 $, TrES-3: $ J-K = 0.407 $) and thus contributes about 1\% of the flux to the red edge of the GTC/OSIRIS spectrum. The estimated modification in the radius ratio is therefore 0.0009. We neglect this contribution because it is smaller than our final uncertainties in the spectrum. Furthermore, the unaffected broad band measurements are in agreement with the GTC data which supports our strategy.

\subsection{Light curve fitting}
To determine the planetary radius, several model-fitting runs were performed for each individual light curve. The photometric uncertainties determined by SExtractor yield a reduced $ \chi^2 $ that is always slightly greater than unity, which indicates underestimated photometric uncertainties. We enlarged the error bars for each light curve by a common factor to produce $ \chi^2 = 1 $ \citep{mallonn_2015b}. We run a first fit to compute the so-called $ \beta $-factor, introduced by \citet{winn_2008b}, which takes systematic noise in the light curves into account \citep{pont_2006}. For the best-fitting model of each individual light curve we calculated the standard deviation of the unbinned residuals $ \sigma_1 $. These residuals were binned into $ M $ bins, each with $ N $ points and the standard deviation was calculated again for the binned residuals $ \sigma_N $. For uncorrelated (white) noise we would expect
\begin{align}
        \sigma_N = \frac{\sigma_1}{\sqrt{N}} \sqrt{\frac{M}{M-1}}.
\end{align}
In the presence of correlated (red) noise, $ \sigma_N $ is larger than expected by a factor $ \beta $. There is a dependence on the bin size and the relevant time scale for transit photometry which is the duration of ingress/egress \citep{winn_2008b}. We binned up to 24 minutes for a total transit duration of about 82 minutes and applied the $ \beta $-factor to the photometric uncertainties if $ \beta > 1 $.
\begin{figure}[tp]
        % flatex input: [./images/16bins_matthias/lc.tex]
% GNUPLOT: LaTeX picture with Postscript
\begingroup
  \makeatletter
  \providecommand\color[2][]{%
    \GenericError{(gnuplot) \space\space\space\@spaces}{%
      Package color not loaded in conjunction with
      terminal option `colourtext'%
    }{See the gnuplot documentation for explanation.%
    }{Either use 'blacktext' in gnuplot or load the package
      color.sty in LaTeX.}%
    \renewcommand\color[2][]{}%
  }%
  \providecommand\includegraphics[2][]{%
    \GenericError{(gnuplot) \space\space\space\@spaces}{%
      Package graphicx or graphics not loaded%
    }{See the gnuplot documentation for explanation.%
    }{The gnuplot epslatex terminal needs graphicx.sty or graphics.sty.}%
    \renewcommand\includegraphics[2][]{}%
  }%
  \providecommand\rotatebox[2]{#2}%
  \@ifundefined{ifGPcolor}{%
    \newif\ifGPcolor
    \GPcolortrue
  }{}%
  \@ifundefined{ifGPblacktext}{%
    \newif\ifGPblacktext
    \GPblacktexttrue
  }{}%
  % define a \g@addto@macro without @ in the name:
  \let\gplgaddtomacro\g@addto@macro
  % define empty templates for all commands taking text:
  \gdef\gplbacktext{}%
  \gdef\gplfronttext{}%
  \makeatother
  \ifGPblacktext
    % no textcolor at all
    \def\colorrgb#1{}%
    \def\colorgray#1{}%
  \else
    % gray or color?
    \ifGPcolor
      \def\colorrgb#1{\color[rgb]{#1}}%
      \def\colorgray#1{\color[gray]{#1}}%
      \expandafter\def\csname LTw\endcsname{\color{white}}%
      \expandafter\def\csname LTb\endcsname{\color{black}}%
      \expandafter\def\csname LTa\endcsname{\color{black}}%
      \expandafter\def\csname LT0\endcsname{\color[rgb]{1,0,0}}%
      \expandafter\def\csname LT1\endcsname{\color[rgb]{0,1,0}}%
      \expandafter\def\csname LT2\endcsname{\color[rgb]{0,0,1}}%
      \expandafter\def\csname LT3\endcsname{\color[rgb]{1,0,1}}%
      \expandafter\def\csname LT4\endcsname{\color[rgb]{0,1,1}}%
      \expandafter\def\csname LT5\endcsname{\color[rgb]{1,1,0}}%
      \expandafter\def\csname LT6\endcsname{\color[rgb]{0,0,0}}%
      \expandafter\def\csname LT7\endcsname{\color[rgb]{1,0.3,0}}%
      \expandafter\def\csname LT8\endcsname{\color[rgb]{0.5,0.5,0.5}}%
    \else
      % gray
      \def\colorrgb#1{\color{black}}%
      \def\colorgray#1{\color[gray]{#1}}%
      \expandafter\def\csname LTw\endcsname{\color{white}}%
      \expandafter\def\csname LTb\endcsname{\color{black}}%
      \expandafter\def\csname LTa\endcsname{\color{black}}%
      \expandafter\def\csname LT0\endcsname{\color{black}}%
      \expandafter\def\csname LT1\endcsname{\color{black}}%
      \expandafter\def\csname LT2\endcsname{\color{black}}%
      \expandafter\def\csname LT3\endcsname{\color{black}}%
      \expandafter\def\csname LT4\endcsname{\color{black}}%
      \expandafter\def\csname LT5\endcsname{\color{black}}%
      \expandafter\def\csname LT6\endcsname{\color{black}}%
      \expandafter\def\csname LT7\endcsname{\color{black}}%
      \expandafter\def\csname LT8\endcsname{\color{black}}%
    \fi
  \fi
    \setlength{\unitlength}{0.0500bp}%
    \ifx\gptboxheight\undefined%
      \newlength{\gptboxheight}%
      \newlength{\gptboxwidth}%
      \newsavebox{\gptboxtext}%
    \fi%
    \setlength{\fboxrule}{0.5pt}%
    \setlength{\fboxsep}{1pt}%
\begin{picture}(4988.00,9534.00)%
    \gplgaddtomacro\gplbacktext{%
      \csname LTb\endcsname%
      \put(815,9367){\makebox(0,0)[r]{\strut{}$-0.01$}}%
      \put(815,9054){\makebox(0,0)[r]{\strut{}$0$}}%
      \put(815,8741){\makebox(0,0)[r]{\strut{}$0.01$}}%
      \put(815,8428){\makebox(0,0)[r]{\strut{}$0.02$}}%
      \put(815,8115){\makebox(0,0)[r]{\strut{}$0.03$}}%
      \put(1146,7645){\makebox(0,0){\strut{}}}%
      \put(1446,7645){\makebox(0,0){\strut{}}}%
      \put(1745,7645){\makebox(0,0){\strut{}}}%
      \put(1027,7998){\makebox(0,0)[l]{\strut{}5525}}%
    }%
    \gplgaddtomacro\gplfronttext{%
      \csname LTb\endcsname%
      \put(150,8694){\rotatebox{-270}{\makebox(0,0){\strut{}$\Delta m$}}}%
    }%
    \gplgaddtomacro\gplbacktext{%
      \csname LTb\endcsname%
      \put(815,7715){\makebox(0,0)[r]{\strut{}$-0.005$}}%
      \put(815,7567){\makebox(0,0)[r]{\strut{}$0$}}%
      \put(815,7418){\makebox(0,0)[r]{\strut{}$0.005$}}%
      \put(1146,7049){\makebox(0,0){\strut{}}}%
      \put(1446,7049){\makebox(0,0){\strut{}}}%
      \put(1745,7049){\makebox(0,0){\strut{}}}%
    }%
    \gplgaddtomacro\gplfronttext{%
      \csname LTb\endcsname%
      \put(150,7566){\rotatebox{-270}{\makebox(0,0){\strut{}O-C}}}%
    }%
    \gplgaddtomacro\gplbacktext{%
      \csname LTb\endcsname%
      \put(1813,9367){\makebox(0,0)[r]{\strut{}}}%
      \put(1813,9054){\makebox(0,0)[r]{\strut{}}}%
      \put(1813,8741){\makebox(0,0)[r]{\strut{}}}%
      \put(1813,8428){\makebox(0,0)[r]{\strut{}}}%
      \put(1813,8115){\makebox(0,0)[r]{\strut{}}}%
      \put(2144,7645){\makebox(0,0){\strut{}}}%
      \put(2444,7645){\makebox(0,0){\strut{}}}%
      \put(2743,7645){\makebox(0,0){\strut{}}}%
      \put(2025,7998){\makebox(0,0)[l]{\strut{}5775}}%
    }%
    \gplgaddtomacro\gplfronttext{%
    }%
    \gplgaddtomacro\gplbacktext{%
      \csname LTb\endcsname%
      \put(1813,7715){\makebox(0,0)[r]{\strut{}}}%
      \put(1813,7567){\makebox(0,0)[r]{\strut{}}}%
      \put(1813,7418){\makebox(0,0)[r]{\strut{}}}%
      \put(2144,7049){\makebox(0,0){\strut{}}}%
      \put(2444,7049){\makebox(0,0){\strut{}}}%
      \put(2743,7049){\makebox(0,0){\strut{}}}%
    }%
    \gplgaddtomacro\gplfronttext{%
    }%
    \gplgaddtomacro\gplbacktext{%
      \csname LTb\endcsname%
      \put(2810,9367){\makebox(0,0)[r]{\strut{}}}%
      \put(2810,9054){\makebox(0,0)[r]{\strut{}}}%
      \put(2810,8741){\makebox(0,0)[r]{\strut{}}}%
      \put(2810,8428){\makebox(0,0)[r]{\strut{}}}%
      \put(2810,8115){\makebox(0,0)[r]{\strut{}}}%
      \put(3141,7645){\makebox(0,0){\strut{}}}%
      \put(3441,7645){\makebox(0,0){\strut{}}}%
      \put(3740,7645){\makebox(0,0){\strut{}}}%
      \put(3022,7998){\makebox(0,0)[l]{\strut{}6025}}%
    }%
    \gplgaddtomacro\gplfronttext{%
    }%
    \gplgaddtomacro\gplbacktext{%
      \csname LTb\endcsname%
      \put(2810,7715){\makebox(0,0)[r]{\strut{}}}%
      \put(2810,7567){\makebox(0,0)[r]{\strut{}}}%
      \put(2810,7418){\makebox(0,0)[r]{\strut{}}}%
      \put(3141,7049){\makebox(0,0){\strut{}}}%
      \put(3441,7049){\makebox(0,0){\strut{}}}%
      \put(3740,7049){\makebox(0,0){\strut{}}}%
    }%
    \gplgaddtomacro\gplfronttext{%
    }%
    \gplgaddtomacro\gplbacktext{%
      \csname LTb\endcsname%
      \put(3808,9367){\makebox(0,0)[r]{\strut{}}}%
      \put(3808,9054){\makebox(0,0)[r]{\strut{}}}%
      \put(3808,8741){\makebox(0,0)[r]{\strut{}}}%
      \put(3808,8428){\makebox(0,0)[r]{\strut{}}}%
      \put(3808,8115){\makebox(0,0)[r]{\strut{}}}%
      \put(4139,7645){\makebox(0,0){\strut{}}}%
      \put(4439,7645){\makebox(0,0){\strut{}}}%
      \put(4738,7645){\makebox(0,0){\strut{}}}%
      \put(4020,7998){\makebox(0,0)[l]{\strut{}6275}}%
    }%
    \gplgaddtomacro\gplfronttext{%
    }%
    \gplgaddtomacro\gplbacktext{%
      \csname LTb\endcsname%
      \put(3808,7715){\makebox(0,0)[r]{\strut{}}}%
      \put(3808,7567){\makebox(0,0)[r]{\strut{}}}%
      \put(3808,7418){\makebox(0,0)[r]{\strut{}}}%
      \put(4139,7049){\makebox(0,0){\strut{}}}%
      \put(4439,7049){\makebox(0,0){\strut{}}}%
      \put(4738,7049){\makebox(0,0){\strut{}}}%
    }%
    \gplgaddtomacro\gplfronttext{%
    }%
    \gplgaddtomacro\gplbacktext{%
      \csname LTb\endcsname%
      \put(815,7111){\makebox(0,0)[r]{\strut{}$-0.01$}}%
      \put(815,6796){\makebox(0,0)[r]{\strut{}$0$}}%
      \put(815,6482){\makebox(0,0)[r]{\strut{}$0.01$}}%
      \put(815,6167){\makebox(0,0)[r]{\strut{}$0.02$}}%
      \put(815,5853){\makebox(0,0)[r]{\strut{}$0.03$}}%
      \put(1146,5381){\makebox(0,0){\strut{}}}%
      \put(1446,5381){\makebox(0,0){\strut{}}}%
      \put(1745,5381){\makebox(0,0){\strut{}}}%
      \put(1027,5734){\makebox(0,0)[l]{\strut{}6525}}%
    }%
    \gplgaddtomacro\gplfronttext{%
      \csname LTb\endcsname%
      \put(150,6434){\rotatebox{-270}{\makebox(0,0){\strut{}$\Delta m$}}}%
    }%
    \gplgaddtomacro\gplbacktext{%
      \csname LTb\endcsname%
      \put(815,5451){\makebox(0,0)[r]{\strut{}$-0.005$}}%
      \put(815,5303){\makebox(0,0)[r]{\strut{}$0$}}%
      \put(815,5154){\makebox(0,0)[r]{\strut{}$0.005$}}%
      \put(1146,4785){\makebox(0,0){\strut{}}}%
      \put(1446,4785){\makebox(0,0){\strut{}}}%
      \put(1745,4785){\makebox(0,0){\strut{}}}%
    }%
    \gplgaddtomacro\gplfronttext{%
      \csname LTb\endcsname%
      \put(150,5302){\rotatebox{-270}{\makebox(0,0){\strut{}O-C}}}%
    }%
    \gplgaddtomacro\gplbacktext{%
      \csname LTb\endcsname%
      \put(1813,7111){\makebox(0,0)[r]{\strut{}}}%
      \put(1813,6796){\makebox(0,0)[r]{\strut{}}}%
      \put(1813,6482){\makebox(0,0)[r]{\strut{}}}%
      \put(1813,6167){\makebox(0,0)[r]{\strut{}}}%
      \put(1813,5853){\makebox(0,0)[r]{\strut{}}}%
      \put(2144,5381){\makebox(0,0){\strut{}}}%
      \put(2444,5381){\makebox(0,0){\strut{}}}%
      \put(2743,5381){\makebox(0,0){\strut{}}}%
      \put(2025,5734){\makebox(0,0)[l]{\strut{}6775}}%
    }%
    \gplgaddtomacro\gplfronttext{%
    }%
    \gplgaddtomacro\gplbacktext{%
      \csname LTb\endcsname%
      \put(1813,5451){\makebox(0,0)[r]{\strut{}}}%
      \put(1813,5303){\makebox(0,0)[r]{\strut{}}}%
      \put(1813,5154){\makebox(0,0)[r]{\strut{}}}%
      \put(2144,4785){\makebox(0,0){\strut{}}}%
      \put(2444,4785){\makebox(0,0){\strut{}}}%
      \put(2743,4785){\makebox(0,0){\strut{}}}%
    }%
    \gplgaddtomacro\gplfronttext{%
    }%
    \gplgaddtomacro\gplbacktext{%
      \csname LTb\endcsname%
      \put(2810,7111){\makebox(0,0)[r]{\strut{}}}%
      \put(2810,6796){\makebox(0,0)[r]{\strut{}}}%
      \put(2810,6482){\makebox(0,0)[r]{\strut{}}}%
      \put(2810,6167){\makebox(0,0)[r]{\strut{}}}%
      \put(2810,5853){\makebox(0,0)[r]{\strut{}}}%
      \put(3141,5381){\makebox(0,0){\strut{}}}%
      \put(3441,5381){\makebox(0,0){\strut{}}}%
      \put(3740,5381){\makebox(0,0){\strut{}}}%
      \put(3022,5734){\makebox(0,0)[l]{\strut{}7025}}%
    }%
    \gplgaddtomacro\gplfronttext{%
    }%
    \gplgaddtomacro\gplbacktext{%
      \csname LTb\endcsname%
      \put(2810,5451){\makebox(0,0)[r]{\strut{}}}%
      \put(2810,5303){\makebox(0,0)[r]{\strut{}}}%
      \put(2810,5154){\makebox(0,0)[r]{\strut{}}}%
      \put(3141,4785){\makebox(0,0){\strut{}}}%
      \put(3441,4785){\makebox(0,0){\strut{}}}%
      \put(3740,4785){\makebox(0,0){\strut{}}}%
    }%
    \gplgaddtomacro\gplfronttext{%
    }%
    \gplgaddtomacro\gplbacktext{%
      \csname LTb\endcsname%
      \put(3808,7111){\makebox(0,0)[r]{\strut{}}}%
      \put(3808,6796){\makebox(0,0)[r]{\strut{}}}%
      \put(3808,6482){\makebox(0,0)[r]{\strut{}}}%
      \put(3808,6167){\makebox(0,0)[r]{\strut{}}}%
      \put(3808,5853){\makebox(0,0)[r]{\strut{}}}%
      \put(4139,5381){\makebox(0,0){\strut{}}}%
      \put(4439,5381){\makebox(0,0){\strut{}}}%
      \put(4738,5381){\makebox(0,0){\strut{}}}%
      \put(4020,5734){\makebox(0,0)[l]{\strut{}7275}}%
    }%
    \gplgaddtomacro\gplfronttext{%
    }%
    \gplgaddtomacro\gplbacktext{%
      \csname LTb\endcsname%
      \put(3808,5451){\makebox(0,0)[r]{\strut{}}}%
      \put(3808,5303){\makebox(0,0)[r]{\strut{}}}%
      \put(3808,5154){\makebox(0,0)[r]{\strut{}}}%
      \put(4139,4785){\makebox(0,0){\strut{}}}%
      \put(4439,4785){\makebox(0,0){\strut{}}}%
      \put(4738,4785){\makebox(0,0){\strut{}}}%
    }%
    \gplgaddtomacro\gplfronttext{%
    }%
    \gplgaddtomacro\gplbacktext{%
      \csname LTb\endcsname%
      \put(815,4847){\makebox(0,0)[r]{\strut{}$-0.01$}}%
      \put(815,4532){\makebox(0,0)[r]{\strut{}$0$}}%
      \put(815,4217){\makebox(0,0)[r]{\strut{}$0.01$}}%
      \put(815,3902){\makebox(0,0)[r]{\strut{}$0.02$}}%
      \put(815,3588){\makebox(0,0)[r]{\strut{}$0.03$}}%
      \put(1146,3116){\makebox(0,0){\strut{}}}%
      \put(1446,3116){\makebox(0,0){\strut{}}}%
      \put(1745,3116){\makebox(0,0){\strut{}}}%
      \put(1027,3469){\makebox(0,0)[l]{\strut{}7525}}%
    }%
    \gplgaddtomacro\gplfronttext{%
      \csname LTb\endcsname%
      \put(150,4170){\rotatebox{-270}{\makebox(0,0){\strut{}$\Delta m$}}}%
    }%
    \gplgaddtomacro\gplbacktext{%
      \csname LTb\endcsname%
      \put(815,3187){\makebox(0,0)[r]{\strut{}$-0.005$}}%
      \put(815,3039){\makebox(0,0)[r]{\strut{}$0$}}%
      \put(815,2890){\makebox(0,0)[r]{\strut{}$0.005$}}%
      \put(1146,2521){\makebox(0,0){\strut{}}}%
      \put(1446,2521){\makebox(0,0){\strut{}}}%
      \put(1745,2521){\makebox(0,0){\strut{}}}%
    }%
    \gplgaddtomacro\gplfronttext{%
      \csname LTb\endcsname%
      \put(150,3038){\rotatebox{-270}{\makebox(0,0){\strut{}O-C}}}%
    }%
    \gplgaddtomacro\gplbacktext{%
      \csname LTb\endcsname%
      \put(1813,4847){\makebox(0,0)[r]{\strut{}}}%
      \put(1813,4532){\makebox(0,0)[r]{\strut{}}}%
      \put(1813,4217){\makebox(0,0)[r]{\strut{}}}%
      \put(1813,3902){\makebox(0,0)[r]{\strut{}}}%
      \put(1813,3588){\makebox(0,0)[r]{\strut{}}}%
      \put(2144,3116){\makebox(0,0){\strut{}}}%
      \put(2444,3116){\makebox(0,0){\strut{}}}%
      \put(2743,3116){\makebox(0,0){\strut{}}}%
      \put(2025,3469){\makebox(0,0)[l]{\strut{}7775}}%
    }%
    \gplgaddtomacro\gplfronttext{%
    }%
    \gplgaddtomacro\gplbacktext{%
      \csname LTb\endcsname%
      \put(1813,3187){\makebox(0,0)[r]{\strut{}}}%
      \put(1813,3039){\makebox(0,0)[r]{\strut{}}}%
      \put(1813,2890){\makebox(0,0)[r]{\strut{}}}%
      \put(2144,2521){\makebox(0,0){\strut{}}}%
      \put(2444,2521){\makebox(0,0){\strut{}}}%
      \put(2743,2521){\makebox(0,0){\strut{}}}%
    }%
    \gplgaddtomacro\gplfronttext{%
    }%
    \gplgaddtomacro\gplbacktext{%
      \csname LTb\endcsname%
      \put(2810,4847){\makebox(0,0)[r]{\strut{}}}%
      \put(2810,4532){\makebox(0,0)[r]{\strut{}}}%
      \put(2810,4217){\makebox(0,0)[r]{\strut{}}}%
      \put(2810,3902){\makebox(0,0)[r]{\strut{}}}%
      \put(2810,3588){\makebox(0,0)[r]{\strut{}}}%
      \put(3141,3116){\makebox(0,0){\strut{}}}%
      \put(3441,3116){\makebox(0,0){\strut{}}}%
      \put(3740,3116){\makebox(0,0){\strut{}}}%
      \put(3022,3469){\makebox(0,0)[l]{\strut{}8025}}%
    }%
    \gplgaddtomacro\gplfronttext{%
    }%
    \gplgaddtomacro\gplbacktext{%
      \csname LTb\endcsname%
      \put(2810,3187){\makebox(0,0)[r]{\strut{}}}%
      \put(2810,3039){\makebox(0,0)[r]{\strut{}}}%
      \put(2810,2890){\makebox(0,0)[r]{\strut{}}}%
      \put(3141,2521){\makebox(0,0){\strut{}}}%
      \put(3441,2521){\makebox(0,0){\strut{}}}%
      \put(3740,2521){\makebox(0,0){\strut{}}}%
    }%
    \gplgaddtomacro\gplfronttext{%
    }%
    \gplgaddtomacro\gplbacktext{%
      \csname LTb\endcsname%
      \put(3808,4847){\makebox(0,0)[r]{\strut{}}}%
      \put(3808,4532){\makebox(0,0)[r]{\strut{}}}%
      \put(3808,4217){\makebox(0,0)[r]{\strut{}}}%
      \put(3808,3902){\makebox(0,0)[r]{\strut{}}}%
      \put(3808,3588){\makebox(0,0)[r]{\strut{}}}%
      \put(4139,3116){\makebox(0,0){\strut{}}}%
      \put(4439,3116){\makebox(0,0){\strut{}}}%
      \put(4738,3116){\makebox(0,0){\strut{}}}%
      \put(4020,3469){\makebox(0,0)[l]{\strut{}8275}}%
    }%
    \gplgaddtomacro\gplfronttext{%
    }%
    \gplgaddtomacro\gplbacktext{%
      \csname LTb\endcsname%
      \put(3808,3187){\makebox(0,0)[r]{\strut{}}}%
      \put(3808,3039){\makebox(0,0)[r]{\strut{}}}%
      \put(3808,2890){\makebox(0,0)[r]{\strut{}}}%
      \put(4139,2521){\makebox(0,0){\strut{}}}%
      \put(4439,2521){\makebox(0,0){\strut{}}}%
      \put(4738,2521){\makebox(0,0){\strut{}}}%
    }%
    \gplgaddtomacro\gplfronttext{%
    }%
    \gplgaddtomacro\gplbacktext{%
      \csname LTb\endcsname%
      \put(815,2583){\makebox(0,0)[r]{\strut{}$-0.01$}}%
      \put(815,2268){\makebox(0,0)[r]{\strut{}$0$}}%
      \put(815,1953){\makebox(0,0)[r]{\strut{}$0.01$}}%
      \put(815,1638){\makebox(0,0)[r]{\strut{}$0.02$}}%
      \put(815,1324){\makebox(0,0)[r]{\strut{}$0.03$}}%
      \put(1146,852){\makebox(0,0){\strut{}}}%
      \put(1446,852){\makebox(0,0){\strut{}}}%
      \put(1745,852){\makebox(0,0){\strut{}}}%
      \put(1027,1205){\makebox(0,0)[l]{\strut{}8525}}%
    }%
    \gplgaddtomacro\gplfronttext{%
      \csname LTb\endcsname%
      \put(150,1906){\rotatebox{-270}{\makebox(0,0){\strut{}$\Delta m$}}}%
    }%
    \gplgaddtomacro\gplbacktext{%
      \csname LTb\endcsname%
      \put(815,923){\makebox(0,0)[r]{\strut{}$-0.005$}}%
      \put(815,774){\makebox(0,0)[r]{\strut{}$0$}}%
      \put(815,625){\makebox(0,0)[r]{\strut{}$0.005$}}%
      \put(1146,256){\makebox(0,0){\strut{}$-0.03$}}%
      \put(1446,256){\makebox(0,0){\strut{}$0$}}%
      \put(1745,256){\makebox(0,0){\strut{}$0.03$}}%
    }%
    \gplgaddtomacro\gplfronttext{%
      \csname LTb\endcsname%
      \put(150,774){\rotatebox{-270}{\makebox(0,0){\strut{}O-C}}}%
      \put(1445,-74){\makebox(0,0){\strut{}Phase}}%
    }%
    \gplgaddtomacro\gplbacktext{%
      \csname LTb\endcsname%
      \put(1813,2583){\makebox(0,0)[r]{\strut{}}}%
      \put(1813,2268){\makebox(0,0)[r]{\strut{}}}%
      \put(1813,1953){\makebox(0,0)[r]{\strut{}}}%
      \put(1813,1638){\makebox(0,0)[r]{\strut{}}}%
      \put(1813,1324){\makebox(0,0)[r]{\strut{}}}%
      \put(2144,852){\makebox(0,0){\strut{}}}%
      \put(2444,852){\makebox(0,0){\strut{}}}%
      \put(2743,852){\makebox(0,0){\strut{}}}%
      \put(2025,1205){\makebox(0,0)[l]{\strut{}8775}}%
    }%
    \gplgaddtomacro\gplfronttext{%
    }%
    \gplgaddtomacro\gplbacktext{%
      \csname LTb\endcsname%
      \put(1813,923){\makebox(0,0)[r]{\strut{}}}%
      \put(1813,774){\makebox(0,0)[r]{\strut{}}}%
      \put(1813,625){\makebox(0,0)[r]{\strut{}}}%
      \put(2144,256){\makebox(0,0){\strut{}$-0.03$}}%
      \put(2444,256){\makebox(0,0){\strut{}$0$}}%
      \put(2743,256){\makebox(0,0){\strut{}$0.03$}}%
    }%
    \gplgaddtomacro\gplfronttext{%
      \csname LTb\endcsname%
      \put(2443,-74){\makebox(0,0){\strut{}Phase}}%
    }%
    \gplgaddtomacro\gplbacktext{%
      \csname LTb\endcsname%
      \put(2810,2583){\makebox(0,0)[r]{\strut{}}}%
      \put(2810,2268){\makebox(0,0)[r]{\strut{}}}%
      \put(2810,1953){\makebox(0,0)[r]{\strut{}}}%
      \put(2810,1638){\makebox(0,0)[r]{\strut{}}}%
      \put(2810,1324){\makebox(0,0)[r]{\strut{}}}%
      \put(3141,852){\makebox(0,0){\strut{}}}%
      \put(3441,852){\makebox(0,0){\strut{}}}%
      \put(3740,852){\makebox(0,0){\strut{}}}%
      \put(3022,1205){\makebox(0,0)[l]{\strut{}9025}}%
    }%
    \gplgaddtomacro\gplfronttext{%
    }%
    \gplgaddtomacro\gplbacktext{%
      \csname LTb\endcsname%
      \put(2810,923){\makebox(0,0)[r]{\strut{}}}%
      \put(2810,774){\makebox(0,0)[r]{\strut{}}}%
      \put(2810,625){\makebox(0,0)[r]{\strut{}}}%
      \put(3141,256){\makebox(0,0){\strut{}$-0.03$}}%
      \put(3441,256){\makebox(0,0){\strut{}$0$}}%
      \put(3740,256){\makebox(0,0){\strut{}$0.03$}}%
    }%
    \gplgaddtomacro\gplfronttext{%
      \csname LTb\endcsname%
      \put(3440,-74){\makebox(0,0){\strut{}Phase}}%
    }%
    \gplgaddtomacro\gplbacktext{%
      \csname LTb\endcsname%
      \put(3808,2583){\makebox(0,0)[r]{\strut{}}}%
      \put(3808,2268){\makebox(0,0)[r]{\strut{}}}%
      \put(3808,1953){\makebox(0,0)[r]{\strut{}}}%
      \put(3808,1638){\makebox(0,0)[r]{\strut{}}}%
      \put(3808,1324){\makebox(0,0)[r]{\strut{}}}%
      \put(4139,852){\makebox(0,0){\strut{}}}%
      \put(4439,852){\makebox(0,0){\strut{}}}%
      \put(4738,852){\makebox(0,0){\strut{}}}%
      \put(4020,1205){\makebox(0,0)[l]{\strut{}9275}}%
    }%
    \gplgaddtomacro\gplfronttext{%
    }%
    \gplgaddtomacro\gplbacktext{%
      \csname LTb\endcsname%
      \put(3808,923){\makebox(0,0)[r]{\strut{}}}%
      \put(3808,774){\makebox(0,0)[r]{\strut{}}}%
      \put(3808,625){\makebox(0,0)[r]{\strut{}}}%
      \put(4139,256){\makebox(0,0){\strut{}$-0.03$}}%
      \put(4439,256){\makebox(0,0){\strut{}$0$}}%
      \put(4738,256){\makebox(0,0){\strut{}$0.03$}}%
    }%
    \gplgaddtomacro\gplfronttext{%
      \csname LTb\endcsname%
      \put(4438,-74){\makebox(0,0){\strut{}Phase}}%
    }%
    \gplbacktext
    \put(0,0){\includegraphics{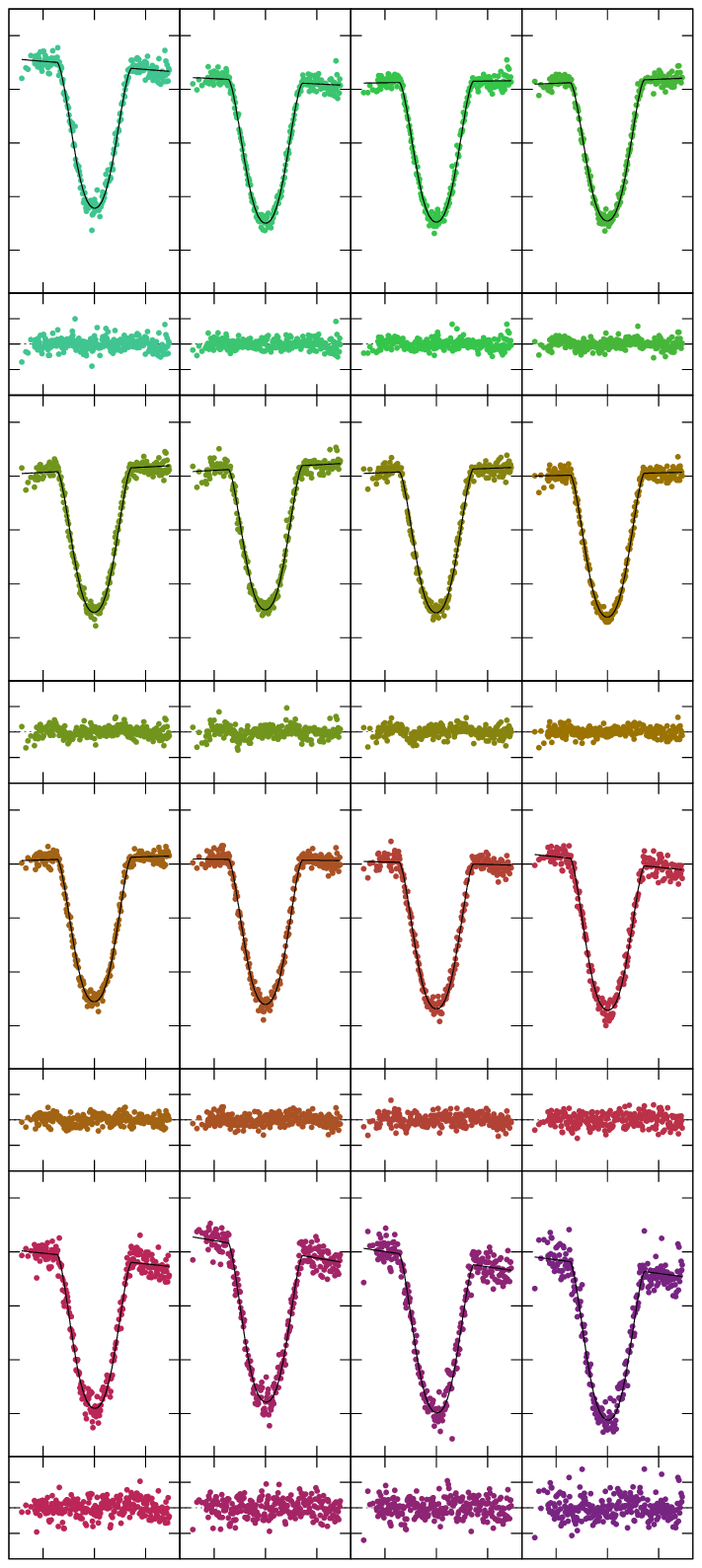}}%
    \gplfronttext
  \end{picture}%
\endgroup

% flatex input end: [./images/16bins_matthias/lc.tex]

        \caption{Sixteen channel GTC light curves. \emph{Upper panels:} Light curves with best fit model in black. Labels indicate the middle of the spectral bin, each with a spectral width of $ \unit[250]{\AA} $. \emph{Lower panels:} Residuals of best fit model.}
        \label{fig:16bins_matthias}
\end{figure}

In a second model-fitting run we corrected for smooth trends of the light curves. Ground-based differential photometry light curves of transit events typically show such a trend over time caused by a colour difference between the exoplanet host star and the comparison stars. To determine whether a linear or a quadratic detrending polynomial in time is a sufficient description for the trend, we used the Bayesian information criterion $ BIC $ \citep{schwarz_1978}. The $ BIC $ penalises free parameters and therefore balances an increase in the likelihood with the number of parameters used in the model.
The model with the lowest $ BIC $ is preferred. In most of the analysed broad band light curves the trend is well described by a linear trend (see Table~\ref{tab:photo-all-broad}). All light curves obtained at the GTC are well described by a linear trend.
\begin{figure}[htp]
        % flatex input: [./images/broad_band_lc/lc1.tex]
% GNUPLOT: LaTeX picture with Postscript
\begingroup
  \makeatletter
  \providecommand\color[2][]{%
    \GenericError{(gnuplot) \space\space\space\@spaces}{%
      Package color not loaded in conjunction with
      terminal option `colourtext'%
    }{See the gnuplot documentation for explanation.%
    }{Either use 'blacktext' in gnuplot or load the package
      color.sty in LaTeX.}%
    \renewcommand\color[2][]{}%
  }%
  \providecommand\includegraphics[2][]{%
    \GenericError{(gnuplot) \space\space\space\@spaces}{%
      Package graphicx or graphics not loaded%
    }{See the gnuplot documentation for explanation.%
    }{The gnuplot epslatex terminal needs graphicx.sty or graphics.sty.}%
    \renewcommand\includegraphics[2][]{}%
  }%
  \providecommand\rotatebox[2]{#2}%
  \@ifundefined{ifGPcolor}{%
    \newif\ifGPcolor
    \GPcolortrue
  }{}%
  \@ifundefined{ifGPblacktext}{%
    \newif\ifGPblacktext
    \GPblacktexttrue
  }{}%
  % define a \g@addto@macro without @ in the name:
  \let\gplgaddtomacro\g@addto@macro
  % define empty templates for all commands taking text:
  \gdef\gplbacktext{}%
  \gdef\gplfronttext{}%
  \makeatother
  \ifGPblacktext
    % no textcolor at all
    \def\colorrgb#1{}%
    \def\colorgray#1{}%
  \else
    % gray or color?
    \ifGPcolor
      \def\colorrgb#1{\color[rgb]{#1}}%
      \def\colorgray#1{\color[gray]{#1}}%
      \expandafter\def\csname LTw\endcsname{\color{white}}%
      \expandafter\def\csname LTb\endcsname{\color{black}}%
      \expandafter\def\csname LTa\endcsname{\color{black}}%
      \expandafter\def\csname LT0\endcsname{\color[rgb]{1,0,0}}%
      \expandafter\def\csname LT1\endcsname{\color[rgb]{0,1,0}}%
      \expandafter\def\csname LT2\endcsname{\color[rgb]{0,0,1}}%
      \expandafter\def\csname LT3\endcsname{\color[rgb]{1,0,1}}%
      \expandafter\def\csname LT4\endcsname{\color[rgb]{0,1,1}}%
      \expandafter\def\csname LT5\endcsname{\color[rgb]{1,1,0}}%
      \expandafter\def\csname LT6\endcsname{\color[rgb]{0,0,0}}%
      \expandafter\def\csname LT7\endcsname{\color[rgb]{1,0.3,0}}%
      \expandafter\def\csname LT8\endcsname{\color[rgb]{0.5,0.5,0.5}}%
    \else
      % gray
      \def\colorrgb#1{\color{black}}%
      \def\colorgray#1{\color[gray]{#1}}%
      \expandafter\def\csname LTw\endcsname{\color{white}}%
      \expandafter\def\csname LTb\endcsname{\color{black}}%
      \expandafter\def\csname LTa\endcsname{\color{black}}%
      \expandafter\def\csname LT0\endcsname{\color{black}}%
      \expandafter\def\csname LT1\endcsname{\color{black}}%
      \expandafter\def\csname LT2\endcsname{\color{black}}%
      \expandafter\def\csname LT3\endcsname{\color{black}}%
      \expandafter\def\csname LT4\endcsname{\color{black}}%
      \expandafter\def\csname LT5\endcsname{\color{black}}%
      \expandafter\def\csname LT6\endcsname{\color{black}}%
      \expandafter\def\csname LT7\endcsname{\color{black}}%
      \expandafter\def\csname LT8\endcsname{\color{black}}%
    \fi
  \fi
    \setlength{\unitlength}{0.0500bp}%
    \ifx\gptboxheight\undefined%
      \newlength{\gptboxheight}%
      \newlength{\gptboxwidth}%
      \newsavebox{\gptboxtext}%
    \fi%
    \setlength{\fboxrule}{0.5pt}%
    \setlength{\fboxsep}{1pt}%
\begin{picture}(4988.00,12472.00)%
    \gplgaddtomacro\gplbacktext{%
      \csname LTb\endcsname%
      \put(815,12295){\makebox(0,0)[r]{\strut{}$-0.01$}}%
      \put(815,11968){\makebox(0,0)[r]{\strut{}$0$}}%
      \put(815,11641){\makebox(0,0)[r]{\strut{}$0.01$}}%
      \put(815,11314){\makebox(0,0)[r]{\strut{}$0.02$}}%
      \put(815,10987){\makebox(0,0)[r]{\strut{}$0.03$}}%
      \put(1134,10505){\makebox(0,0){\strut{}}}%
      \put(1446,10505){\makebox(0,0){\strut{}}}%
      \put(1757,10505){\makebox(0,0){\strut{}}}%
      \put(1047,10898){\makebox(0,0)[l]{\strut{}1}}%
    }%
    \gplgaddtomacro\gplfronttext{%
      \csname LTb\endcsname%
      \put(150,11591){\rotatebox{-270}{\makebox(0,0){\strut{}$\Delta m$}}}%
    }%
    \gplgaddtomacro\gplbacktext{%
      \csname LTb\endcsname%
      \put(815,10569){\makebox(0,0)[r]{\strut{}$-0.01$}}%
      \put(815,10414){\makebox(0,0)[r]{\strut{}$0$}}%
      \put(815,10258){\makebox(0,0)[r]{\strut{}$0.01$}}%
      \put(1134,9882){\makebox(0,0){\strut{}}}%
      \put(1446,9882){\makebox(0,0){\strut{}}}%
      \put(1757,9882){\makebox(0,0){\strut{}}}%
    }%
    \gplgaddtomacro\gplfronttext{%
      \csname LTb\endcsname%
      \put(150,10413){\rotatebox{-270}{\makebox(0,0){\strut{}O-C}}}%
    }%
    \gplgaddtomacro\gplbacktext{%
      \csname LTb\endcsname%
      \put(1813,12295){\makebox(0,0)[r]{\strut{}}}%
      \put(1813,11968){\makebox(0,0)[r]{\strut{}}}%
      \put(1813,11641){\makebox(0,0)[r]{\strut{}}}%
      \put(1813,11314){\makebox(0,0)[r]{\strut{}}}%
      \put(1813,10987){\makebox(0,0)[r]{\strut{}}}%
      \put(2132,10505){\makebox(0,0){\strut{}}}%
      \put(2444,10505){\makebox(0,0){\strut{}}}%
      \put(2755,10505){\makebox(0,0){\strut{}}}%
      \put(2045,10898){\makebox(0,0)[l]{\strut{}2}}%
    }%
    \gplgaddtomacro\gplfronttext{%
    }%
    \gplgaddtomacro\gplbacktext{%
      \csname LTb\endcsname%
      \put(1813,10569){\makebox(0,0)[r]{\strut{}}}%
      \put(1813,10414){\makebox(0,0)[r]{\strut{}}}%
      \put(1813,10258){\makebox(0,0)[r]{\strut{}}}%
      \put(2132,9882){\makebox(0,0){\strut{}}}%
      \put(2444,9882){\makebox(0,0){\strut{}}}%
      \put(2755,9882){\makebox(0,0){\strut{}}}%
    }%
    \gplgaddtomacro\gplfronttext{%
    }%
    \gplgaddtomacro\gplbacktext{%
      \csname LTb\endcsname%
      \put(2810,12295){\makebox(0,0)[r]{\strut{}}}%
      \put(2810,11968){\makebox(0,0)[r]{\strut{}}}%
      \put(2810,11641){\makebox(0,0)[r]{\strut{}}}%
      \put(2810,11314){\makebox(0,0)[r]{\strut{}}}%
      \put(2810,10987){\makebox(0,0)[r]{\strut{}}}%
      \put(3129,10505){\makebox(0,0){\strut{}}}%
      \put(3441,10505){\makebox(0,0){\strut{}}}%
      \put(3752,10505){\makebox(0,0){\strut{}}}%
      \put(3042,10898){\makebox(0,0)[l]{\strut{}3}}%
    }%
    \gplgaddtomacro\gplfronttext{%
    }%
    \gplgaddtomacro\gplbacktext{%
      \csname LTb\endcsname%
      \put(2810,10569){\makebox(0,0)[r]{\strut{}}}%
      \put(2810,10414){\makebox(0,0)[r]{\strut{}}}%
      \put(2810,10258){\makebox(0,0)[r]{\strut{}}}%
      \put(3129,9882){\makebox(0,0){\strut{}}}%
      \put(3441,9882){\makebox(0,0){\strut{}}}%
      \put(3752,9882){\makebox(0,0){\strut{}}}%
    }%
    \gplgaddtomacro\gplfronttext{%
    }%
    \gplgaddtomacro\gplbacktext{%
      \csname LTb\endcsname%
      \put(3808,12295){\makebox(0,0)[r]{\strut{}}}%
      \put(3808,11968){\makebox(0,0)[r]{\strut{}}}%
      \put(3808,11641){\makebox(0,0)[r]{\strut{}}}%
      \put(3808,11314){\makebox(0,0)[r]{\strut{}}}%
      \put(3808,10987){\makebox(0,0)[r]{\strut{}}}%
      \put(4127,10505){\makebox(0,0){\strut{}}}%
      \put(4439,10505){\makebox(0,0){\strut{}}}%
      \put(4750,10505){\makebox(0,0){\strut{}}}%
      \put(4040,10898){\makebox(0,0)[l]{\strut{}4}}%
    }%
    \gplgaddtomacro\gplfronttext{%
    }%
    \gplgaddtomacro\gplbacktext{%
      \csname LTb\endcsname%
      \put(3808,10569){\makebox(0,0)[r]{\strut{}}}%
      \put(3808,10414){\makebox(0,0)[r]{\strut{}}}%
      \put(3808,10258){\makebox(0,0)[r]{\strut{}}}%
      \put(4127,9882){\makebox(0,0){\strut{}}}%
      \put(4439,9882){\makebox(0,0){\strut{}}}%
      \put(4750,9882){\makebox(0,0){\strut{}}}%
    }%
    \gplgaddtomacro\gplfronttext{%
    }%
    \gplgaddtomacro\gplbacktext{%
      \csname LTb\endcsname%
      \put(815,9936){\makebox(0,0)[r]{\strut{}$-0.01$}}%
      \put(815,9607){\makebox(0,0)[r]{\strut{}$0$}}%
      \put(815,9278){\makebox(0,0)[r]{\strut{}$0.01$}}%
      \put(815,8949){\makebox(0,0)[r]{\strut{}$0.02$}}%
      \put(815,8619){\makebox(0,0)[r]{\strut{}$0.03$}}%
      \put(1134,8136){\makebox(0,0){\strut{}}}%
      \put(1446,8136){\makebox(0,0){\strut{}}}%
      \put(1757,8136){\makebox(0,0){\strut{}}}%
      \put(1047,8531){\makebox(0,0)[l]{\strut{}5}}%
    }%
    \gplgaddtomacro\gplfronttext{%
      \csname LTb\endcsname%
      \put(150,9228){\rotatebox{-270}{\makebox(0,0){\strut{}$\Delta m$}}}%
    }%
    \gplgaddtomacro\gplbacktext{%
      \csname LTb\endcsname%
      \put(815,8199){\makebox(0,0)[r]{\strut{}$-0.01$}}%
      \put(815,8044){\makebox(0,0)[r]{\strut{}$0$}}%
      \put(815,7888){\makebox(0,0)[r]{\strut{}$0.01$}}%
      \put(1134,7512){\makebox(0,0){\strut{}}}%
      \put(1446,7512){\makebox(0,0){\strut{}}}%
      \put(1757,7512){\makebox(0,0){\strut{}}}%
    }%
    \gplgaddtomacro\gplfronttext{%
      \csname LTb\endcsname%
      \put(150,8043){\rotatebox{-270}{\makebox(0,0){\strut{}O-C}}}%
    }%
    \gplgaddtomacro\gplbacktext{%
      \csname LTb\endcsname%
      \put(1813,9936){\makebox(0,0)[r]{\strut{}}}%
      \put(1813,9607){\makebox(0,0)[r]{\strut{}}}%
      \put(1813,9278){\makebox(0,0)[r]{\strut{}}}%
      \put(1813,8949){\makebox(0,0)[r]{\strut{}}}%
      \put(1813,8619){\makebox(0,0)[r]{\strut{}}}%
      \put(2132,8136){\makebox(0,0){\strut{}}}%
      \put(2444,8136){\makebox(0,0){\strut{}}}%
      \put(2755,8136){\makebox(0,0){\strut{}}}%
      \put(2045,8531){\makebox(0,0)[l]{\strut{}6}}%
    }%
    \gplgaddtomacro\gplfronttext{%
    }%
    \gplgaddtomacro\gplbacktext{%
      \csname LTb\endcsname%
      \put(1813,8199){\makebox(0,0)[r]{\strut{}}}%
      \put(1813,8044){\makebox(0,0)[r]{\strut{}}}%
      \put(1813,7888){\makebox(0,0)[r]{\strut{}}}%
      \put(2132,7512){\makebox(0,0){\strut{}}}%
      \put(2444,7512){\makebox(0,0){\strut{}}}%
      \put(2755,7512){\makebox(0,0){\strut{}}}%
    }%
    \gplgaddtomacro\gplfronttext{%
    }%
    \gplgaddtomacro\gplbacktext{%
      \csname LTb\endcsname%
      \put(2810,9936){\makebox(0,0)[r]{\strut{}}}%
      \put(2810,9607){\makebox(0,0)[r]{\strut{}}}%
      \put(2810,9278){\makebox(0,0)[r]{\strut{}}}%
      \put(2810,8949){\makebox(0,0)[r]{\strut{}}}%
      \put(2810,8619){\makebox(0,0)[r]{\strut{}}}%
      \put(3129,8136){\makebox(0,0){\strut{}}}%
      \put(3441,8136){\makebox(0,0){\strut{}}}%
      \put(3752,8136){\makebox(0,0){\strut{}}}%
      \put(3042,8531){\makebox(0,0)[l]{\strut{}7}}%
    }%
    \gplgaddtomacro\gplfronttext{%
    }%
    \gplgaddtomacro\gplbacktext{%
      \csname LTb\endcsname%
      \put(2810,8199){\makebox(0,0)[r]{\strut{}}}%
      \put(2810,8044){\makebox(0,0)[r]{\strut{}}}%
      \put(2810,7888){\makebox(0,0)[r]{\strut{}}}%
      \put(3129,7512){\makebox(0,0){\strut{}}}%
      \put(3441,7512){\makebox(0,0){\strut{}}}%
      \put(3752,7512){\makebox(0,0){\strut{}}}%
    }%
    \gplgaddtomacro\gplfronttext{%
    }%
    \gplgaddtomacro\gplbacktext{%
      \csname LTb\endcsname%
      \put(3808,9936){\makebox(0,0)[r]{\strut{}}}%
      \put(3808,9607){\makebox(0,0)[r]{\strut{}}}%
      \put(3808,9278){\makebox(0,0)[r]{\strut{}}}%
      \put(3808,8949){\makebox(0,0)[r]{\strut{}}}%
      \put(3808,8619){\makebox(0,0)[r]{\strut{}}}%
      \put(4127,8136){\makebox(0,0){\strut{}}}%
      \put(4439,8136){\makebox(0,0){\strut{}}}%
      \put(4750,8136){\makebox(0,0){\strut{}}}%
      \put(4040,8531){\makebox(0,0)[l]{\strut{}8}}%
    }%
    \gplgaddtomacro\gplfronttext{%
    }%
    \gplgaddtomacro\gplbacktext{%
      \csname LTb\endcsname%
      \put(3808,8199){\makebox(0,0)[r]{\strut{}}}%
      \put(3808,8044){\makebox(0,0)[r]{\strut{}}}%
      \put(3808,7888){\makebox(0,0)[r]{\strut{}}}%
      \put(4127,7512){\makebox(0,0){\strut{}}}%
      \put(4439,7512){\makebox(0,0){\strut{}}}%
      \put(4750,7512){\makebox(0,0){\strut{}}}%
    }%
    \gplgaddtomacro\gplfronttext{%
    }%
    \gplgaddtomacro\gplbacktext{%
      \csname LTb\endcsname%
      \put(815,7567){\makebox(0,0)[r]{\strut{}$-0.01$}}%
      \put(815,7238){\makebox(0,0)[r]{\strut{}$0$}}%
      \put(815,6908){\makebox(0,0)[r]{\strut{}$0.01$}}%
      \put(815,6579){\makebox(0,0)[r]{\strut{}$0.02$}}%
      \put(815,6250){\makebox(0,0)[r]{\strut{}$0.03$}}%
      \put(1134,5766){\makebox(0,0){\strut{}}}%
      \put(1446,5766){\makebox(0,0){\strut{}}}%
      \put(1757,5766){\makebox(0,0){\strut{}}}%
      \put(1047,6161){\makebox(0,0)[l]{\strut{}9}}%
    }%
    \gplgaddtomacro\gplfronttext{%
      \csname LTb\endcsname%
      \put(150,6859){\rotatebox{-270}{\makebox(0,0){\strut{}$\Delta m$}}}%
    }%
    \gplgaddtomacro\gplbacktext{%
      \csname LTb\endcsname%
      \put(815,5830){\makebox(0,0)[r]{\strut{}$-0.01$}}%
      \put(815,5674){\makebox(0,0)[r]{\strut{}$0$}}%
      \put(815,5518){\makebox(0,0)[r]{\strut{}$0.01$}}%
      \put(1134,5142){\makebox(0,0){\strut{}}}%
      \put(1446,5142){\makebox(0,0){\strut{}}}%
      \put(1757,5142){\makebox(0,0){\strut{}}}%
    }%
    \gplgaddtomacro\gplfronttext{%
      \csname LTb\endcsname%
      \put(150,5674){\rotatebox{-270}{\makebox(0,0){\strut{}O-C}}}%
    }%
    \gplgaddtomacro\gplbacktext{%
      \csname LTb\endcsname%
      \put(1813,7567){\makebox(0,0)[r]{\strut{}}}%
      \put(1813,7238){\makebox(0,0)[r]{\strut{}}}%
      \put(1813,6908){\makebox(0,0)[r]{\strut{}}}%
      \put(1813,6579){\makebox(0,0)[r]{\strut{}}}%
      \put(1813,6250){\makebox(0,0)[r]{\strut{}}}%
      \put(2132,5766){\makebox(0,0){\strut{}}}%
      \put(2444,5766){\makebox(0,0){\strut{}}}%
      \put(2755,5766){\makebox(0,0){\strut{}}}%
      \put(2045,6161){\makebox(0,0)[l]{\strut{}10}}%
    }%
    \gplgaddtomacro\gplfronttext{%
    }%
    \gplgaddtomacro\gplbacktext{%
      \csname LTb\endcsname%
      \put(1813,5830){\makebox(0,0)[r]{\strut{}}}%
      \put(1813,5674){\makebox(0,0)[r]{\strut{}}}%
      \put(1813,5518){\makebox(0,0)[r]{\strut{}}}%
      \put(2132,5142){\makebox(0,0){\strut{}}}%
      \put(2444,5142){\makebox(0,0){\strut{}}}%
      \put(2755,5142){\makebox(0,0){\strut{}}}%
    }%
    \gplgaddtomacro\gplfronttext{%
    }%
    \gplgaddtomacro\gplbacktext{%
      \csname LTb\endcsname%
      \put(2810,7567){\makebox(0,0)[r]{\strut{}}}%
      \put(2810,7238){\makebox(0,0)[r]{\strut{}}}%
      \put(2810,6908){\makebox(0,0)[r]{\strut{}}}%
      \put(2810,6579){\makebox(0,0)[r]{\strut{}}}%
      \put(2810,6250){\makebox(0,0)[r]{\strut{}}}%
      \put(3129,5766){\makebox(0,0){\strut{}}}%
      \put(3441,5766){\makebox(0,0){\strut{}}}%
      \put(3752,5766){\makebox(0,0){\strut{}}}%
      \put(3042,6161){\makebox(0,0)[l]{\strut{}11}}%
    }%
    \gplgaddtomacro\gplfronttext{%
    }%
    \gplgaddtomacro\gplbacktext{%
      \csname LTb\endcsname%
      \put(2810,5830){\makebox(0,0)[r]{\strut{}}}%
      \put(2810,5674){\makebox(0,0)[r]{\strut{}}}%
      \put(2810,5518){\makebox(0,0)[r]{\strut{}}}%
      \put(3129,5142){\makebox(0,0){\strut{}}}%
      \put(3441,5142){\makebox(0,0){\strut{}}}%
      \put(3752,5142){\makebox(0,0){\strut{}}}%
    }%
    \gplgaddtomacro\gplfronttext{%
    }%
    \gplgaddtomacro\gplbacktext{%
      \csname LTb\endcsname%
      \put(3808,7567){\makebox(0,0)[r]{\strut{}}}%
      \put(3808,7238){\makebox(0,0)[r]{\strut{}}}%
      \put(3808,6908){\makebox(0,0)[r]{\strut{}}}%
      \put(3808,6579){\makebox(0,0)[r]{\strut{}}}%
      \put(3808,6250){\makebox(0,0)[r]{\strut{}}}%
      \put(4127,5766){\makebox(0,0){\strut{}}}%
      \put(4439,5766){\makebox(0,0){\strut{}}}%
      \put(4750,5766){\makebox(0,0){\strut{}}}%
      \put(4040,6161){\makebox(0,0)[l]{\strut{}12}}%
    }%
    \gplgaddtomacro\gplfronttext{%
    }%
    \gplgaddtomacro\gplbacktext{%
      \csname LTb\endcsname%
      \put(3808,5830){\makebox(0,0)[r]{\strut{}}}%
      \put(3808,5674){\makebox(0,0)[r]{\strut{}}}%
      \put(3808,5518){\makebox(0,0)[r]{\strut{}}}%
      \put(4127,5142){\makebox(0,0){\strut{}}}%
      \put(4439,5142){\makebox(0,0){\strut{}}}%
      \put(4750,5142){\makebox(0,0){\strut{}}}%
    }%
    \gplgaddtomacro\gplfronttext{%
    }%
    \gplgaddtomacro\gplbacktext{%
      \csname LTb\endcsname%
      \put(815,5197){\makebox(0,0)[r]{\strut{}$-0.01$}}%
      \put(815,4868){\makebox(0,0)[r]{\strut{}$0$}}%
      \put(815,4538){\makebox(0,0)[r]{\strut{}$0.01$}}%
      \put(815,4209){\makebox(0,0)[r]{\strut{}$0.02$}}%
      \put(815,3880){\makebox(0,0)[r]{\strut{}$0.03$}}%
      \put(1134,3396){\makebox(0,0){\strut{}}}%
      \put(1446,3396){\makebox(0,0){\strut{}}}%
      \put(1757,3396){\makebox(0,0){\strut{}}}%
      \put(1047,3791){\makebox(0,0)[l]{\strut{}13}}%
    }%
    \gplgaddtomacro\gplfronttext{%
      \csname LTb\endcsname%
      \put(150,4489){\rotatebox{-270}{\makebox(0,0){\strut{}$\Delta m$}}}%
    }%
    \gplgaddtomacro\gplbacktext{%
      \csname LTb\endcsname%
      \put(815,3460){\makebox(0,0)[r]{\strut{}$-0.01$}}%
      \put(815,3305){\makebox(0,0)[r]{\strut{}$0$}}%
      \put(815,3149){\makebox(0,0)[r]{\strut{}$0.01$}}%
      \put(1134,2773){\makebox(0,0){\strut{}}}%
      \put(1446,2773){\makebox(0,0){\strut{}}}%
      \put(1757,2773){\makebox(0,0){\strut{}}}%
    }%
    \gplgaddtomacro\gplfronttext{%
      \csname LTb\endcsname%
      \put(150,3304){\rotatebox{-270}{\makebox(0,0){\strut{}O-C}}}%
    }%
    \gplgaddtomacro\gplbacktext{%
      \csname LTb\endcsname%
      \put(1813,5197){\makebox(0,0)[r]{\strut{}}}%
      \put(1813,4868){\makebox(0,0)[r]{\strut{}}}%
      \put(1813,4538){\makebox(0,0)[r]{\strut{}}}%
      \put(1813,4209){\makebox(0,0)[r]{\strut{}}}%
      \put(1813,3880){\makebox(0,0)[r]{\strut{}}}%
      \put(2132,3396){\makebox(0,0){\strut{}}}%
      \put(2444,3396){\makebox(0,0){\strut{}}}%
      \put(2755,3396){\makebox(0,0){\strut{}}}%
      \put(2045,3791){\makebox(0,0)[l]{\strut{}14}}%
    }%
    \gplgaddtomacro\gplfronttext{%
    }%
    \gplgaddtomacro\gplbacktext{%
      \csname LTb\endcsname%
      \put(1813,3460){\makebox(0,0)[r]{\strut{}}}%
      \put(1813,3305){\makebox(0,0)[r]{\strut{}}}%
      \put(1813,3149){\makebox(0,0)[r]{\strut{}}}%
      \put(2132,2773){\makebox(0,0){\strut{}}}%
      \put(2444,2773){\makebox(0,0){\strut{}}}%
      \put(2755,2773){\makebox(0,0){\strut{}}}%
    }%
    \gplgaddtomacro\gplfronttext{%
    }%
    \gplgaddtomacro\gplbacktext{%
      \csname LTb\endcsname%
      \put(2810,5197){\makebox(0,0)[r]{\strut{}}}%
      \put(2810,4868){\makebox(0,0)[r]{\strut{}}}%
      \put(2810,4538){\makebox(0,0)[r]{\strut{}}}%
      \put(2810,4209){\makebox(0,0)[r]{\strut{}}}%
      \put(2810,3880){\makebox(0,0)[r]{\strut{}}}%
      \put(3129,3396){\makebox(0,0){\strut{}}}%
      \put(3441,3396){\makebox(0,0){\strut{}}}%
      \put(3752,3396){\makebox(0,0){\strut{}}}%
      \put(3042,3791){\makebox(0,0)[l]{\strut{}15}}%
    }%
    \gplgaddtomacro\gplfronttext{%
    }%
    \gplgaddtomacro\gplbacktext{%
      \csname LTb\endcsname%
      \put(2810,3460){\makebox(0,0)[r]{\strut{}}}%
      \put(2810,3305){\makebox(0,0)[r]{\strut{}}}%
      \put(2810,3149){\makebox(0,0)[r]{\strut{}}}%
      \put(3129,2773){\makebox(0,0){\strut{}}}%
      \put(3441,2773){\makebox(0,0){\strut{}}}%
      \put(3752,2773){\makebox(0,0){\strut{}}}%
    }%
    \gplgaddtomacro\gplfronttext{%
    }%
    \gplgaddtomacro\gplbacktext{%
      \csname LTb\endcsname%
      \put(3808,5197){\makebox(0,0)[r]{\strut{}}}%
      \put(3808,4868){\makebox(0,0)[r]{\strut{}}}%
      \put(3808,4538){\makebox(0,0)[r]{\strut{}}}%
      \put(3808,4209){\makebox(0,0)[r]{\strut{}}}%
      \put(3808,3880){\makebox(0,0)[r]{\strut{}}}%
      \put(4127,3396){\makebox(0,0){\strut{}}}%
      \put(4439,3396){\makebox(0,0){\strut{}}}%
      \put(4750,3396){\makebox(0,0){\strut{}}}%
      \put(4040,3791){\makebox(0,0)[l]{\strut{}16}}%
    }%
    \gplgaddtomacro\gplfronttext{%
    }%
    \gplgaddtomacro\gplbacktext{%
      \csname LTb\endcsname%
      \put(3808,3460){\makebox(0,0)[r]{\strut{}}}%
      \put(3808,3305){\makebox(0,0)[r]{\strut{}}}%
      \put(3808,3149){\makebox(0,0)[r]{\strut{}}}%
      \put(4127,2773){\makebox(0,0){\strut{}}}%
      \put(4439,2773){\makebox(0,0){\strut{}}}%
      \put(4750,2773){\makebox(0,0){\strut{}}}%
    }%
    \gplgaddtomacro\gplfronttext{%
    }%
    \gplgaddtomacro\gplbacktext{%
      \csname LTb\endcsname%
      \put(815,2828){\makebox(0,0)[r]{\strut{}$-0.01$}}%
      \put(815,2499){\makebox(0,0)[r]{\strut{}$0$}}%
      \put(815,2169){\makebox(0,0)[r]{\strut{}$0.01$}}%
      \put(815,1840){\makebox(0,0)[r]{\strut{}$0.02$}}%
      \put(815,1511){\makebox(0,0)[r]{\strut{}$0.03$}}%
      \put(1134,1027){\makebox(0,0){\strut{}}}%
      \put(1446,1027){\makebox(0,0){\strut{}}}%
      \put(1757,1027){\makebox(0,0){\strut{}}}%
      \put(1047,1422){\makebox(0,0)[l]{\strut{}17}}%
    }%
    \gplgaddtomacro\gplfronttext{%
      \csname LTb\endcsname%
      \put(150,2120){\rotatebox{-270}{\makebox(0,0){\strut{}$\Delta m$}}}%
    }%
    \gplgaddtomacro\gplbacktext{%
      \csname LTb\endcsname%
      \put(815,1091){\makebox(0,0)[r]{\strut{}$-0.01$}}%
      \put(815,935){\makebox(0,0)[r]{\strut{}$0$}}%
      \put(815,779){\makebox(0,0)[r]{\strut{}$0.01$}}%
      \put(1134,403){\makebox(0,0){\strut{}$-0.05$}}%
      \put(1446,403){\makebox(0,0){\strut{}$0$}}%
      \put(1757,403){\makebox(0,0){\strut{}$0.05$}}%
    }%
    \gplgaddtomacro\gplfronttext{%
      \csname LTb\endcsname%
      \put(150,935){\rotatebox{-270}{\makebox(0,0){\strut{}O-C}}}%
      \put(1445,73){\makebox(0,0){\strut{}Phase}}%
    }%
    \gplgaddtomacro\gplbacktext{%
      \csname LTb\endcsname%
      \put(1813,2828){\makebox(0,0)[r]{\strut{}}}%
      \put(1813,2499){\makebox(0,0)[r]{\strut{}}}%
      \put(1813,2169){\makebox(0,0)[r]{\strut{}}}%
      \put(1813,1840){\makebox(0,0)[r]{\strut{}}}%
      \put(1813,1511){\makebox(0,0)[r]{\strut{}}}%
      \put(2132,1027){\makebox(0,0){\strut{}}}%
      \put(2444,1027){\makebox(0,0){\strut{}}}%
      \put(2755,1027){\makebox(0,0){\strut{}}}%
      \put(2045,1422){\makebox(0,0)[l]{\strut{}18}}%
    }%
    \gplgaddtomacro\gplfronttext{%
    }%
    \gplgaddtomacro\gplbacktext{%
      \csname LTb\endcsname%
      \put(1813,1091){\makebox(0,0)[r]{\strut{}}}%
      \put(1813,935){\makebox(0,0)[r]{\strut{}}}%
      \put(1813,779){\makebox(0,0)[r]{\strut{}}}%
      \put(2132,403){\makebox(0,0){\strut{}$-0.05$}}%
      \put(2444,403){\makebox(0,0){\strut{}$0$}}%
      \put(2755,403){\makebox(0,0){\strut{}$0.05$}}%
    }%
    \gplgaddtomacro\gplfronttext{%
      \csname LTb\endcsname%
      \put(2443,73){\makebox(0,0){\strut{}Phase}}%
    }%
    \gplgaddtomacro\gplbacktext{%
      \csname LTb\endcsname%
      \put(2810,2828){\makebox(0,0)[r]{\strut{}}}%
      \put(2810,2499){\makebox(0,0)[r]{\strut{}}}%
      \put(2810,2169){\makebox(0,0)[r]{\strut{}}}%
      \put(2810,1840){\makebox(0,0)[r]{\strut{}}}%
      \put(2810,1511){\makebox(0,0)[r]{\strut{}}}%
      \put(3129,1027){\makebox(0,0){\strut{}}}%
      \put(3441,1027){\makebox(0,0){\strut{}}}%
      \put(3752,1027){\makebox(0,0){\strut{}}}%
      \put(3042,1422){\makebox(0,0)[l]{\strut{}19}}%
    }%
    \gplgaddtomacro\gplfronttext{%
    }%
    \gplgaddtomacro\gplbacktext{%
      \csname LTb\endcsname%
      \put(2810,1091){\makebox(0,0)[r]{\strut{}}}%
      \put(2810,935){\makebox(0,0)[r]{\strut{}}}%
      \put(2810,779){\makebox(0,0)[r]{\strut{}}}%
      \put(3129,403){\makebox(0,0){\strut{}$-0.05$}}%
      \put(3441,403){\makebox(0,0){\strut{}$0$}}%
      \put(3752,403){\makebox(0,0){\strut{}$0.05$}}%
    }%
    \gplgaddtomacro\gplfronttext{%
      \csname LTb\endcsname%
      \put(3440,73){\makebox(0,0){\strut{}Phase}}%
    }%
    \gplgaddtomacro\gplbacktext{%
      \csname LTb\endcsname%
      \put(3808,2828){\makebox(0,0)[r]{\strut{}}}%
      \put(3808,2499){\makebox(0,0)[r]{\strut{}}}%
      \put(3808,2169){\makebox(0,0)[r]{\strut{}}}%
      \put(3808,1840){\makebox(0,0)[r]{\strut{}}}%
      \put(3808,1511){\makebox(0,0)[r]{\strut{}}}%
      \put(4127,1027){\makebox(0,0){\strut{}}}%
      \put(4439,1027){\makebox(0,0){\strut{}}}%
      \put(4750,1027){\makebox(0,0){\strut{}}}%
      \put(4040,1422){\makebox(0,0)[l]{\strut{}20}}%
    }%
    \gplgaddtomacro\gplfronttext{%
    }%
    \gplgaddtomacro\gplbacktext{%
      \csname LTb\endcsname%
      \put(3808,1091){\makebox(0,0)[r]{\strut{}}}%
      \put(3808,935){\makebox(0,0)[r]{\strut{}}}%
      \put(3808,779){\makebox(0,0)[r]{\strut{}}}%
      \put(4127,403){\makebox(0,0){\strut{}$-0.05$}}%
      \put(4439,403){\makebox(0,0){\strut{}$0$}}%
      \put(4750,403){\makebox(0,0){\strut{}$0.05$}}%
    }%
    \gplgaddtomacro\gplfronttext{%
      \csname LTb\endcsname%
      \put(4438,73){\makebox(0,0){\strut{}Phase}}%
    }%
    \gplbacktext
    \put(0,0){\includegraphics{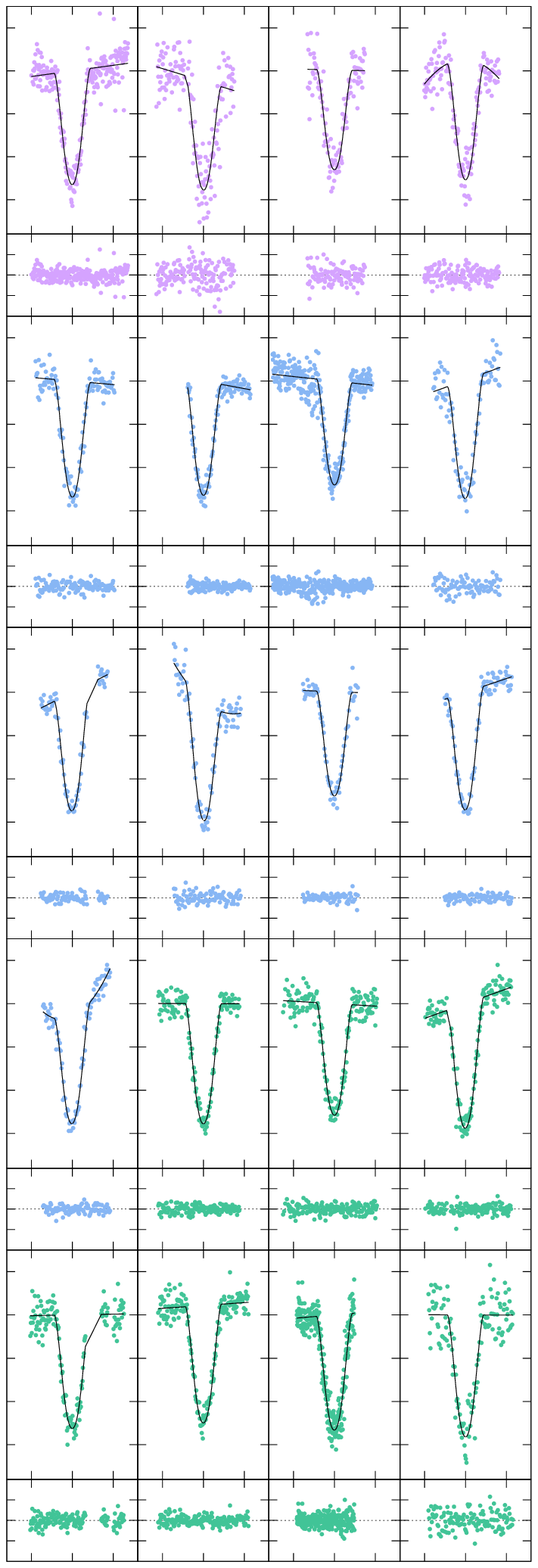}}%
    \gplfronttext
  \end{picture}%
\endgroup

% flatex input end: [./images/broad_band_lc/lc1.tex]

        \caption{Broad band light curves I. Light curves are sorted by the observation number given in Table~\ref{tab:photo-all-broad}. \emph{Upper panels:} Differential magnitude as a function of orbital phase. The black line shows the best fit model. \emph{Lower panels:} Residuals of the best fit model.}
        \label{fig:broad-band1}
\end{figure}
\begin{figure}[h!]
        % flatex input: [./images/broad_band_lc/lc2.tex]
% GNUPLOT: LaTeX picture with Postscript
\begingroup
  \makeatletter
  \providecommand\color[2][]{%
    \GenericError{(gnuplot) \space\space\space\@spaces}{%
      Package color not loaded in conjunction with
      terminal option `colourtext'%
    }{See the gnuplot documentation for explanation.%
    }{Either use 'blacktext' in gnuplot or load the package
      color.sty in LaTeX.}%
    \renewcommand\color[2][]{}%
  }%
  \providecommand\includegraphics[2][]{%
    \GenericError{(gnuplot) \space\space\space\@spaces}{%
      Package graphicx or graphics not loaded%
    }{See the gnuplot documentation for explanation.%
    }{The gnuplot epslatex terminal needs graphicx.sty or graphics.sty.}%
    \renewcommand\includegraphics[2][]{}%
  }%
  \providecommand\rotatebox[2]{#2}%
  \@ifundefined{ifGPcolor}{%
    \newif\ifGPcolor
    \GPcolortrue
  }{}%
  \@ifundefined{ifGPblacktext}{%
    \newif\ifGPblacktext
    \GPblacktexttrue
  }{}%
  % define a \g@addto@macro without @ in the name:
  \let\gplgaddtomacro\g@addto@macro
  % define empty templates for all commands taking text:
  \gdef\gplbacktext{}%
  \gdef\gplfronttext{}%
  \makeatother
  \ifGPblacktext
    % no textcolor at all
    \def\colorrgb#1{}%
    \def\colorgray#1{}%
  \else
    % gray or color?
    \ifGPcolor
      \def\colorrgb#1{\color[rgb]{#1}}%
      \def\colorgray#1{\color[gray]{#1}}%
      \expandafter\def\csname LTw\endcsname{\color{white}}%
      \expandafter\def\csname LTb\endcsname{\color{black}}%
      \expandafter\def\csname LTa\endcsname{\color{black}}%
      \expandafter\def\csname LT0\endcsname{\color[rgb]{1,0,0}}%
      \expandafter\def\csname LT1\endcsname{\color[rgb]{0,1,0}}%
      \expandafter\def\csname LT2\endcsname{\color[rgb]{0,0,1}}%
      \expandafter\def\csname LT3\endcsname{\color[rgb]{1,0,1}}%
      \expandafter\def\csname LT4\endcsname{\color[rgb]{0,1,1}}%
      \expandafter\def\csname LT5\endcsname{\color[rgb]{1,1,0}}%
      \expandafter\def\csname LT6\endcsname{\color[rgb]{0,0,0}}%
      \expandafter\def\csname LT7\endcsname{\color[rgb]{1,0.3,0}}%
      \expandafter\def\csname LT8\endcsname{\color[rgb]{0.5,0.5,0.5}}%
    \else
      % gray
      \def\colorrgb#1{\color{black}}%
      \def\colorgray#1{\color[gray]{#1}}%
      \expandafter\def\csname LTw\endcsname{\color{white}}%
      \expandafter\def\csname LTb\endcsname{\color{black}}%
      \expandafter\def\csname LTa\endcsname{\color{black}}%
      \expandafter\def\csname LT0\endcsname{\color{black}}%
      \expandafter\def\csname LT1\endcsname{\color{black}}%
      \expandafter\def\csname LT2\endcsname{\color{black}}%
      \expandafter\def\csname LT3\endcsname{\color{black}}%
      \expandafter\def\csname LT4\endcsname{\color{black}}%
      \expandafter\def\csname LT5\endcsname{\color{black}}%
      \expandafter\def\csname LT6\endcsname{\color{black}}%
      \expandafter\def\csname LT7\endcsname{\color{black}}%
      \expandafter\def\csname LT8\endcsname{\color{black}}%
    \fi
  \fi
    \setlength{\unitlength}{0.0500bp}%
    \ifx\gptboxheight\undefined%
      \newlength{\gptboxheight}%
      \newlength{\gptboxwidth}%
      \newsavebox{\gptboxtext}%
    \fi%
    \setlength{\fboxrule}{0.5pt}%
    \setlength{\fboxsep}{1pt}%
\begin{picture}(4988.00,12472.00)%
    \gplgaddtomacro\gplbacktext{%
      \csname LTb\endcsname%
      \put(815,12295){\makebox(0,0)[r]{\strut{}$-0.01$}}%
      \put(815,11968){\makebox(0,0)[r]{\strut{}$0$}}%
      \put(815,11641){\makebox(0,0)[r]{\strut{}$0.01$}}%
      \put(815,11314){\makebox(0,0)[r]{\strut{}$0.02$}}%
      \put(815,10987){\makebox(0,0)[r]{\strut{}$0.03$}}%
      \put(1134,10505){\makebox(0,0){\strut{}}}%
      \put(1446,10505){\makebox(0,0){\strut{}}}%
      \put(1757,10505){\makebox(0,0){\strut{}}}%
      \put(1047,10898){\makebox(0,0)[l]{\strut{}21}}%
    }%
    \gplgaddtomacro\gplfronttext{%
      \csname LTb\endcsname%
      \put(150,11591){\rotatebox{-270}{\makebox(0,0){\strut{}$\Delta m$}}}%
    }%
    \gplgaddtomacro\gplbacktext{%
      \csname LTb\endcsname%
      \put(815,10569){\makebox(0,0)[r]{\strut{}$-0.01$}}%
      \put(815,10414){\makebox(0,0)[r]{\strut{}$0$}}%
      \put(815,10258){\makebox(0,0)[r]{\strut{}$0.01$}}%
      \put(1134,9882){\makebox(0,0){\strut{}}}%
      \put(1446,9882){\makebox(0,0){\strut{}}}%
      \put(1757,9882){\makebox(0,0){\strut{}}}%
    }%
    \gplgaddtomacro\gplfronttext{%
      \csname LTb\endcsname%
      \put(150,10413){\rotatebox{-270}{\makebox(0,0){\strut{}O-C}}}%
    }%
    \gplgaddtomacro\gplbacktext{%
      \csname LTb\endcsname%
      \put(1813,12295){\makebox(0,0)[r]{\strut{}}}%
      \put(1813,11968){\makebox(0,0)[r]{\strut{}}}%
      \put(1813,11641){\makebox(0,0)[r]{\strut{}}}%
      \put(1813,11314){\makebox(0,0)[r]{\strut{}}}%
      \put(1813,10987){\makebox(0,0)[r]{\strut{}}}%
      \put(2132,10505){\makebox(0,0){\strut{}}}%
      \put(2444,10505){\makebox(0,0){\strut{}}}%
      \put(2755,10505){\makebox(0,0){\strut{}}}%
      \put(2045,10898){\makebox(0,0)[l]{\strut{}22}}%
    }%
    \gplgaddtomacro\gplfronttext{%
    }%
    \gplgaddtomacro\gplbacktext{%
      \csname LTb\endcsname%
      \put(1813,10569){\makebox(0,0)[r]{\strut{}}}%
      \put(1813,10414){\makebox(0,0)[r]{\strut{}}}%
      \put(1813,10258){\makebox(0,0)[r]{\strut{}}}%
      \put(2132,9882){\makebox(0,0){\strut{}}}%
      \put(2444,9882){\makebox(0,0){\strut{}}}%
      \put(2755,9882){\makebox(0,0){\strut{}}}%
    }%
    \gplgaddtomacro\gplfronttext{%
    }%
    \gplgaddtomacro\gplbacktext{%
      \csname LTb\endcsname%
      \put(2810,12295){\makebox(0,0)[r]{\strut{}}}%
      \put(2810,11968){\makebox(0,0)[r]{\strut{}}}%
      \put(2810,11641){\makebox(0,0)[r]{\strut{}}}%
      \put(2810,11314){\makebox(0,0)[r]{\strut{}}}%
      \put(2810,10987){\makebox(0,0)[r]{\strut{}}}%
      \put(3129,10505){\makebox(0,0){\strut{}}}%
      \put(3441,10505){\makebox(0,0){\strut{}}}%
      \put(3752,10505){\makebox(0,0){\strut{}}}%
      \put(3042,10898){\makebox(0,0)[l]{\strut{}23}}%
    }%
    \gplgaddtomacro\gplfronttext{%
    }%
    \gplgaddtomacro\gplbacktext{%
      \csname LTb\endcsname%
      \put(2810,10569){\makebox(0,0)[r]{\strut{}}}%
      \put(2810,10414){\makebox(0,0)[r]{\strut{}}}%
      \put(2810,10258){\makebox(0,0)[r]{\strut{}}}%
      \put(3129,9882){\makebox(0,0){\strut{}}}%
      \put(3441,9882){\makebox(0,0){\strut{}}}%
      \put(3752,9882){\makebox(0,0){\strut{}}}%
    }%
    \gplgaddtomacro\gplfronttext{%
    }%
    \gplgaddtomacro\gplbacktext{%
      \csname LTb\endcsname%
      \put(3808,12295){\makebox(0,0)[r]{\strut{}}}%
      \put(3808,11968){\makebox(0,0)[r]{\strut{}}}%
      \put(3808,11641){\makebox(0,0)[r]{\strut{}}}%
      \put(3808,11314){\makebox(0,0)[r]{\strut{}}}%
      \put(3808,10987){\makebox(0,0)[r]{\strut{}}}%
      \put(4127,10505){\makebox(0,0){\strut{}}}%
      \put(4439,10505){\makebox(0,0){\strut{}}}%
      \put(4750,10505){\makebox(0,0){\strut{}}}%
      \put(4040,10898){\makebox(0,0)[l]{\strut{}24}}%
    }%
    \gplgaddtomacro\gplfronttext{%
    }%
    \gplgaddtomacro\gplbacktext{%
      \csname LTb\endcsname%
      \put(3808,10569){\makebox(0,0)[r]{\strut{}}}%
      \put(3808,10414){\makebox(0,0)[r]{\strut{}}}%
      \put(3808,10258){\makebox(0,0)[r]{\strut{}}}%
      \put(4127,9882){\makebox(0,0){\strut{}}}%
      \put(4439,9882){\makebox(0,0){\strut{}}}%
      \put(4750,9882){\makebox(0,0){\strut{}}}%
    }%
    \gplgaddtomacro\gplfronttext{%
    }%
    \gplgaddtomacro\gplbacktext{%
      \csname LTb\endcsname%
      \put(815,9936){\makebox(0,0)[r]{\strut{}$-0.01$}}%
      \put(815,9607){\makebox(0,0)[r]{\strut{}$0$}}%
      \put(815,9278){\makebox(0,0)[r]{\strut{}$0.01$}}%
      \put(815,8949){\makebox(0,0)[r]{\strut{}$0.02$}}%
      \put(815,8619){\makebox(0,0)[r]{\strut{}$0.03$}}%
      \put(1134,8136){\makebox(0,0){\strut{}}}%
      \put(1446,8136){\makebox(0,0){\strut{}}}%
      \put(1757,8136){\makebox(0,0){\strut{}}}%
      \put(1047,8531){\makebox(0,0)[l]{\strut{}25}}%
    }%
    \gplgaddtomacro\gplfronttext{%
      \csname LTb\endcsname%
      \put(150,9228){\rotatebox{-270}{\makebox(0,0){\strut{}$\Delta m$}}}%
    }%
    \gplgaddtomacro\gplbacktext{%
      \csname LTb\endcsname%
      \put(815,8199){\makebox(0,0)[r]{\strut{}$-0.01$}}%
      \put(815,8044){\makebox(0,0)[r]{\strut{}$0$}}%
      \put(815,7888){\makebox(0,0)[r]{\strut{}$0.01$}}%
      \put(1134,7512){\makebox(0,0){\strut{}}}%
      \put(1446,7512){\makebox(0,0){\strut{}}}%
      \put(1757,7512){\makebox(0,0){\strut{}}}%
    }%
    \gplgaddtomacro\gplfronttext{%
      \csname LTb\endcsname%
      \put(150,8043){\rotatebox{-270}{\makebox(0,0){\strut{}O-C}}}%
    }%
    \gplgaddtomacro\gplbacktext{%
      \csname LTb\endcsname%
      \put(1813,9936){\makebox(0,0)[r]{\strut{}}}%
      \put(1813,9607){\makebox(0,0)[r]{\strut{}}}%
      \put(1813,9278){\makebox(0,0)[r]{\strut{}}}%
      \put(1813,8949){\makebox(0,0)[r]{\strut{}}}%
      \put(1813,8619){\makebox(0,0)[r]{\strut{}}}%
      \put(2132,8136){\makebox(0,0){\strut{}}}%
      \put(2444,8136){\makebox(0,0){\strut{}}}%
      \put(2755,8136){\makebox(0,0){\strut{}}}%
      \put(2045,8531){\makebox(0,0)[l]{\strut{}26}}%
    }%
    \gplgaddtomacro\gplfronttext{%
    }%
    \gplgaddtomacro\gplbacktext{%
      \csname LTb\endcsname%
      \put(1813,8199){\makebox(0,0)[r]{\strut{}}}%
      \put(1813,8044){\makebox(0,0)[r]{\strut{}}}%
      \put(1813,7888){\makebox(0,0)[r]{\strut{}}}%
      \put(2132,7512){\makebox(0,0){\strut{}}}%
      \put(2444,7512){\makebox(0,0){\strut{}}}%
      \put(2755,7512){\makebox(0,0){\strut{}}}%
    }%
    \gplgaddtomacro\gplfronttext{%
    }%
    \gplgaddtomacro\gplbacktext{%
      \csname LTb\endcsname%
      \put(2810,9936){\makebox(0,0)[r]{\strut{}}}%
      \put(2810,9607){\makebox(0,0)[r]{\strut{}}}%
      \put(2810,9278){\makebox(0,0)[r]{\strut{}}}%
      \put(2810,8949){\makebox(0,0)[r]{\strut{}}}%
      \put(2810,8619){\makebox(0,0)[r]{\strut{}}}%
      \put(3129,8136){\makebox(0,0){\strut{}}}%
      \put(3441,8136){\makebox(0,0){\strut{}}}%
      \put(3752,8136){\makebox(0,0){\strut{}}}%
      \put(3042,8531){\makebox(0,0)[l]{\strut{}27}}%
    }%
    \gplgaddtomacro\gplfronttext{%
    }%
    \gplgaddtomacro\gplbacktext{%
      \csname LTb\endcsname%
      \put(2810,8199){\makebox(0,0)[r]{\strut{}}}%
      \put(2810,8044){\makebox(0,0)[r]{\strut{}}}%
      \put(2810,7888){\makebox(0,0)[r]{\strut{}}}%
      \put(3129,7512){\makebox(0,0){\strut{}}}%
      \put(3441,7512){\makebox(0,0){\strut{}}}%
      \put(3752,7512){\makebox(0,0){\strut{}}}%
    }%
    \gplgaddtomacro\gplfronttext{%
    }%
    \gplgaddtomacro\gplbacktext{%
      \csname LTb\endcsname%
      \put(3808,9936){\makebox(0,0)[r]{\strut{}}}%
      \put(3808,9607){\makebox(0,0)[r]{\strut{}}}%
      \put(3808,9278){\makebox(0,0)[r]{\strut{}}}%
      \put(3808,8949){\makebox(0,0)[r]{\strut{}}}%
      \put(3808,8619){\makebox(0,0)[r]{\strut{}}}%
      \put(4127,8136){\makebox(0,0){\strut{}}}%
      \put(4439,8136){\makebox(0,0){\strut{}}}%
      \put(4750,8136){\makebox(0,0){\strut{}}}%
      \put(4040,8531){\makebox(0,0)[l]{\strut{}28}}%
    }%
    \gplgaddtomacro\gplfronttext{%
    }%
    \gplgaddtomacro\gplbacktext{%
      \csname LTb\endcsname%
      \put(3808,8199){\makebox(0,0)[r]{\strut{}}}%
      \put(3808,8044){\makebox(0,0)[r]{\strut{}}}%
      \put(3808,7888){\makebox(0,0)[r]{\strut{}}}%
      \put(4127,7512){\makebox(0,0){\strut{}}}%
      \put(4439,7512){\makebox(0,0){\strut{}}}%
      \put(4750,7512){\makebox(0,0){\strut{}}}%
    }%
    \gplgaddtomacro\gplfronttext{%
    }%
    \gplgaddtomacro\gplbacktext{%
      \csname LTb\endcsname%
      \put(815,7567){\makebox(0,0)[r]{\strut{}$-0.01$}}%
      \put(815,7238){\makebox(0,0)[r]{\strut{}$0$}}%
      \put(815,6908){\makebox(0,0)[r]{\strut{}$0.01$}}%
      \put(815,6579){\makebox(0,0)[r]{\strut{}$0.02$}}%
      \put(815,6250){\makebox(0,0)[r]{\strut{}$0.03$}}%
      \put(1134,5766){\makebox(0,0){\strut{}}}%
      \put(1446,5766){\makebox(0,0){\strut{}}}%
      \put(1757,5766){\makebox(0,0){\strut{}}}%
      \put(1047,6161){\makebox(0,0)[l]{\strut{}29}}%
    }%
    \gplgaddtomacro\gplfronttext{%
      \csname LTb\endcsname%
      \put(150,6859){\rotatebox{-270}{\makebox(0,0){\strut{}$\Delta m$}}}%
    }%
    \gplgaddtomacro\gplbacktext{%
      \csname LTb\endcsname%
      \put(815,5830){\makebox(0,0)[r]{\strut{}$-0.01$}}%
      \put(815,5674){\makebox(0,0)[r]{\strut{}$0$}}%
      \put(815,5518){\makebox(0,0)[r]{\strut{}$0.01$}}%
      \put(1134,5142){\makebox(0,0){\strut{}}}%
      \put(1446,5142){\makebox(0,0){\strut{}}}%
      \put(1757,5142){\makebox(0,0){\strut{}}}%
    }%
    \gplgaddtomacro\gplfronttext{%
      \csname LTb\endcsname%
      \put(150,5674){\rotatebox{-270}{\makebox(0,0){\strut{}O-C}}}%
    }%
    \gplgaddtomacro\gplbacktext{%
      \csname LTb\endcsname%
      \put(1813,7567){\makebox(0,0)[r]{\strut{}}}%
      \put(1813,7238){\makebox(0,0)[r]{\strut{}}}%
      \put(1813,6908){\makebox(0,0)[r]{\strut{}}}%
      \put(1813,6579){\makebox(0,0)[r]{\strut{}}}%
      \put(1813,6250){\makebox(0,0)[r]{\strut{}}}%
      \put(2132,5766){\makebox(0,0){\strut{}}}%
      \put(2444,5766){\makebox(0,0){\strut{}}}%
      \put(2755,5766){\makebox(0,0){\strut{}}}%
      \put(2045,6161){\makebox(0,0)[l]{\strut{}30}}%
    }%
    \gplgaddtomacro\gplfronttext{%
    }%
    \gplgaddtomacro\gplbacktext{%
      \csname LTb\endcsname%
      \put(1813,5830){\makebox(0,0)[r]{\strut{}}}%
      \put(1813,5674){\makebox(0,0)[r]{\strut{}}}%
      \put(1813,5518){\makebox(0,0)[r]{\strut{}}}%
      \put(2132,5142){\makebox(0,0){\strut{}}}%
      \put(2444,5142){\makebox(0,0){\strut{}}}%
      \put(2755,5142){\makebox(0,0){\strut{}}}%
    }%
    \gplgaddtomacro\gplfronttext{%
    }%
    \gplgaddtomacro\gplbacktext{%
      \csname LTb\endcsname%
      \put(2810,7567){\makebox(0,0)[r]{\strut{}}}%
      \put(2810,7238){\makebox(0,0)[r]{\strut{}}}%
      \put(2810,6908){\makebox(0,0)[r]{\strut{}}}%
      \put(2810,6579){\makebox(0,0)[r]{\strut{}}}%
      \put(2810,6250){\makebox(0,0)[r]{\strut{}}}%
      \put(3129,5766){\makebox(0,0){\strut{}}}%
      \put(3441,5766){\makebox(0,0){\strut{}}}%
      \put(3752,5766){\makebox(0,0){\strut{}}}%
      \put(3042,6161){\makebox(0,0)[l]{\strut{}31}}%
    }%
    \gplgaddtomacro\gplfronttext{%
    }%
    \gplgaddtomacro\gplbacktext{%
      \csname LTb\endcsname%
      \put(2810,5830){\makebox(0,0)[r]{\strut{}}}%
      \put(2810,5674){\makebox(0,0)[r]{\strut{}}}%
      \put(2810,5518){\makebox(0,0)[r]{\strut{}}}%
      \put(3129,5142){\makebox(0,0){\strut{}}}%
      \put(3441,5142){\makebox(0,0){\strut{}}}%
      \put(3752,5142){\makebox(0,0){\strut{}}}%
    }%
    \gplgaddtomacro\gplfronttext{%
    }%
    \gplgaddtomacro\gplbacktext{%
      \csname LTb\endcsname%
      \put(3808,7567){\makebox(0,0)[r]{\strut{}}}%
      \put(3808,7238){\makebox(0,0)[r]{\strut{}}}%
      \put(3808,6908){\makebox(0,0)[r]{\strut{}}}%
      \put(3808,6579){\makebox(0,0)[r]{\strut{}}}%
      \put(3808,6250){\makebox(0,0)[r]{\strut{}}}%
      \put(4127,5766){\makebox(0,0){\strut{}}}%
      \put(4439,5766){\makebox(0,0){\strut{}}}%
      \put(4750,5766){\makebox(0,0){\strut{}}}%
      \put(4040,6161){\makebox(0,0)[l]{\strut{}32}}%
    }%
    \gplgaddtomacro\gplfronttext{%
    }%
    \gplgaddtomacro\gplbacktext{%
      \csname LTb\endcsname%
      \put(3808,5830){\makebox(0,0)[r]{\strut{}}}%
      \put(3808,5674){\makebox(0,0)[r]{\strut{}}}%
      \put(3808,5518){\makebox(0,0)[r]{\strut{}}}%
      \put(4127,5142){\makebox(0,0){\strut{}}}%
      \put(4439,5142){\makebox(0,0){\strut{}}}%
      \put(4750,5142){\makebox(0,0){\strut{}}}%
    }%
    \gplgaddtomacro\gplfronttext{%
    }%
    \gplgaddtomacro\gplbacktext{%
      \csname LTb\endcsname%
      \put(815,5197){\makebox(0,0)[r]{\strut{}$-0.01$}}%
      \put(815,4868){\makebox(0,0)[r]{\strut{}$0$}}%
      \put(815,4538){\makebox(0,0)[r]{\strut{}$0.01$}}%
      \put(815,4209){\makebox(0,0)[r]{\strut{}$0.02$}}%
      \put(815,3880){\makebox(0,0)[r]{\strut{}$0.03$}}%
      \put(1134,3396){\makebox(0,0){\strut{}}}%
      \put(1446,3396){\makebox(0,0){\strut{}}}%
      \put(1757,3396){\makebox(0,0){\strut{}}}%
      \put(1047,3791){\makebox(0,0)[l]{\strut{}33}}%
    }%
    \gplgaddtomacro\gplfronttext{%
      \csname LTb\endcsname%
      \put(150,4489){\rotatebox{-270}{\makebox(0,0){\strut{}$\Delta m$}}}%
    }%
    \gplgaddtomacro\gplbacktext{%
      \csname LTb\endcsname%
      \put(815,3460){\makebox(0,0)[r]{\strut{}$-0.01$}}%
      \put(815,3305){\makebox(0,0)[r]{\strut{}$0$}}%
      \put(815,3149){\makebox(0,0)[r]{\strut{}$0.01$}}%
      \put(1134,2773){\makebox(0,0){\strut{}}}%
      \put(1446,2773){\makebox(0,0){\strut{}}}%
      \put(1757,2773){\makebox(0,0){\strut{}}}%
    }%
    \gplgaddtomacro\gplfronttext{%
      \csname LTb\endcsname%
      \put(150,3304){\rotatebox{-270}{\makebox(0,0){\strut{}O-C}}}%
    }%
    \gplgaddtomacro\gplbacktext{%
      \csname LTb\endcsname%
      \put(1813,5197){\makebox(0,0)[r]{\strut{}}}%
      \put(1813,4868){\makebox(0,0)[r]{\strut{}}}%
      \put(1813,4538){\makebox(0,0)[r]{\strut{}}}%
      \put(1813,4209){\makebox(0,0)[r]{\strut{}}}%
      \put(1813,3880){\makebox(0,0)[r]{\strut{}}}%
      \put(2132,3396){\makebox(0,0){\strut{}}}%
      \put(2444,3396){\makebox(0,0){\strut{}}}%
      \put(2755,3396){\makebox(0,0){\strut{}}}%
      \put(2045,3791){\makebox(0,0)[l]{\strut{}34}}%
    }%
    \gplgaddtomacro\gplfronttext{%
    }%
    \gplgaddtomacro\gplbacktext{%
      \csname LTb\endcsname%
      \put(1813,3460){\makebox(0,0)[r]{\strut{}}}%
      \put(1813,3305){\makebox(0,0)[r]{\strut{}}}%
      \put(1813,3149){\makebox(0,0)[r]{\strut{}}}%
      \put(2132,2773){\makebox(0,0){\strut{}}}%
      \put(2444,2773){\makebox(0,0){\strut{}}}%
      \put(2755,2773){\makebox(0,0){\strut{}}}%
    }%
    \gplgaddtomacro\gplfronttext{%
    }%
    \gplgaddtomacro\gplbacktext{%
      \csname LTb\endcsname%
      \put(2810,5197){\makebox(0,0)[r]{\strut{}}}%
      \put(2810,4868){\makebox(0,0)[r]{\strut{}}}%
      \put(2810,4538){\makebox(0,0)[r]{\strut{}}}%
      \put(2810,4209){\makebox(0,0)[r]{\strut{}}}%
      \put(2810,3880){\makebox(0,0)[r]{\strut{}}}%
      \put(3129,3396){\makebox(0,0){\strut{}}}%
      \put(3441,3396){\makebox(0,0){\strut{}}}%
      \put(3752,3396){\makebox(0,0){\strut{}}}%
      \put(3042,3791){\makebox(0,0)[l]{\strut{}35}}%
    }%
    \gplgaddtomacro\gplfronttext{%
    }%
    \gplgaddtomacro\gplbacktext{%
      \csname LTb\endcsname%
      \put(2810,3460){\makebox(0,0)[r]{\strut{}}}%
      \put(2810,3305){\makebox(0,0)[r]{\strut{}}}%
      \put(2810,3149){\makebox(0,0)[r]{\strut{}}}%
      \put(3129,2773){\makebox(0,0){\strut{}}}%
      \put(3441,2773){\makebox(0,0){\strut{}}}%
      \put(3752,2773){\makebox(0,0){\strut{}}}%
    }%
    \gplgaddtomacro\gplfronttext{%
    }%
    \gplgaddtomacro\gplbacktext{%
      \csname LTb\endcsname%
      \put(3808,5197){\makebox(0,0)[r]{\strut{}}}%
      \put(3808,4868){\makebox(0,0)[r]{\strut{}}}%
      \put(3808,4538){\makebox(0,0)[r]{\strut{}}}%
      \put(3808,4209){\makebox(0,0)[r]{\strut{}}}%
      \put(3808,3880){\makebox(0,0)[r]{\strut{}}}%
      \put(4127,3396){\makebox(0,0){\strut{}}}%
      \put(4439,3396){\makebox(0,0){\strut{}}}%
      \put(4750,3396){\makebox(0,0){\strut{}}}%
      \put(4040,3791){\makebox(0,0)[l]{\strut{}36}}%
    }%
    \gplgaddtomacro\gplfronttext{%
    }%
    \gplgaddtomacro\gplbacktext{%
      \csname LTb\endcsname%
      \put(3808,3460){\makebox(0,0)[r]{\strut{}}}%
      \put(3808,3305){\makebox(0,0)[r]{\strut{}}}%
      \put(3808,3149){\makebox(0,0)[r]{\strut{}}}%
      \put(4127,2773){\makebox(0,0){\strut{}}}%
      \put(4439,2773){\makebox(0,0){\strut{}}}%
      \put(4750,2773){\makebox(0,0){\strut{}}}%
    }%
    \gplgaddtomacro\gplfronttext{%
    }%
    \gplgaddtomacro\gplbacktext{%
      \csname LTb\endcsname%
      \put(815,2828){\makebox(0,0)[r]{\strut{}$-0.01$}}%
      \put(815,2499){\makebox(0,0)[r]{\strut{}$0$}}%
      \put(815,2169){\makebox(0,0)[r]{\strut{}$0.01$}}%
      \put(815,1840){\makebox(0,0)[r]{\strut{}$0.02$}}%
      \put(815,1511){\makebox(0,0)[r]{\strut{}$0.03$}}%
      \put(1134,1027){\makebox(0,0){\strut{}}}%
      \put(1446,1027){\makebox(0,0){\strut{}}}%
      \put(1757,1027){\makebox(0,0){\strut{}}}%
      \put(1047,1422){\makebox(0,0)[l]{\strut{}37}}%
    }%
    \gplgaddtomacro\gplfronttext{%
      \csname LTb\endcsname%
      \put(150,2120){\rotatebox{-270}{\makebox(0,0){\strut{}$\Delta m$}}}%
    }%
    \gplgaddtomacro\gplbacktext{%
      \csname LTb\endcsname%
      \put(815,1091){\makebox(0,0)[r]{\strut{}$-0.01$}}%
      \put(815,935){\makebox(0,0)[r]{\strut{}$0$}}%
      \put(815,779){\makebox(0,0)[r]{\strut{}$0.01$}}%
      \put(1134,403){\makebox(0,0){\strut{}$-0.05$}}%
      \put(1446,403){\makebox(0,0){\strut{}$0$}}%
      \put(1757,403){\makebox(0,0){\strut{}$0.05$}}%
    }%
    \gplgaddtomacro\gplfronttext{%
      \csname LTb\endcsname%
      \put(150,935){\rotatebox{-270}{\makebox(0,0){\strut{}O-C}}}%
      \put(1445,73){\makebox(0,0){\strut{}Phase}}%
    }%
    \gplgaddtomacro\gplbacktext{%
      \csname LTb\endcsname%
      \put(1813,2828){\makebox(0,0)[r]{\strut{}}}%
      \put(1813,2499){\makebox(0,0)[r]{\strut{}}}%
      \put(1813,2169){\makebox(0,0)[r]{\strut{}}}%
      \put(1813,1840){\makebox(0,0)[r]{\strut{}}}%
      \put(1813,1511){\makebox(0,0)[r]{\strut{}}}%
      \put(2132,1027){\makebox(0,0){\strut{}}}%
      \put(2444,1027){\makebox(0,0){\strut{}}}%
      \put(2755,1027){\makebox(0,0){\strut{}}}%
      \put(2045,1422){\makebox(0,0)[l]{\strut{}38}}%
    }%
    \gplgaddtomacro\gplfronttext{%
    }%
    \gplgaddtomacro\gplbacktext{%
      \csname LTb\endcsname%
      \put(1813,1091){\makebox(0,0)[r]{\strut{}}}%
      \put(1813,935){\makebox(0,0)[r]{\strut{}}}%
      \put(1813,779){\makebox(0,0)[r]{\strut{}}}%
      \put(2132,403){\makebox(0,0){\strut{}$-0.05$}}%
      \put(2444,403){\makebox(0,0){\strut{}$0$}}%
      \put(2755,403){\makebox(0,0){\strut{}$0.05$}}%
    }%
    \gplgaddtomacro\gplfronttext{%
      \csname LTb\endcsname%
      \put(2443,73){\makebox(0,0){\strut{}Phase}}%
    }%
    \gplgaddtomacro\gplbacktext{%
      \csname LTb\endcsname%
      \put(2810,2828){\makebox(0,0)[r]{\strut{}}}%
      \put(2810,2499){\makebox(0,0)[r]{\strut{}}}%
      \put(2810,2169){\makebox(0,0)[r]{\strut{}}}%
      \put(2810,1840){\makebox(0,0)[r]{\strut{}}}%
      \put(2810,1511){\makebox(0,0)[r]{\strut{}}}%
      \put(3129,1027){\makebox(0,0){\strut{}}}%
      \put(3441,1027){\makebox(0,0){\strut{}}}%
      \put(3752,1027){\makebox(0,0){\strut{}}}%
      \put(3042,1422){\makebox(0,0)[l]{\strut{}39}}%
    }%
    \gplgaddtomacro\gplfronttext{%
    }%
    \gplgaddtomacro\gplbacktext{%
      \csname LTb\endcsname%
      \put(2810,1091){\makebox(0,0)[r]{\strut{}}}%
      \put(2810,935){\makebox(0,0)[r]{\strut{}}}%
      \put(2810,779){\makebox(0,0)[r]{\strut{}}}%
      \put(3129,403){\makebox(0,0){\strut{}$-0.05$}}%
      \put(3441,403){\makebox(0,0){\strut{}$0$}}%
      \put(3752,403){\makebox(0,0){\strut{}$0.05$}}%
    }%
    \gplgaddtomacro\gplfronttext{%
      \csname LTb\endcsname%
      \put(3440,73){\makebox(0,0){\strut{}Phase}}%
    }%
    \gplgaddtomacro\gplbacktext{%
      \csname LTb\endcsname%
      \put(3808,2828){\makebox(0,0)[r]{\strut{}}}%
      \put(3808,2499){\makebox(0,0)[r]{\strut{}}}%
      \put(3808,2169){\makebox(0,0)[r]{\strut{}}}%
      \put(3808,1840){\makebox(0,0)[r]{\strut{}}}%
      \put(3808,1511){\makebox(0,0)[r]{\strut{}}}%
      \put(4127,1027){\makebox(0,0){\strut{}}}%
      \put(4439,1027){\makebox(0,0){\strut{}}}%
      \put(4750,1027){\makebox(0,0){\strut{}}}%
      \put(4040,1422){\makebox(0,0)[l]{\strut{}40}}%
    }%
    \gplgaddtomacro\gplfronttext{%
    }%
    \gplgaddtomacro\gplbacktext{%
      \csname LTb\endcsname%
      \put(3808,1091){\makebox(0,0)[r]{\strut{}}}%
      \put(3808,935){\makebox(0,0)[r]{\strut{}}}%
      \put(3808,779){\makebox(0,0)[r]{\strut{}}}%
      \put(4127,403){\makebox(0,0){\strut{}$-0.05$}}%
      \put(4439,403){\makebox(0,0){\strut{}$0$}}%
      \put(4750,403){\makebox(0,0){\strut{}$0.05$}}%
    }%
    \gplgaddtomacro\gplfronttext{%
      \csname LTb\endcsname%
      \put(4438,73){\makebox(0,0){\strut{}Phase}}%
    }%
    \gplbacktext
    \put(0,0){\includegraphics{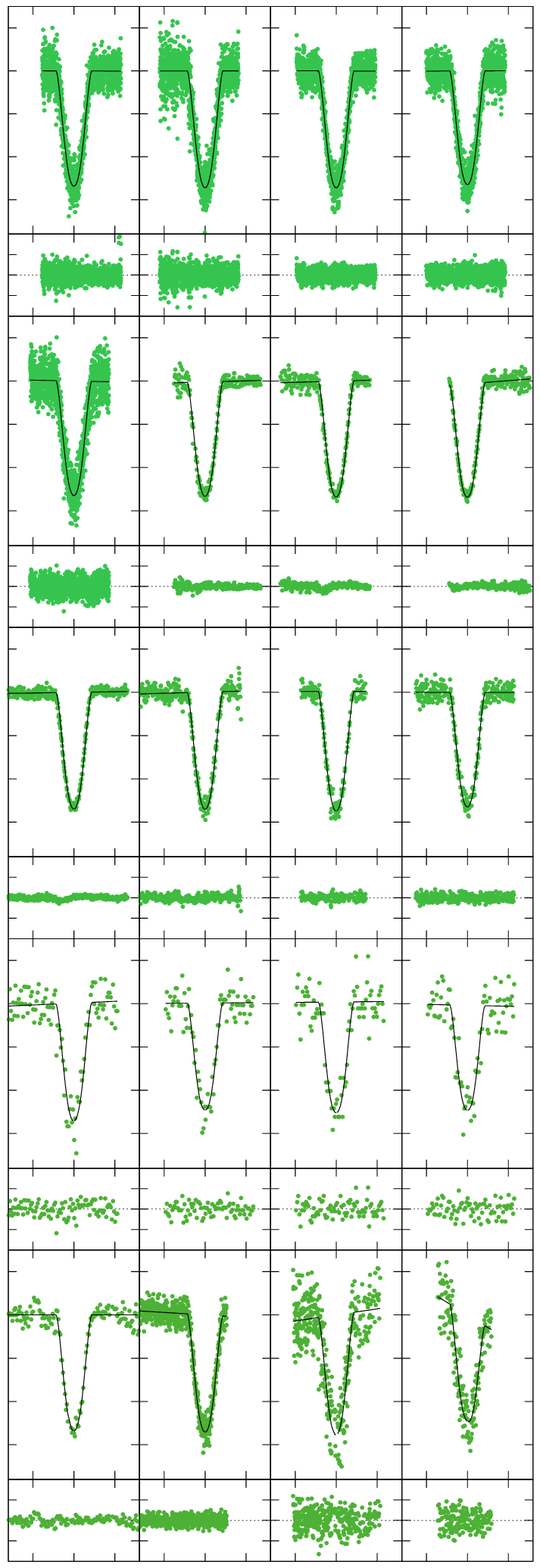}}%
    \gplfronttext
  \end{picture}%
\endgroup

% flatex input end: [./images/broad_band_lc/lc2.tex]

        \caption{Broad band light curves II.  Light curves are sorted by the observation number given in Table~\ref{tab:photo-all-broad}. This plot is a continuation of Fig.~\ref{fig:broad-band1}.}
        \label{fig:broad-band2}
\end{figure}
\begin{figure}[h!]
        % flatex input: [./images/broad_band_lc/lc3.tex]
% GNUPLOT: LaTeX picture with Postscript
\begingroup
  \makeatletter
  \providecommand\color[2][]{%
    \GenericError{(gnuplot) \space\space\space\@spaces}{%
      Package color not loaded in conjunction with
      terminal option `colourtext'%
    }{See the gnuplot documentation for explanation.%
    }{Either use 'blacktext' in gnuplot or load the package
      color.sty in LaTeX.}%
    \renewcommand\color[2][]{}%
  }%
  \providecommand\includegraphics[2][]{%
    \GenericError{(gnuplot) \space\space\space\@spaces}{%
      Package graphicx or graphics not loaded%
    }{See the gnuplot documentation for explanation.%
    }{The gnuplot epslatex terminal needs graphicx.sty or graphics.sty.}%
    \renewcommand\includegraphics[2][]{}%
  }%
  \providecommand\rotatebox[2]{#2}%
  \@ifundefined{ifGPcolor}{%
    \newif\ifGPcolor
    \GPcolortrue
  }{}%
  \@ifundefined{ifGPblacktext}{%
    \newif\ifGPblacktext
    \GPblacktexttrue
  }{}%
  % define a \g@addto@macro without @ in the name:
  \let\gplgaddtomacro\g@addto@macro
  % define empty templates for all commands taking text:
  \gdef\gplbacktext{}%
  \gdef\gplfronttext{}%
  \makeatother
  \ifGPblacktext
    % no textcolor at all
    \def\colorrgb#1{}%
    \def\colorgray#1{}%
  \else
    % gray or color?
    \ifGPcolor
      \def\colorrgb#1{\color[rgb]{#1}}%
      \def\colorgray#1{\color[gray]{#1}}%
      \expandafter\def\csname LTw\endcsname{\color{white}}%
      \expandafter\def\csname LTb\endcsname{\color{black}}%
      \expandafter\def\csname LTa\endcsname{\color{black}}%
      \expandafter\def\csname LT0\endcsname{\color[rgb]{1,0,0}}%
      \expandafter\def\csname LT1\endcsname{\color[rgb]{0,1,0}}%
      \expandafter\def\csname LT2\endcsname{\color[rgb]{0,0,1}}%
      \expandafter\def\csname LT3\endcsname{\color[rgb]{1,0,1}}%
      \expandafter\def\csname LT4\endcsname{\color[rgb]{0,1,1}}%
      \expandafter\def\csname LT5\endcsname{\color[rgb]{1,1,0}}%
      \expandafter\def\csname LT6\endcsname{\color[rgb]{0,0,0}}%
      \expandafter\def\csname LT7\endcsname{\color[rgb]{1,0.3,0}}%
      \expandafter\def\csname LT8\endcsname{\color[rgb]{0.5,0.5,0.5}}%
    \else
      % gray
      \def\colorrgb#1{\color{black}}%
      \def\colorgray#1{\color[gray]{#1}}%
      \expandafter\def\csname LTw\endcsname{\color{white}}%
      \expandafter\def\csname LTb\endcsname{\color{black}}%
      \expandafter\def\csname LTa\endcsname{\color{black}}%
      \expandafter\def\csname LT0\endcsname{\color{black}}%
      \expandafter\def\csname LT1\endcsname{\color{black}}%
      \expandafter\def\csname LT2\endcsname{\color{black}}%
      \expandafter\def\csname LT3\endcsname{\color{black}}%
      \expandafter\def\csname LT4\endcsname{\color{black}}%
      \expandafter\def\csname LT5\endcsname{\color{black}}%
      \expandafter\def\csname LT6\endcsname{\color{black}}%
      \expandafter\def\csname LT7\endcsname{\color{black}}%
      \expandafter\def\csname LT8\endcsname{\color{black}}%
    \fi
  \fi
    \setlength{\unitlength}{0.0500bp}%
    \ifx\gptboxheight\undefined%
      \newlength{\gptboxheight}%
      \newlength{\gptboxwidth}%
      \newsavebox{\gptboxtext}%
    \fi%
    \setlength{\fboxrule}{0.5pt}%
    \setlength{\fboxsep}{1pt}%
\begin{picture}(4988.00,7482.00)%
    \gplgaddtomacro\gplbacktext{%
      \csname LTb\endcsname%
      \put(815,7309){\makebox(0,0)[r]{\strut{}$-0.01$}}%
      \put(815,6980){\makebox(0,0)[r]{\strut{}$0$}}%
      \put(815,6651){\makebox(0,0)[r]{\strut{}$0.01$}}%
      \put(815,6323){\makebox(0,0)[r]{\strut{}$0.02$}}%
      \put(815,5994){\makebox(0,0)[r]{\strut{}$0.03$}}%
      \put(1134,5511){\makebox(0,0){\strut{}}}%
      \put(1446,5511){\makebox(0,0){\strut{}}}%
      \put(1757,5511){\makebox(0,0){\strut{}}}%
      \put(1047,5905){\makebox(0,0)[l]{\strut{}41}}%
    }%
    \gplgaddtomacro\gplfronttext{%
      \csname LTb\endcsname%
      \put(150,6602){\rotatebox{-270}{\makebox(0,0){\strut{}$\Delta m$}}}%
    }%
    \gplgaddtomacro\gplbacktext{%
      \csname LTb\endcsname%
      \put(815,5575){\makebox(0,0)[r]{\strut{}$-0.01$}}%
      \put(815,5420){\makebox(0,0)[r]{\strut{}$0$}}%
      \put(815,5265){\makebox(0,0)[r]{\strut{}$0.01$}}%
      \put(1134,4890){\makebox(0,0){\strut{}}}%
      \put(1446,4890){\makebox(0,0){\strut{}}}%
      \put(1757,4890){\makebox(0,0){\strut{}}}%
    }%
    \gplgaddtomacro\gplfronttext{%
      \csname LTb\endcsname%
      \put(150,5420){\rotatebox{-270}{\makebox(0,0){\strut{}O-C}}}%
    }%
    \gplgaddtomacro\gplbacktext{%
      \csname LTb\endcsname%
      \put(1813,7309){\makebox(0,0)[r]{\strut{}}}%
      \put(1813,6980){\makebox(0,0)[r]{\strut{}}}%
      \put(1813,6651){\makebox(0,0)[r]{\strut{}}}%
      \put(1813,6323){\makebox(0,0)[r]{\strut{}}}%
      \put(1813,5994){\makebox(0,0)[r]{\strut{}}}%
      \put(2132,5511){\makebox(0,0){\strut{}}}%
      \put(2444,5511){\makebox(0,0){\strut{}}}%
      \put(2755,5511){\makebox(0,0){\strut{}}}%
      \put(2045,5905){\makebox(0,0)[l]{\strut{}42}}%
    }%
    \gplgaddtomacro\gplfronttext{%
    }%
    \gplgaddtomacro\gplbacktext{%
      \csname LTb\endcsname%
      \put(1813,5575){\makebox(0,0)[r]{\strut{}}}%
      \put(1813,5420){\makebox(0,0)[r]{\strut{}}}%
      \put(1813,5265){\makebox(0,0)[r]{\strut{}}}%
      \put(2132,4890){\makebox(0,0){\strut{}}}%
      \put(2444,4890){\makebox(0,0){\strut{}}}%
      \put(2755,4890){\makebox(0,0){\strut{}}}%
    }%
    \gplgaddtomacro\gplfronttext{%
    }%
    \gplgaddtomacro\gplbacktext{%
      \csname LTb\endcsname%
      \put(2810,7309){\makebox(0,0)[r]{\strut{}}}%
      \put(2810,6980){\makebox(0,0)[r]{\strut{}}}%
      \put(2810,6651){\makebox(0,0)[r]{\strut{}}}%
      \put(2810,6323){\makebox(0,0)[r]{\strut{}}}%
      \put(2810,5994){\makebox(0,0)[r]{\strut{}}}%
      \put(3129,5511){\makebox(0,0){\strut{}}}%
      \put(3441,5511){\makebox(0,0){\strut{}}}%
      \put(3752,5511){\makebox(0,0){\strut{}}}%
      \put(3042,5905){\makebox(0,0)[l]{\strut{}43}}%
    }%
    \gplgaddtomacro\gplfronttext{%
    }%
    \gplgaddtomacro\gplbacktext{%
      \csname LTb\endcsname%
      \put(2810,5575){\makebox(0,0)[r]{\strut{}}}%
      \put(2810,5420){\makebox(0,0)[r]{\strut{}}}%
      \put(2810,5265){\makebox(0,0)[r]{\strut{}}}%
      \put(3129,4890){\makebox(0,0){\strut{}}}%
      \put(3441,4890){\makebox(0,0){\strut{}}}%
      \put(3752,4890){\makebox(0,0){\strut{}}}%
    }%
    \gplgaddtomacro\gplfronttext{%
    }%
    \gplgaddtomacro\gplbacktext{%
      \csname LTb\endcsname%
      \put(3808,7309){\makebox(0,0)[r]{\strut{}}}%
      \put(3808,6980){\makebox(0,0)[r]{\strut{}}}%
      \put(3808,6651){\makebox(0,0)[r]{\strut{}}}%
      \put(3808,6323){\makebox(0,0)[r]{\strut{}}}%
      \put(3808,5994){\makebox(0,0)[r]{\strut{}}}%
      \put(4127,5511){\makebox(0,0){\strut{}}}%
      \put(4439,5511){\makebox(0,0){\strut{}}}%
      \put(4750,5511){\makebox(0,0){\strut{}}}%
      \put(4040,5905){\makebox(0,0)[l]{\strut{}44}}%
    }%
    \gplgaddtomacro\gplfronttext{%
    }%
    \gplgaddtomacro\gplbacktext{%
      \csname LTb\endcsname%
      \put(3808,5575){\makebox(0,0)[r]{\strut{}}}%
      \put(3808,5420){\makebox(0,0)[r]{\strut{}}}%
      \put(3808,5265){\makebox(0,0)[r]{\strut{}}}%
      \put(4127,4890){\makebox(0,0){\strut{}}}%
      \put(4439,4890){\makebox(0,0){\strut{}}}%
      \put(4750,4890){\makebox(0,0){\strut{}}}%
    }%
    \gplgaddtomacro\gplfronttext{%
    }%
    \gplgaddtomacro\gplbacktext{%
      \csname LTb\endcsname%
      \put(815,4944){\makebox(0,0)[r]{\strut{}$-0.01$}}%
      \put(815,4614){\makebox(0,0)[r]{\strut{}$0$}}%
      \put(815,4284){\makebox(0,0)[r]{\strut{}$0.01$}}%
      \put(815,3953){\makebox(0,0)[r]{\strut{}$0.02$}}%
      \put(815,3623){\makebox(0,0)[r]{\strut{}$0.03$}}%
      \put(1134,3139){\makebox(0,0){\strut{}}}%
      \put(1446,3139){\makebox(0,0){\strut{}}}%
      \put(1757,3139){\makebox(0,0){\strut{}}}%
      \put(1047,3534){\makebox(0,0)[l]{\strut{}45}}%
    }%
    \gplgaddtomacro\gplfronttext{%
      \csname LTb\endcsname%
      \put(150,4234){\rotatebox{-270}{\makebox(0,0){\strut{}$\Delta m$}}}%
    }%
    \gplgaddtomacro\gplbacktext{%
      \csname LTb\endcsname%
      \put(815,3203){\makebox(0,0)[r]{\strut{}$-0.01$}}%
      \put(815,3048){\makebox(0,0)[r]{\strut{}$0$}}%
      \put(815,2893){\makebox(0,0)[r]{\strut{}$0.01$}}%
      \put(1134,2518){\makebox(0,0){\strut{}}}%
      \put(1446,2518){\makebox(0,0){\strut{}}}%
      \put(1757,2518){\makebox(0,0){\strut{}}}%
    }%
    \gplgaddtomacro\gplfronttext{%
      \csname LTb\endcsname%
      \put(150,3048){\rotatebox{-270}{\makebox(0,0){\strut{}O-C}}}%
    }%
    \gplgaddtomacro\gplbacktext{%
      \csname LTb\endcsname%
      \put(1813,4944){\makebox(0,0)[r]{\strut{}}}%
      \put(1813,4614){\makebox(0,0)[r]{\strut{}}}%
      \put(1813,4284){\makebox(0,0)[r]{\strut{}}}%
      \put(1813,3953){\makebox(0,0)[r]{\strut{}}}%
      \put(1813,3623){\makebox(0,0)[r]{\strut{}}}%
      \put(2132,3139){\makebox(0,0){\strut{}}}%
      \put(2444,3139){\makebox(0,0){\strut{}}}%
      \put(2755,3139){\makebox(0,0){\strut{}}}%
      \put(2045,3534){\makebox(0,0)[l]{\strut{}46}}%
    }%
    \gplgaddtomacro\gplfronttext{%
    }%
    \gplgaddtomacro\gplbacktext{%
      \csname LTb\endcsname%
      \put(1813,3203){\makebox(0,0)[r]{\strut{}}}%
      \put(1813,3048){\makebox(0,0)[r]{\strut{}}}%
      \put(1813,2893){\makebox(0,0)[r]{\strut{}}}%
      \put(2132,2518){\makebox(0,0){\strut{}}}%
      \put(2444,2518){\makebox(0,0){\strut{}}}%
      \put(2755,2518){\makebox(0,0){\strut{}}}%
    }%
    \gplgaddtomacro\gplfronttext{%
    }%
    \gplgaddtomacro\gplbacktext{%
      \csname LTb\endcsname%
      \put(2810,4944){\makebox(0,0)[r]{\strut{}}}%
      \put(2810,4614){\makebox(0,0)[r]{\strut{}}}%
      \put(2810,4284){\makebox(0,0)[r]{\strut{}}}%
      \put(2810,3953){\makebox(0,0)[r]{\strut{}}}%
      \put(2810,3623){\makebox(0,0)[r]{\strut{}}}%
      \put(3129,3139){\makebox(0,0){\strut{}}}%
      \put(3441,3139){\makebox(0,0){\strut{}}}%
      \put(3752,3139){\makebox(0,0){\strut{}}}%
      \put(3042,3534){\makebox(0,0)[l]{\strut{}47}}%
    }%
    \gplgaddtomacro\gplfronttext{%
    }%
    \gplgaddtomacro\gplbacktext{%
      \csname LTb\endcsname%
      \put(2810,3203){\makebox(0,0)[r]{\strut{}}}%
      \put(2810,3048){\makebox(0,0)[r]{\strut{}}}%
      \put(2810,2893){\makebox(0,0)[r]{\strut{}}}%
      \put(3129,2518){\makebox(0,0){\strut{}}}%
      \put(3441,2518){\makebox(0,0){\strut{}}}%
      \put(3752,2518){\makebox(0,0){\strut{}}}%
    }%
    \gplgaddtomacro\gplfronttext{%
    }%
    \gplgaddtomacro\gplbacktext{%
      \csname LTb\endcsname%
      \put(3808,4944){\makebox(0,0)[r]{\strut{}}}%
      \put(3808,4614){\makebox(0,0)[r]{\strut{}}}%
      \put(3808,4284){\makebox(0,0)[r]{\strut{}}}%
      \put(3808,3953){\makebox(0,0)[r]{\strut{}}}%
      \put(3808,3623){\makebox(0,0)[r]{\strut{}}}%
      \put(4127,3139){\makebox(0,0){\strut{}}}%
      \put(4439,3139){\makebox(0,0){\strut{}}}%
      \put(4750,3139){\makebox(0,0){\strut{}}}%
      \put(4040,3534){\makebox(0,0)[l]{\strut{}48}}%
    }%
    \gplgaddtomacro\gplfronttext{%
    }%
    \gplgaddtomacro\gplbacktext{%
      \csname LTb\endcsname%
      \put(3808,3203){\makebox(0,0)[r]{\strut{}}}%
      \put(3808,3048){\makebox(0,0)[r]{\strut{}}}%
      \put(3808,2893){\makebox(0,0)[r]{\strut{}}}%
      \put(4127,2518){\makebox(0,0){\strut{}$-0.05$}}%
      \put(4439,2518){\makebox(0,0){\strut{}$0$}}%
      \put(4750,2518){\makebox(0,0){\strut{}$0.05$}}%
    }%
    \gplgaddtomacro\gplfronttext{%
      \csname LTb\endcsname%
      \put(4438,2298){\makebox(0,0){\strut{}Phase}}%
    }%
    \gplgaddtomacro\gplbacktext{%
      \csname LTb\endcsname%
      \put(815,2573){\makebox(0,0)[r]{\strut{}$-0.01$}}%
      \put(815,2242){\makebox(0,0)[r]{\strut{}$0$}}%
      \put(815,1912){\makebox(0,0)[r]{\strut{}$0.01$}}%
      \put(815,1582){\makebox(0,0)[r]{\strut{}$0.02$}}%
      \put(815,1251){\makebox(0,0)[r]{\strut{}$0.03$}}%
      \put(1134,767){\makebox(0,0){\strut{}}}%
      \put(1446,767){\makebox(0,0){\strut{}}}%
      \put(1757,767){\makebox(0,0){\strut{}}}%
      \put(1047,1162){\makebox(0,0)[l]{\strut{}49}}%
    }%
    \gplgaddtomacro\gplfronttext{%
      \csname LTb\endcsname%
      \put(150,1862){\rotatebox{-270}{\makebox(0,0){\strut{}$\Delta m$}}}%
    }%
    \gplgaddtomacro\gplbacktext{%
      \csname LTb\endcsname%
      \put(815,834){\makebox(0,0)[r]{\strut{}$-0.01$}}%
      \put(815,681){\makebox(0,0)[r]{\strut{}$0$}}%
      \put(815,527){\makebox(0,0)[r]{\strut{}$0.01$}}%
      \put(1134,154){\makebox(0,0){\strut{}$-0.05$}}%
      \put(1446,154){\makebox(0,0){\strut{}$0$}}%
      \put(1757,154){\makebox(0,0){\strut{}$0.05$}}%
    }%
    \gplgaddtomacro\gplfronttext{%
      \csname LTb\endcsname%
      \put(150,680){\rotatebox{-270}{\makebox(0,0){\strut{}O-C}}}%
      \put(1445,-66){\makebox(0,0){\strut{}Phase}}%
    }%
    \gplgaddtomacro\gplbacktext{%
      \csname LTb\endcsname%
      \put(1813,2573){\makebox(0,0)[r]{\strut{}}}%
      \put(1813,2242){\makebox(0,0)[r]{\strut{}}}%
      \put(1813,1912){\makebox(0,0)[r]{\strut{}}}%
      \put(1813,1582){\makebox(0,0)[r]{\strut{}}}%
      \put(1813,1251){\makebox(0,0)[r]{\strut{}}}%
      \put(2132,767){\makebox(0,0){\strut{}}}%
      \put(2444,767){\makebox(0,0){\strut{}}}%
      \put(2755,767){\makebox(0,0){\strut{}}}%
      \put(2045,1162){\makebox(0,0)[l]{\strut{}50}}%
    }%
    \gplgaddtomacro\gplfronttext{%
    }%
    \gplgaddtomacro\gplbacktext{%
      \csname LTb\endcsname%
      \put(1813,834){\makebox(0,0)[r]{\strut{}}}%
      \put(1813,681){\makebox(0,0)[r]{\strut{}}}%
      \put(1813,527){\makebox(0,0)[r]{\strut{}}}%
      \put(2132,154){\makebox(0,0){\strut{}$-0.05$}}%
      \put(2444,154){\makebox(0,0){\strut{}$0$}}%
      \put(2755,154){\makebox(0,0){\strut{}$0.05$}}%
    }%
    \gplgaddtomacro\gplfronttext{%
      \csname LTb\endcsname%
      \put(2443,-66){\makebox(0,0){\strut{}Phase}}%
    }%
    \gplgaddtomacro\gplbacktext{%
      \csname LTb\endcsname%
      \put(2810,2573){\makebox(0,0)[r]{\strut{}}}%
      \put(2810,2242){\makebox(0,0)[r]{\strut{}}}%
      \put(2810,1912){\makebox(0,0)[r]{\strut{}}}%
      \put(2810,1582){\makebox(0,0)[r]{\strut{}}}%
      \put(2810,1251){\makebox(0,0)[r]{\strut{}}}%
      \put(3129,767){\makebox(0,0){\strut{}}}%
      \put(3441,767){\makebox(0,0){\strut{}}}%
      \put(3752,767){\makebox(0,0){\strut{}}}%
      \put(3042,1162){\makebox(0,0)[l]{\strut{}51}}%
    }%
    \gplgaddtomacro\gplfronttext{%
    }%
    \gplgaddtomacro\gplbacktext{%
      \csname LTb\endcsname%
      \put(2810,834){\makebox(0,0)[r]{\strut{}}}%
      \put(2810,681){\makebox(0,0)[r]{\strut{}}}%
      \put(2810,527){\makebox(0,0)[r]{\strut{}}}%
      \put(3129,154){\makebox(0,0){\strut{}$-0.05$}}%
      \put(3441,154){\makebox(0,0){\strut{}$0$}}%
      \put(3752,154){\makebox(0,0){\strut{}$0.05$}}%
    }%
    \gplgaddtomacro\gplfronttext{%
      \csname LTb\endcsname%
      \put(3440,-66){\makebox(0,0){\strut{}Phase}}%
    }%
    \gplbacktext
    \put(0,0){\includegraphics{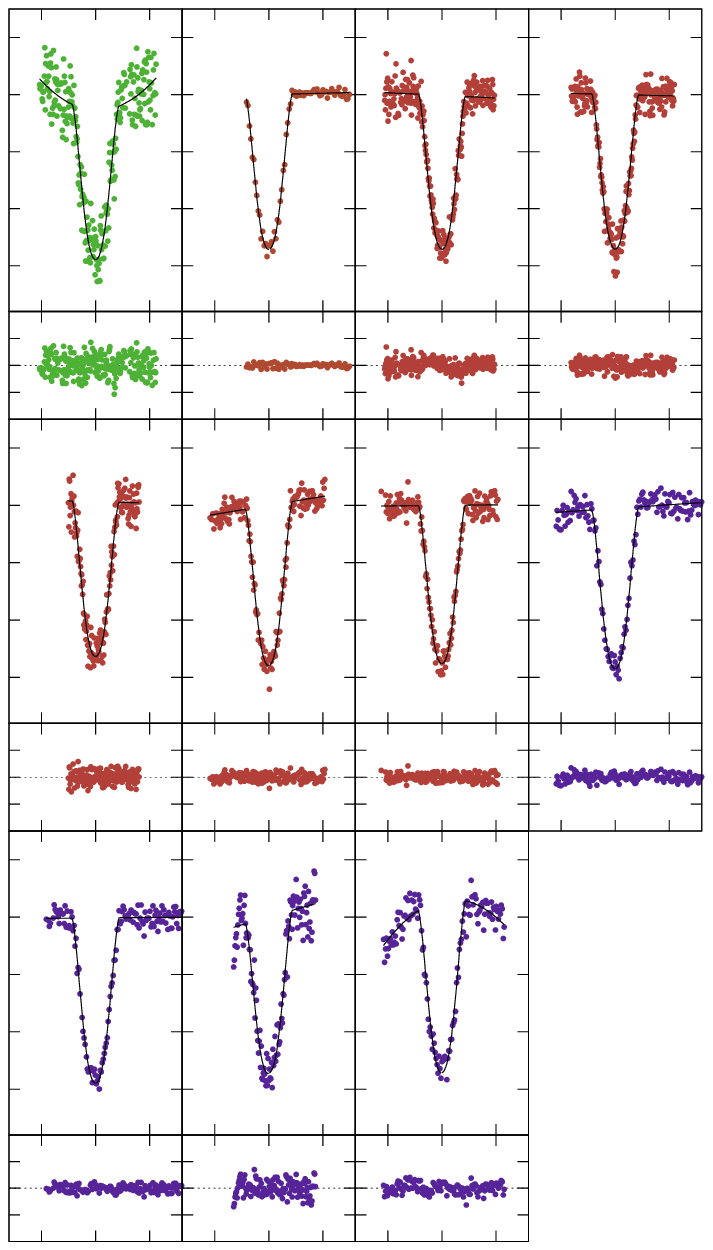}}%
    \gplfronttext
  \end{picture}%
\endgroup

% flatex input end: [./images/broad_band_lc/lc3.tex]

        \caption{Broad band light curves III.  Light curves are sorted by the observation number given in Table~\ref{tab:photo-all-broad}. This plot is a continuation of Fig.~\ref{fig:broad-band2}.}
        \label{fig:broad-band3}
\end{figure}

\begin{figure*}[htp]
        % flatex input: [./images/broad_band_ratio/broad_band.tex]
% GNUPLOT: LaTeX picture with Postscript
\begingroup
  \makeatletter
  \providecommand\color[2][]{%
    \GenericError{(gnuplot) \space\space\space\@spaces}{%
      Package color not loaded in conjunction with
      terminal option `colourtext'%
    }{See the gnuplot documentation for explanation.%
    }{Either use 'blacktext' in gnuplot or load the package
      color.sty in LaTeX.}%
    \renewcommand\color[2][]{}%
  }%
  \providecommand\includegraphics[2][]{%
    \GenericError{(gnuplot) \space\space\space\@spaces}{%
      Package graphicx or graphics not loaded%
    }{See the gnuplot documentation for explanation.%
    }{The gnuplot epslatex terminal needs graphicx.sty or graphics.sty.}%
    \renewcommand\includegraphics[2][]{}%
  }%
  \providecommand\rotatebox[2]{#2}%
  \@ifundefined{ifGPcolor}{%
    \newif\ifGPcolor
    \GPcolortrue
  }{}%
  \@ifundefined{ifGPblacktext}{%
    \newif\ifGPblacktext
    \GPblacktexttrue
  }{}%
  % define a \g@addto@macro without @ in the name:
  \let\gplgaddtomacro\g@addto@macro
  % define empty templates for all commands taking text:
  \gdef\gplbacktext{}%
  \gdef\gplfronttext{}%
  \makeatother
  \ifGPblacktext
    % no textcolor at all
    \def\colorrgb#1{}%
    \def\colorgray#1{}%
  \else
    % gray or color?
    \ifGPcolor
      \def\colorrgb#1{\color[rgb]{#1}}%
      \def\colorgray#1{\color[gray]{#1}}%
      \expandafter\def\csname LTw\endcsname{\color{white}}%
      \expandafter\def\csname LTb\endcsname{\color{black}}%
      \expandafter\def\csname LTa\endcsname{\color{black}}%
      \expandafter\def\csname LT0\endcsname{\color[rgb]{1,0,0}}%
      \expandafter\def\csname LT1\endcsname{\color[rgb]{0,1,0}}%
      \expandafter\def\csname LT2\endcsname{\color[rgb]{0,0,1}}%
      \expandafter\def\csname LT3\endcsname{\color[rgb]{1,0,1}}%
      \expandafter\def\csname LT4\endcsname{\color[rgb]{0,1,1}}%
      \expandafter\def\csname LT5\endcsname{\color[rgb]{1,1,0}}%
      \expandafter\def\csname LT6\endcsname{\color[rgb]{0,0,0}}%
      \expandafter\def\csname LT7\endcsname{\color[rgb]{1,0.3,0}}%
      \expandafter\def\csname LT8\endcsname{\color[rgb]{0.5,0.5,0.5}}%
    \else
      % gray
      \def\colorrgb#1{\color{black}}%
      \def\colorgray#1{\color[gray]{#1}}%
      \expandafter\def\csname LTw\endcsname{\color{white}}%
      \expandafter\def\csname LTb\endcsname{\color{black}}%
      \expandafter\def\csname LTa\endcsname{\color{black}}%
      \expandafter\def\csname LT0\endcsname{\color{black}}%
      \expandafter\def\csname LT1\endcsname{\color{black}}%
      \expandafter\def\csname LT2\endcsname{\color{black}}%
      \expandafter\def\csname LT3\endcsname{\color{black}}%
      \expandafter\def\csname LT4\endcsname{\color{black}}%
      \expandafter\def\csname LT5\endcsname{\color{black}}%
      \expandafter\def\csname LT6\endcsname{\color{black}}%
      \expandafter\def\csname LT7\endcsname{\color{black}}%
      \expandafter\def\csname LT8\endcsname{\color{black}}%
    \fi
  \fi
    \setlength{\unitlength}{0.0500bp}%
    \ifx\gptboxheight\undefined%
      \newlength{\gptboxheight}%
      \newlength{\gptboxwidth}%
      \newsavebox{\gptboxtext}%
    \fi%
    \setlength{\fboxrule}{0.5pt}%
    \setlength{\fboxsep}{1pt}%
\begin{picture}(10204.00,3458.00)%
    \gplgaddtomacro\gplbacktext{%
      \csname LTb\endcsname%
      \put(1078,704){\makebox(0,0)[r]{\strut{}0.145}}%
      \put(1078,981){\makebox(0,0)[r]{\strut{}0.150}}%
      \put(1078,1257){\makebox(0,0)[r]{\strut{}0.155}}%
      \put(1078,1534){\makebox(0,0)[r]{\strut{}0.160}}%
      \put(1078,1810){\makebox(0,0)[r]{\strut{}0.165}}%
      \put(1078,2087){\makebox(0,0)[r]{\strut{}0.170}}%
      \put(1078,2363){\makebox(0,0)[r]{\strut{}0.175}}%
      \put(1078,2640){\makebox(0,0)[r]{\strut{}0.180}}%
      \put(1078,2916){\makebox(0,0)[r]{\strut{}0.185}}%
      \put(1078,3193){\makebox(0,0)[r]{\strut{}0.190}}%
      \put(1210,484){\makebox(0,0){\strut{}$0$}}%
      \put(2065,484){\makebox(0,0){\strut{}$5$}}%
      \put(2920,484){\makebox(0,0){\strut{}$10$}}%
      \put(3774,484){\makebox(0,0){\strut{}$15$}}%
      \put(4629,484){\makebox(0,0){\strut{}$20$}}%
      \put(5484,484){\makebox(0,0){\strut{}$25$}}%
      \put(6339,484){\makebox(0,0){\strut{}$30$}}%
      \put(7194,484){\makebox(0,0){\strut{}$35$}}%
      \put(8048,484){\makebox(0,0){\strut{}$40$}}%
      \put(8903,484){\makebox(0,0){\strut{}$45$}}%
      \put(9758,484){\makebox(0,0){\strut{}$50$}}%
      \colorrgb{0.75,0.75,0.75}%
      \put(1552,981){\makebox(0,0){\strut{}\tiny U}}%
      \put(2749,981){\makebox(0,0){\strut{}\tiny B}}%
      \put(4116,981){\makebox(0,0){\strut{}\tiny V}}%
      \put(5142,981){\makebox(0,0){\strut{}\tiny RISE}}%
      \put(6253,981){\makebox(0,0){\strut{}\tiny r'}}%
      \put(7536,981){\makebox(0,0){\strut{}\tiny R}}%
      \put(8390,981){\makebox(0,0){\strut{}\tiny 792}}%
      \put(8903,981){\makebox(0,0){\strut{}\tiny I}}%
      \put(9673,981){\makebox(0,0){\strut{}\tiny z'}}%
    }%
    \gplgaddtomacro\gplfronttext{%
      \csname LTb\endcsname%
      \put(176,1948){\rotatebox{-270}{\makebox(0,0){\strut{}$\frac{R_p}{R_*}$}}}%
      \put(5655,154){\makebox(0,0){\strut{}obs. number}}%
    }%
    \gplbacktext
    \put(0,0){\includegraphics{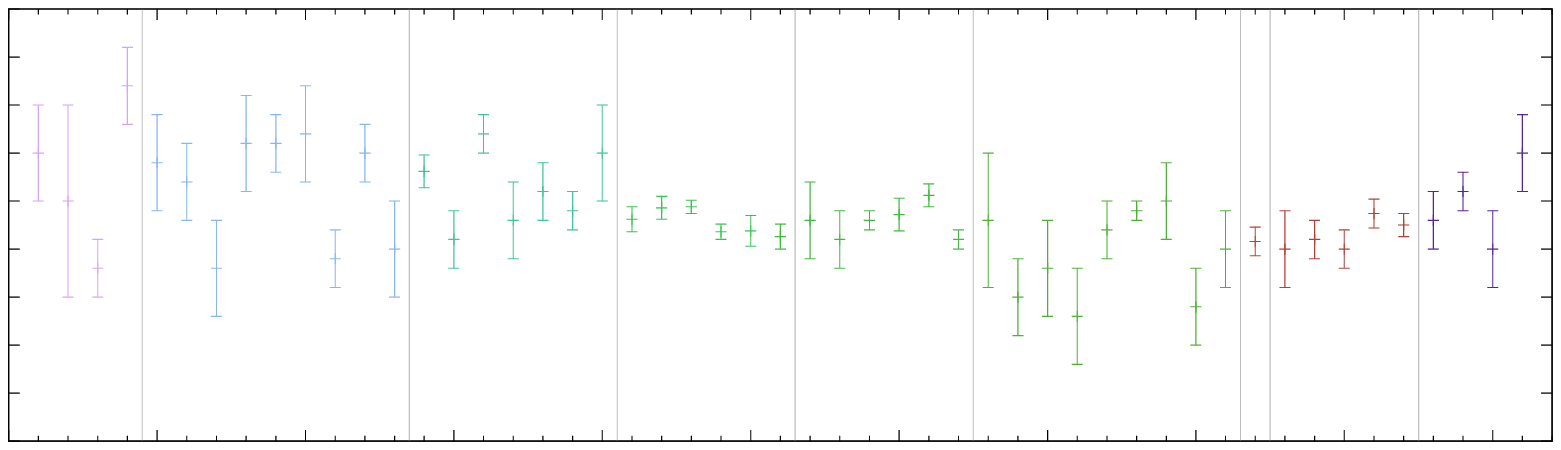}}%
    \gplfronttext
  \end{picture}%
\endgroup

% flatex input end: [./images/broad_band_ratio/broad_band.tex]

%\end{figure*}
        \caption{Radius ratios of broad band observations. They are sorted by the observation number given in Table~\ref{tab:photo-obs}. All vertical error-bars indicate the  $ \unit[1]{\sigma} $ uncertainty.}
        \label{fig:broad-band-all}
\end{figure*}

We performed a final fit with the obtained enlarged uncertainties and the determined detrending polynomial. Because no cosmic ray correction was applied in the reduction of the GTC-spectra, we applied a $ 5 \sigma $-clipping to the residuals of the light curves after a first fit with JKTEBOP.
The final model fitting contains the planetary radius and the polynomial coefficients as free parameters. The parameter uncertainties were obtained with JKTEBOP using both a Monte Carlo simulation with 5000 steps and a residual permutation method. For each spectral bin, we adopted the method that yields the largest uncertainty in the radius. The results of the light curve analysis are listed in Table~\ref{tab:photo-all-broad}. 
Figure~\ref{fig:16bins_matthias} shows modelled light curves of the GTC/OSIRIS spectra. Figures~\ref{fig:broad-band1}, \ref{fig:broad-band2}, and \ref{fig:broad-band3} show the broad band light curves. An overview of the measured radius ratios for the broad band light curves is presented in Fig.~\ref{fig:broad-band-all}.
% flatex input: [tab-broad-result-new.tex]
\begin{table*}[h!]
\caption{Results of the individual broad band observations. Columns are the photometric band, source reference (if applicable), the date of observation, as well as the radius ratio, $ \beta $-factor and used order of the detrending polynomial over time. The last column indicates the observation number.} 
\label{tab:photo-all-broad}
\centering
\begin{tabular}{cccr@{ $ \pm $ }lccc}
        \toprule
        Band & Reference & Date of obs. & \multicolumn{2}{c}{$ R_P/R_* $} & $\beta$ & detrend & \# \\
        \midrule
\textit{U}  & 1 & 2016-04-25 & 0.175 & 0.005 & 1.17 & linear &  1\\
& 2 & 2011-10-14 & 0.17 & 0.01 & 1.21 & linear &  2\\
&  2  & 2011-11-04 & 0.163 & 0.003 & 1.00 & linear &  3\\
&  2  & 2012-03-25 & 0.182 & 0.004 & 1.24 & quadratic &  4\\
\textit{B}  & 3 & 2016-05-02 & 0.174 & 0.005 & 1.08 & linear &  5\\
& 4 & 2007-04-08 & 0.172 & 0.004 & 1.09 & linear &  6\\
& 2 & 2012-04-11 & 0.163 & 0.005 & 1.36 & linear &  7\\
& 5 & 2016-04-12 & 0.176 & 0.005 & 1.13 & linear &  8\\
&  5  & 2016-04-25 & 0.176 & 0.003 & 1.00 & linear &  9\\
&  5  & 2016-05-19 & 0.185 & 0.004 & 1.14 & quadratic &  10\\
&  5  & 2016-06-01 & 0.164 & 0.003 & 1.00 & linear &  11\\
&  5  & 2016-06-05 & 0.175 & 0.003 & 1.08 & linear &  12\\
&  5  & 2016-06-18 & 0.179 & 0.005 & 1.23 & quadratic &  13\\
\textit{V}  & 6 & 2007-04-24 & 0.1731 & 0.0017 & 1.13 & linear &  14\\
& 7 & 2016-03-09 & 0.166 & 0.003 & 1.14 & linear &  15\\
&  7  & 2016-04-28 & 0.177 & 0.002 & 1.06 & linear &  16\\
&  7  & 2016-05-02 & 0.168 & 0.004 & 1.19 & linear &  17\\
&  7  & 2016-05-19 & 0.171 & 0.003 & 1.16 & linear &  18\\
& 2 & 2009-07-04 & 0.169 & 0.002 & 1.04 & linear &  19\\
& 8 & 2016-04-19 & 0.175 & 0.005 & 1.09 & linear &  20\\
\textit{RISE}  & 9 & 2008-03-08 & 0.1681 & 0.0013 & 1.02 & linear &  21\\
&  9  & 2008-05-28 & 0.1693 & 0.0012 & 1.00 & linear &  22\\
&  9  & 2008-06-14 & 0.1694 & 0.0007 & 1.09 & linear &  23\\
&  9  & 2008-07-01 & 0.1668 & 0.0008 & 1.00 & linear &  24\\
&  9  & 2008-07-22 & 0.1669 & 0.0016 & 1.43 & linear &  25\\
\textit{r’}  & 10 & 2009-05-14 & 0.1663 & 0.0013 & 1.08 & linear &  26\\
&  10  & 2010-05-16 & 0.168 & 0.004 & 1.62 & linear &  27\\
&  10  & 2010-10-12 & 0.166 & 0.003 & 1.36 & linear &  28\\
&  10  & 2011-03-24 & 0.168 & 0.001 & 1.55 & linear &  29\\
&  10  & 2011-06-21 & 0.1686 & 0.0017 & 1.40 & linear &  30\\
&  10  & 2011-08-24 & 0.1706 & 0.0012 & 1.11 & linear &  31\\
& 6 & 2008-03-27 & 0.166 & 0.001 & 1.08 & linear &  32\\
\textit{R}  & 11 & 2010-05-25 & 0.168 & 0.007 & 1.10 & linear &  33\\
&  11  & 2010-06-11 & 0.160 & 0.004 & 1.00 & linear &  34\\
&  11  & 2010-06-15 & 0.163 & 0.005 & 1.06 & linear &  35\\
&  11  & 2010-06-28 & 0.158 & 0.005 & 1.04 & linear &  36\\
& 12 & 2010-06-15 & 0.167 & 0.003 & 1.32 & linear &  37\\
& 2 & 2009-06-22 & 0.169 & 0.001 & 1.00 & linear &  38\\
& 13 & 2009-08-01 & 0.170 & 0.004 & 1.08 & linear &  39\\
&  13  & 2010-04-27 & 0.159 & 0.004 & 1.05 & linear &  40\\
&  13  & 2010-06-30 & 0.177 & 0.006 & 1.27 & quadratic &  41\\
\textit{792} & 14 & 2009-08-10 & 0.1658 & 0.0015 & 1.00 & linear &  42\\
\textit{I}  & 6 & 2008-03-09 & 0.165 & 0.004 & 1.43 & linear &  43\\
&  6  & 2008-03-27 & 0.166 & 0.002 & 1.24 & linear &  44\\
&  6  & 2008-04-12 & 0.165 & 0.002 & 1.10 & linear &  45\\
& 7 & 2016-07-18 & 0.1687 & 0.0015 & 1.12 & linear &  46\\
&  7  & 2016-08-25 & 0.1675 & 0.0012 & 1.01 & linear &  47\\
\textit{z'}  & 4 & 2007-03-26 & 0.168 & 0.003 & 1.26 & linear &  48\\
& 6 & 2007-03-25 & 0.171 & 0.002 & 1.11 & linear &  49\\
& 5 & 2016-06-28 & 0.165 & 0.004 & 1.18 & linear &  50\\
&  5  & 2016-07-15 & 0.163 & 0.007 & 1.46 & quadratic &  51\\
        \bottomrule
\end{tabular}
\tablebib{
        (1)~TNG; (2)~\citet{turner_2013}; (3) CAHA; (4) \citet{odonovan_2007}; (5) STELLA; (6) \citet{sozzetti_2009}; (7) T1T; (8) VBO; (9) \citet{gibson_2009}; (10) \citet{kundurthy_2013}; (11) \citet{jiang_2013}; (12) \citet{lee_2011}; (13) \citet{vanko_2013}; (14) \citet{colon_2010}
}
\end{table*}
% flatex input end: [tab-broad-result-new.tex]

%\end{figure*}

Next, we fitted all broad band light curves of one band simultaneously with JKTEBOP using again enlarged error-bars as described above. JKTEBOP allows for a simultaneous fit of multiple transit observations if one common LD function can be applied. The software cannot handle different LDCs to subsamples of the input data, which prohibits the simultaneous fit of multicolour data. However, JKTEBOP is able to assign independent detrending polynomials to the individual light curves, so that the detrending is included in the transit fit.

% flatex input end: [analysis.tex]

% flatex input: [result.tex]
\section{Results}
\subsection{Transit depth as a function of wavelength}
\begin{figure*}[htp]
        % flatex input: [./images/spectrum/trans-spec-all.tex]
% GNUPLOT: LaTeX picture with Postscript
\begingroup
  \makeatletter
  \providecommand\color[2][]{%
    \GenericError{(gnuplot) \space\space\space\@spaces}{%
      Package color not loaded in conjunction with
      terminal option `colourtext'%
    }{See the gnuplot documentation for explanation.%
    }{Either use 'blacktext' in gnuplot or load the package
      color.sty in LaTeX.}%
    \renewcommand\color[2][]{}%
  }%
  \providecommand\includegraphics[2][]{%
    \GenericError{(gnuplot) \space\space\space\@spaces}{%
      Package graphicx or graphics not loaded%
    }{See the gnuplot documentation for explanation.%
    }{The gnuplot epslatex terminal needs graphicx.sty or graphics.sty.}%
    \renewcommand\includegraphics[2][]{}%
  }%
  \providecommand\rotatebox[2]{#2}%
  \@ifundefined{ifGPcolor}{%
    \newif\ifGPcolor
    \GPcolortrue
  }{}%
  \@ifundefined{ifGPblacktext}{%
    \newif\ifGPblacktext
    \GPblacktextfalse
  }{}%
  % define a \g@addto@macro without @ in the name:
  \let\gplgaddtomacro\g@addto@macro
  % define empty templates for all commands taking text:
  \gdef\gplbacktext{}%
  \gdef\gplfronttext{}%
  \makeatother
  \ifGPblacktext
    % no textcolor at all
    \def\colorrgb#1{}%
    \def\colorgray#1{}%
  \else
    % gray or color?
    \ifGPcolor
      \def\colorrgb#1{\color[rgb]{#1}}%
      \def\colorgray#1{\color[gray]{#1}}%
      \expandafter\def\csname LTw\endcsname{\color{white}}%
      \expandafter\def\csname LTb\endcsname{\color{black}}%
      \expandafter\def\csname LTa\endcsname{\color{black}}%
      \expandafter\def\csname LT0\endcsname{\color[rgb]{1,0,0}}%
      \expandafter\def\csname LT1\endcsname{\color[rgb]{0,1,0}}%
      \expandafter\def\csname LT2\endcsname{\color[rgb]{0,0,1}}%
      \expandafter\def\csname LT3\endcsname{\color[rgb]{1,0,1}}%
      \expandafter\def\csname LT4\endcsname{\color[rgb]{0,1,1}}%
      \expandafter\def\csname LT5\endcsname{\color[rgb]{1,1,0}}%
      \expandafter\def\csname LT6\endcsname{\color[rgb]{0,0,0}}%
      \expandafter\def\csname LT7\endcsname{\color[rgb]{1,0.3,0}}%
      \expandafter\def\csname LT8\endcsname{\color[rgb]{0.5,0.5,0.5}}%
    \else
      % gray
      \def\colorrgb#1{\color{black}}%
      \def\colorgray#1{\color[gray]{#1}}%
      \expandafter\def\csname LTw\endcsname{\color{white}}%
      \expandafter\def\csname LTb\endcsname{\color{black}}%
      \expandafter\def\csname LTa\endcsname{\color{black}}%
      \expandafter\def\csname LT0\endcsname{\color{black}}%
      \expandafter\def\csname LT1\endcsname{\color{black}}%
      \expandafter\def\csname LT2\endcsname{\color{black}}%
      \expandafter\def\csname LT3\endcsname{\color{black}}%
      \expandafter\def\csname LT4\endcsname{\color{black}}%
      \expandafter\def\csname LT5\endcsname{\color{black}}%
      \expandafter\def\csname LT6\endcsname{\color{black}}%
      \expandafter\def\csname LT7\endcsname{\color{black}}%
      \expandafter\def\csname LT8\endcsname{\color{black}}%
    \fi
  \fi
    \setlength{\unitlength}{0.0500bp}%
    \ifx\gptboxheight\undefined%
      \newlength{\gptboxheight}%
      \newlength{\gptboxwidth}%
      \newsavebox{\gptboxtext}%
    \fi%
    \setlength{\fboxrule}{0.5pt}%
    \setlength{\fboxsep}{1pt}%
\begin{picture}(10204.00,3458.00)%
    \gplgaddtomacro\gplbacktext{%
      \csname LTb\endcsname%
      \put(1078,704){\makebox(0,0)[r]{\strut{}0.150}}%
      \put(1078,1165){\makebox(0,0)[r]{\strut{}0.155}}%
      \put(1078,1626){\makebox(0,0)[r]{\strut{}0.160}}%
      \put(1078,2087){\makebox(0,0)[r]{\strut{}0.165}}%
      \put(1078,2548){\makebox(0,0)[r]{\strut{}0.170}}%
      \put(1078,3009){\makebox(0,0)[r]{\strut{}0.175}}%
      \put(1210,484){\makebox(0,0){\strut{}$3000$}}%
      \put(2258,484){\makebox(0,0){\strut{}$4000$}}%
      \put(3305,484){\makebox(0,0){\strut{}$5000$}}%
      \put(4353,484){\makebox(0,0){\strut{}$6000$}}%
      \put(5400,484){\makebox(0,0){\strut{}$7000$}}%
      \put(6448,484){\makebox(0,0){\strut{}$8000$}}%
      \put(7495,484){\makebox(0,0){\strut{}$9000$}}%
      \put(8543,484){\makebox(0,0){\strut{}$10000$}}%
      \put(9590,484){\makebox(0,0){\strut{}$11000$}}%
      \put(9722,1995){\makebox(0,0)[l]{\strut{}0}}%
      \put(9722,2386){\makebox(0,0)[l]{\strut{}10}}%
      \put(9722,2778){\makebox(0,0)[l]{\strut{}20}}%
      \put(9722,3170){\makebox(0,0)[l]{\strut{}30}}%
      \put(9722,819){\makebox(0,0)[l]{\strut{}}}%
      \put(9722,1211){\makebox(0,0)[l]{\strut{}}}%
      \put(9722,1603){\makebox(0,0)[l]{\strut{}}}%
      \colorrgb{0.80,0.80,0.80}%
      \put(1897,888){\makebox(0,0){\strut{}\tiny U}}%
      \put(2627,888){\makebox(0,0){\strut{}\tiny B}}%
      \put(3805,888){\makebox(0,0){\strut{}\tiny V}}%
      \colorrgb{0.63,0.71,0.80}%
      \put(4248,888){\makebox(0,0){\strut{}\tiny RISE}}%
      \colorrgb{0.94,0.50,0.50}%
      \put(4566,888){\makebox(0,0){\strut{}\tiny r'}}%
      \colorrgb{0.80,0.80,0.80}%
      \put(4718,888){\makebox(0,0){\strut{}\tiny R}}%
      \put(6510,888){\makebox(0,0){\strut{}\tiny I}}%
      \colorrgb{0.94,0.50,0.50}%
      \put(8192,888){\makebox(0,0){\strut{}\tiny z'}}%
    }%
    \gplgaddtomacro\gplfronttext{%
      \csname LTb\endcsname%
      \put(176,1948){\rotatebox{-270}{\makebox(0,0){\strut{}$\frac{R_p}{R_*}$}}}%
      \put(10161,1948){\rotatebox{-270}{\makebox(0,0){\strut{}H}}}%
      \put(5400,154){\makebox(0,0){\strut{}$\unit[\lambda /]{\AA}$}}%
      \csname LTb\endcsname%
      \put(8603,3020){\makebox(0,0)[r]{\strut{}P16}}%
    }%
    \gplbacktext
    \put(0,0){\includegraphics{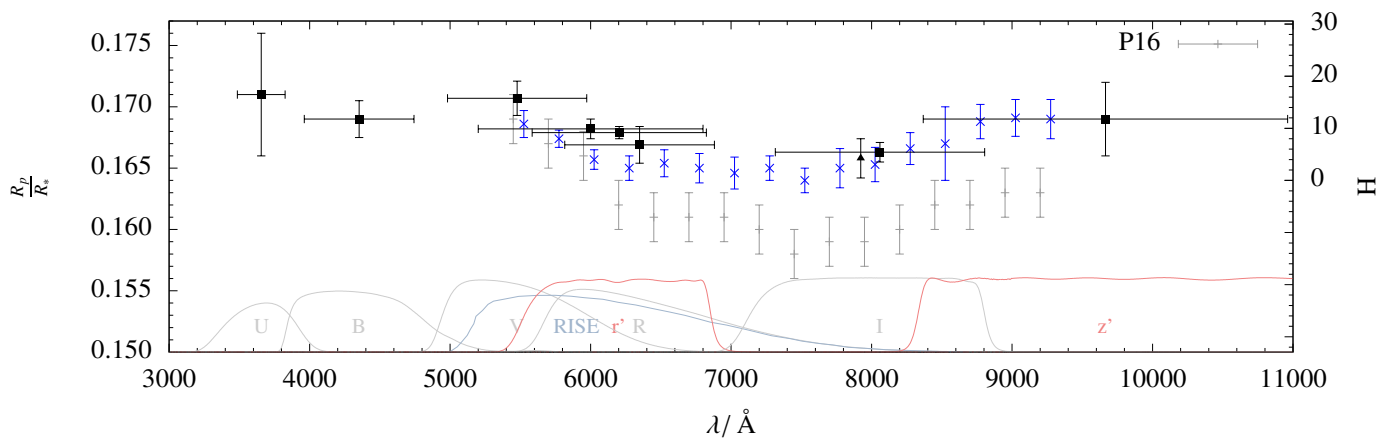}}%
    \gplfronttext
  \end{picture}%
\endgroup

% flatex input end: [./images/spectrum/trans-spec-all.tex]

        \caption{Transmission spectrum of TrES-3\,b. Shown is the GTC/OSIRS spectrum (crosses). The broadband photometry is indicated with squares. The response function for each photometric band is shown at the bottom. Horizontal error-bars indicate the FWHM of the band. The narrow band observations by \citet{colon_2010} are indicated with a triangle. The results of the previous study by P16, their Fig.~9, are shown for comparison with an arbitrary offset of $ \unit[-75]{\AA} $ for clarity. All vertical error-bars indicate the $ \unit[1]{\sigma} $ uncertainty.}
        \label{fig:trans-spec-all}
\end{figure*}

The combined transmission spectrum for TrES-3\,b of broad band observations and the newly analysed GTC/OSIRIS data is shown in Fig.~\ref{fig:trans-spec-all}. Additionally, the Figure contains the GTC results derived by P16. Our work cannot reproduce the strong slope found by P16 of $ 25.9 \pm 4.7 $ scale heights from $ \sim 5500 $ to $ \unit[7500]{\AA} $.
We compare the weighted mean of the three bluest wavelength bins $ \unit[5525]{\AA} $, $ \unit[5775]{\AA,} $ and $ \unit[6025]{\AA} $ to the central bins of $ \unit[7275]{\AA} $, $ \unit[7575]{\AA,} $ and $ \unit[7775]{\AA} $ . In P16, they exhibit a difference of $ 19.5 \pm 2.9 $ scale heights. The weighted mean of the same spectral bins for the reanalysed GTC data of this work show a difference of $ 5.7 \pm 1.7 $ scale heights.
A comparison of the broad band data for similar wavelength ranges shows a difference of $ 4.0 \pm 1.9 $ scale heights between the weighted mean of the \textit{RISE}, \textit{r}' and \textit{R} band and the \textit{I} band.

The broad band data alone do not exhibit a significant wavelength dependence of $ R_p/R_* $. The weighted mean of the nine band passes is $ k = 0.1678 \pm 0.0004 $, and the reduced $ \chi^2 $ of all nine measurements compared to that mean is 1.41, being sufficiently close to unity to be in reasonable agreement with a flat spectrum.
% flatex input: [tab-result-combined.tex]
\begin{table}[t]
\caption{Measured radius ratios of the broad band observations. The columns show the photometric band, their corresponding central wavelength and FWHM, as well as the radius ratio with        $ \unit[1]{\sigma} $ uncertainty.} 
\label{tab:result-broad}
\centering
\begin{tabular}{cr@{ $ \pm $ }lr@{ $ \pm $ }l}
        \toprule
        Band & \multicolumn{2}{c}{$ \unit[\lambda]{/ \AA} $} & \multicolumn{2}{c}{$ R_p / R_* $} \\
        \midrule
        \textit{U} & 3656 & 170 & 0.171 & 0.005 \\
        \textit{B} & 4353 & 390.5 & 0.1690 & 0.0015 \\
        \textit{V} & 5477 & 495.5 & 0.1707 & 0.0014 \\
        \textit{RISE} & 6000 & 800 & 0.1682 & 0.0008 \\
        \textit{r'} & 6204 & 620 & 0.1679 & 0.0005 \\
        \textit{R} & 6349 & 532.8 & 0.1669 & 0.0015 \\
        \textit{792} & 7923 & 41 & 0.1658 & 0.0016 \\
        \textit{I} & 8060 & 745 & 0.1663 & 0.0008 \\
        \textit{z'} & 9665 & 1297 & 0.169 & 0.003 \\
        \bottomrule
\end{tabular}

\vspace{10pt}

\caption{Measured radius ratios for the GTC/OSIRIS spectra. The columns show the centre of the spectral bin, its width, and the radius ratio with$ \unit[1]{\sigma} $ uncertainty and the $ \beta $-factor.} 
\label{tab:result-gtc}
\centering
\begin{tabular}{r@{ $ \pm $ }lr@{ $ \pm $ }lc}
        \toprule
        \multicolumn{2}{c}{$ \unit[\lambda]{/ \AA} $} & \multicolumn{2}{c}{$ R_p / R_* $} & $ \beta $ \\
        \midrule
        5525 & 125 &  0.1686 & 0.0011 & 1.19 \\
        5775 & 125 & 0.1674 & 0.0007 & 1.13 \\
        6025 & 125 & 0.1657 & 0.0008 & 1.26 \\
        6275 & 125 & 0.165 & 0.001 & 1.39 \\
        6525 & 125 & 0.1654 & 0.0011 & 1.42      \\
        6775 & 125 & 0.1650 & 0.0012 & 1.52 \\
        7025 & 125 & 0.1646 & 0.0013 & 1.47 \\
        7275 & 125 & 0.165 & 0.001 & 1.15 \\
        7525 & 125 & 0.164 & 0.001 & 1.30 \\
        7775 & 125 & 0.1650 & 0.0016 & 1.35 \\
        8025 & 125 & 0.1653 & 0.0014 & 1.32 \\
        8275 & 125 & 0.1666 & 0.0013 & 1.24 \\
        8525 & 125 & 0.167 & 0.003 & 1.29 \\
        8775 & 125 & 0.1688 & 0.0014 & 1.21 \\
        9025 & 125 & 0.1691 & 0.0015 & 1.04 \\
        9275 & 125 & 0.1690 & 0.0016 & 1.15 \\
        \bottomrule
\end{tabular}
\end{table}
% flatex input end: [tab-result-combined.tex]

We compared the derived broad band spectrophotometry and spectroscopic transmission spectrum of this work with theoretical models produced with an existing radiative transfer model \citep{munoz_2012,nielsen_2016}. The methodology is similar to that followed in \citet{sedaghati_2016}. The models assume that the planet atmosphere is dominated by H$_{2}$/He, and includes as minor gases H$_{2}$O (volume mixing ratio $ vmr = 10^{-3} $, independent of altitude), Na ($ vmr=2.96 \times 10^{-6} $), and K ($ vmr=2.4 \times 10^{-4} $). The background number density of the gas is determined by solving the hydrostatic balance equation with an altitude-dependent gravity from $ 10^3 $ to $ 10^{-7} $ bar. The temperature is assumed constant throughout the atmosphere and equal to $ \unit[1700]{K} $.
        
\begin{figure*}[htp]
        % flatex input: [./images/atmosphere-models/atmosphere.tex]
% GNUPLOT: LaTeX picture with Postscript
\begingroup
  \makeatletter
  \providecommand\color[2][]{%
    \GenericError{(gnuplot) \space\space\space\@spaces}{%
      Package color not loaded in conjunction with
      terminal option `colourtext'%
    }{See the gnuplot documentation for explanation.%
    }{Either use 'blacktext' in gnuplot or load the package
      color.sty in LaTeX.}%
    \renewcommand\color[2][]{}%
  }%
  \providecommand\includegraphics[2][]{%
    \GenericError{(gnuplot) \space\space\space\@spaces}{%
      Package graphicx or graphics not loaded%
    }{See the gnuplot documentation for explanation.%
    }{The gnuplot epslatex terminal needs graphicx.sty or graphics.sty.}%
    \renewcommand\includegraphics[2][]{}%
  }%
  \providecommand\rotatebox[2]{#2}%
  \@ifundefined{ifGPcolor}{%
    \newif\ifGPcolor
    \GPcolortrue
  }{}%
  \@ifundefined{ifGPblacktext}{%
    \newif\ifGPblacktext
    \GPblacktexttrue
  }{}%
  % define a \g@addto@macro without @ in the name:
  \let\gplgaddtomacro\g@addto@macro
  % define empty templates for all commands taking text:
  \gdef\gplbacktext{}%
  \gdef\gplfronttext{}%
  \makeatother
  \ifGPblacktext
    % no textcolor at all
    \def\colorrgb#1{}%
    \def\colorgray#1{}%
  \else
    % gray or color?
    \ifGPcolor
      \def\colorrgb#1{\color[rgb]{#1}}%
      \def\colorgray#1{\color[gray]{#1}}%
      \expandafter\def\csname LTw\endcsname{\color{white}}%
      \expandafter\def\csname LTb\endcsname{\color{black}}%
      \expandafter\def\csname LTa\endcsname{\color{black}}%
      \expandafter\def\csname LT0\endcsname{\color[rgb]{1,0,0}}%
      \expandafter\def\csname LT1\endcsname{\color[rgb]{0,1,0}}%
      \expandafter\def\csname LT2\endcsname{\color[rgb]{0,0,1}}%
      \expandafter\def\csname LT3\endcsname{\color[rgb]{1,0,1}}%
      \expandafter\def\csname LT4\endcsname{\color[rgb]{0,1,1}}%
      \expandafter\def\csname LT5\endcsname{\color[rgb]{1,1,0}}%
      \expandafter\def\csname LT6\endcsname{\color[rgb]{0,0,0}}%
      \expandafter\def\csname LT7\endcsname{\color[rgb]{1,0.3,0}}%
      \expandafter\def\csname LT8\endcsname{\color[rgb]{0.5,0.5,0.5}}%
    \else
      % gray
      \def\colorrgb#1{\color{black}}%
      \def\colorgray#1{\color[gray]{#1}}%
      \expandafter\def\csname LTw\endcsname{\color{white}}%
      \expandafter\def\csname LTb\endcsname{\color{black}}%
      \expandafter\def\csname LTa\endcsname{\color{black}}%
      \expandafter\def\csname LT0\endcsname{\color{black}}%
      \expandafter\def\csname LT1\endcsname{\color{black}}%
      \expandafter\def\csname LT2\endcsname{\color{black}}%
      \expandafter\def\csname LT3\endcsname{\color{black}}%
      \expandafter\def\csname LT4\endcsname{\color{black}}%
      \expandafter\def\csname LT5\endcsname{\color{black}}%
      \expandafter\def\csname LT6\endcsname{\color{black}}%
      \expandafter\def\csname LT7\endcsname{\color{black}}%
      \expandafter\def\csname LT8\endcsname{\color{black}}%
    \fi
  \fi
    \setlength{\unitlength}{0.0500bp}%
    \ifx\gptboxheight\undefined%
      \newlength{\gptboxheight}%
      \newlength{\gptboxwidth}%
      \newsavebox{\gptboxtext}%
    \fi%
    \setlength{\fboxrule}{0.5pt}%
    \setlength{\fboxsep}{1pt}%
\begin{picture}(10204.00,3458.00)%
    \gplgaddtomacro\gplbacktext{%
      \csname LTb\endcsname%
      \put(1078,704){\makebox(0,0)[r]{\strut{}0.162}}%
      \put(1078,1036){\makebox(0,0)[r]{\strut{}0.164}}%
      \put(1078,1368){\makebox(0,0)[r]{\strut{}0.166}}%
      \put(1078,1700){\makebox(0,0)[r]{\strut{}0.168}}%
      \put(1078,2031){\makebox(0,0)[r]{\strut{}0.170}}%
      \put(1078,2363){\makebox(0,0)[r]{\strut{}0.172}}%
      \put(1078,2695){\makebox(0,0)[r]{\strut{}0.174}}%
      \put(1078,3027){\makebox(0,0)[r]{\strut{}0.176}}%
      \put(1210,484){\makebox(0,0){\strut{}$3000$}}%
      \put(2258,484){\makebox(0,0){\strut{}$4000$}}%
      \put(3305,484){\makebox(0,0){\strut{}$5000$}}%
      \put(4353,484){\makebox(0,0){\strut{}$6000$}}%
      \put(5400,484){\makebox(0,0){\strut{}$7000$}}%
      \put(6448,484){\makebox(0,0){\strut{}$8000$}}%
      \put(7495,484){\makebox(0,0){\strut{}$9000$}}%
      \put(8543,484){\makebox(0,0){\strut{}$10000$}}%
      \put(9590,484){\makebox(0,0){\strut{}$11000$}}%
      \put(9722,1036){\makebox(0,0)[l]{\strut{}$0$}}%
      \put(9722,1388){\makebox(0,0)[l]{\strut{}$5$}}%
      \put(9722,1741){\makebox(0,0)[l]{\strut{}$10$}}%
      \put(9722,2094){\makebox(0,0)[l]{\strut{}$15$}}%
      \put(9722,2446){\makebox(0,0)[l]{\strut{}$20$}}%
      \put(9722,2799){\makebox(0,0)[l]{\strut{}$25$}}%
      \put(9722,3151){\makebox(0,0)[l]{\strut{}$30$}}%
    }%
    \gplgaddtomacro\gplfronttext{%
      \csname LTb\endcsname%
      \put(176,1948){\rotatebox{-270}{\makebox(0,0){\strut{}$\frac{R_p}{R_*}$}}}%
      \put(10161,1948){\rotatebox{-270}{\makebox(0,0){\strut{}H}}}%
      \put(5400,154){\makebox(0,0){\strut{}$\unit[\lambda /]{\AA}$}}%
      \csname LTb\endcsname%
      \put(9118,3020){\makebox(0,0)[r]{\strut{}hazy atmosphere}}%
      \csname LTb\endcsname%
      \put(9118,2800){\makebox(0,0)[r]{\strut{}clear atmosphere}}%
    }%
    \gplbacktext
    \put(0,0){\includegraphics{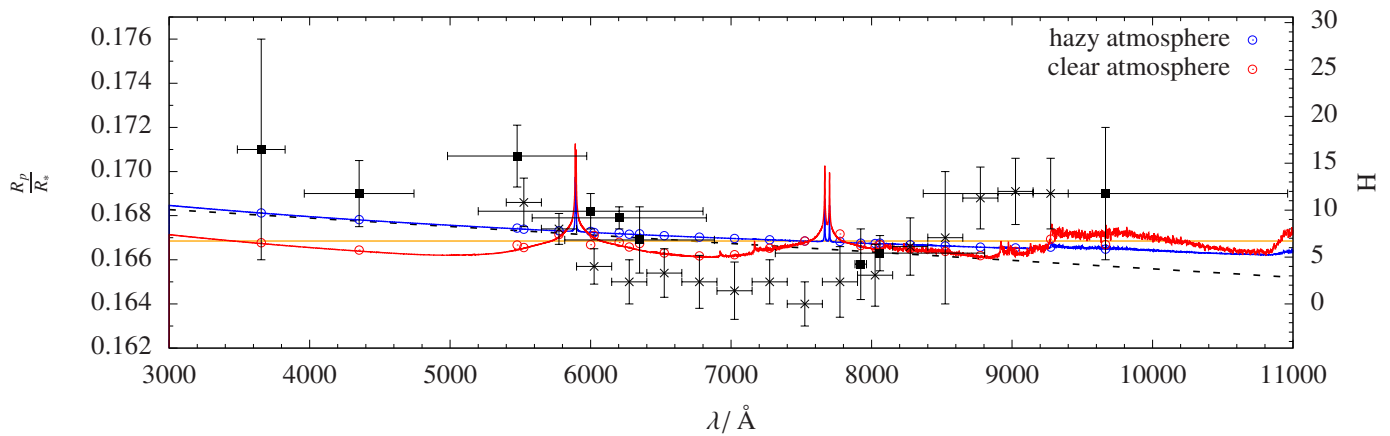}}%
    \gplfronttext
  \end{picture}%
\endgroup

% flatex input end: [./images/atmosphere-models/atmosphere.tex]

        \caption{Model atmospheres for a clear and an atmosphere with haze fitted to the transmission spectrum of TrES-3b. The measurements are plotted in black. The binned model data are indicated as coloured circles. Additionally, a constant (orange solid line) and a linear best fit (dashed black line) are shown.}
        \label{fig:model}
\end{figure*}

We include extinction from the gas in the form of Rayleigh scattering and molecular/atomic absorption, as well as extinction from a postulated haze. The H$_{2}$O linelist of optical properties is from the HITEMP database \citep{rothman_2010}. For the alkalis, we implemented the parametrisation of absorption given in \citet{iro_2005}. For the haze, our nominal (x1) scenario assumes the following variation of the extinction coefficient (in $ \unit{cm^{-1}} $) with altitude at a wavelength of $ \unit[1]{\mu m} $, $ \gamma(z; \lambda=1 \unit{\mu m}) = 4.1 \times 10^{-7} \exp(-z/H_a) $. At other wavelengths, we corrected according to a Rayleigh law, $ \gamma(z; \lambda)= \gamma(z; \lambda=\unit[1]{\mu m}) (1/ \lambda[\unit{\mu m}])^4 $. For the x0 (clear) and x1000 (hazy) scenarios (see Fig.~\ref{fig:model}), we multiplied $ \gamma(z; \lambda) $ by 0 or by 1000, respectively. Altitude $ z $ is measured with respect to the pressure level of $ \unit[10^3]{bar} $. We assumed $ H_a= \unit[250]{km} $ for the scale height of the haze, which is comparable to the scale height of the gas.

The synthetic spectra were binned to the same wavelength channels and filter curves as our measurements. The vertical offset, as the only remaining parameter, was fitted to measurements using the least squares method. For each model we calculated the $ \chi^2 $ value and the probability $ P $ of the $ \chi^2 $ test, that is, the probability that the measurements could result from statistical fluctuations alone in case the model is true. The values are summarised in Table~\ref{tab:fit}.
% flatex input: [tab-fit.tex]
\begin{table}[tp]
\caption{$ \chi^2_{red} $ and one-tailed probability $ P $ for the fitted atmospheric models using the complete data set and the broad band measurements only.} 
\label{tab:fit}
\centering
\begin{tabular}{lcccc}
        \toprule
         & \multicolumn{2}{c}{GTC + broad band} & \multicolumn{2}{c}{broad band} \\
        Model & $ \chi^2_{red} $ & $ P $ & $ \chi^2_{red} $ & $ P $ \\
        \midrule
        clear & 2.34 & 0.000213 & 1.43 & 0.177803 \\
        haze & 2.13 & 0.001029 & 1.47 & 0.160381 \\
        constant & 2.29 & 0.000644 & 1.41 & 0.189385 \\
        linear &  2.23 & 0.000325 & 0.69 & 0.683219 \\
        \bottomrule
\end{tabular}
\end{table}
% flatex input end: [tab-fit.tex]

Both models show smaller variations in the radius ratio than our data and are ruled out by the data according to their $ P $ values. For comparison, we also included a wavelength-independent radius ratio which represents an atmosphere dominated by clouds (flat model) and a linear function with slope of increased absorption towards shorter wavelength. None of the models is in agreement with our combined GTC and broad band data. 
The main cause is the $ R_p/R_* $ variation within the GTC data, and therefore relies on a single transit observation, while the multi-epoche multi-telescope broad band observations agree within their uncertainties to the models (but cannot differentiate between them). Repeated long-slit or multi-object spectroscopic transit observations would be beneficial to confirm the GTC data.


\subsection{Stellar activity}\label{sec:result:acitivity}
Unocculted star spots mimic a negative slope in the optical transmission spectrum \citep{sing_2011,pont_2013,oshagh_2014}.
TrES-3 is classified as a moderately active star according to the CaII H \& K line strengths \citep[$ \log(R_{HK}) = -4.54 $,][]{sozzetti_2009,knutson_2010}, however, to our knowledge no photometric out-of-transit variation was reported yet in the literature. We monitored the host star to deduce an upper limit on the star spot coverage and their effect on the transmission spectrum from its photometric variability. Similar photometric monitoring campaigns for different targets are presented, for example, in \citet{mallonn_2015a} and \citet{mallonn_2016a}.
The Lomb-Scargle periodograms in Fig.~\ref{fig:active-period} show no peaks at a false alarm probability (FAP) of below 0.01, thus no significantly periodic variation was detected (looking for periods of 1 to 100 days). Fitting sine curves to the light curves with frequencies fixed at the highest peaks in the periodograms results in semi-amplitudes below $ \unit[4.5]{mmag} $. We conclude that our monitoring data rule out periodic variations higher than this value.

The modification of the depth of a transit light curve is directly related to the photometric variation caused by unocculted spots. A decrease of 1 per cent in stellar flux increases the derived transit depth by about 1 per cent in relative units \citep{sing_2011}. During our multi-epoch transit observations the stellar flux varied by less than 1 percent. Thus, we derive an upper limit of spot-caused uncertainty in the radius ratio $ k $ of 0.0008. This value approximately equals our best transit fit uncertainties for $ k $. However, the wavelength-dependent effect on the transmission spectrum is about an order of magnitude smaller for solar-type host stars (depending on the temperature contrast of the spot and the quiet photosphere) than this absolute effect \citep{sing_2011,mallonn_2015a}. One scale height for TrES-3\,b equals approximately $ \Delta k \sim 0.0004 $, and the upper limit of a star-spot effect is less than that value. Therefore, we can neglect any influence of potentially present unocculted star spots on the broad band transmission spectrum derived here. It is unlikely that spots significantly modify the results of the GTC data.
% flatex input end: [result.tex]

% flatex input: [dicussion.tex]
\section{Summary and Conclusion}
In this work we investigate the transmission spectrum of the Hot Jupiter TrES-\,3b. We follow-up on a result published by P16 who reported an overly large Rayleigh-like feature.% of $ \sim 30 $~scale heights amplitude.
We reanalyse the GTC spectroscopic transit observation used by P16 and complement these data by a large sample of published and newly acquired broad band observations in nine different pass bands. We use more than 50 individual broad band light curves from the near-UV to the near-infrared. All light curves are analysed homogeneously and simultaneously per filter to avoid systematic error and reduce the final uncertainties.

Due to the almost grazing orbit of TrES-3\,b we did not fit the LDCs but rely on theoretical values. While for the majority of exoplanet host stars the theoretical LDCs agree reasonably well with their measured counterparts \citep{mueller_2013}, there are known cases of discrepant values, for example HD 209458 \citep{claret_2009} and HAT-P-32 \citep{mallonn_2016a}. 

In the transmission spectrum, both the newly analysed GTC data and the multi-colour broad band data rule out the very large feature found in the transmission spectrum by P16. This allows us to disprove the suggested overly large Rayleigh-like feature in our analysis.
In the new analysis the increase of $ 10.8 \pm 3.6 $ scale heights between $ \unit[5525]{\AA} $ and $ \unit[7525]{\AA} $ is of much lower amplitude than for the measurements by P16 with $ 25.9 \pm 4.7 $ scale heights. Since this intriguing feature is based on a single transit observation and disagrees with the \textit{r'} band as our most precise broad band measurement by $ \unit[3]{\sigma} $, we suggest a repetition of the spectroscopic transit observation to verify this signal. From our analysis we conclude the overly large Rayleigh-like feature is not intrinsic to the TrES-3 system.

For comparison we also analysed the GTC stellar spectra reduced by P16 which are  available publicly\footnote{\url{https://github.com/hpparvi/Parviainen-2015-TrES-3b-OSIRIS}}. We extracted the transit light curves from these 1D spectra of TrES-3\,b and comparison star and performed the light curve analysis as described in Section~\ref{sec:analysis}. The resulting transmission spectrum shows a slope very similar to P16 except for being vertically offset compared to P16. This offset might be caused by systematic differences in the limb-darkening coefficients from the different calculations by P16 and \citet{claret_2013}. Since the large variation in the transmission spectrum of P16 is reproduced when we analyse their reduced data, we suggest that the origin of the discrepancy in the GTC transmission spectrum between P16 and this work is not to be found in the analysis, but in the data reduction. The two versions of the data reduction of this work and P16 differ in details of the flux extraction and the wavelength calibration, but it was not possible to clarify what detail was responsible.

Our long-term photometric monitoring rules out a significant modification of the derived transmission spectrum in this work by unocculted star spots. The relative amplitudes induced by stellar activity are smaller than the uncertainties of our radius ratio measurements. Therefore, unocculted spots cannot introduce a spectral slope in the transmission spectrum at the current level of measurement precision.

Still the measurements cannot distinguish between individual atmospheric model spectra because TrES-3\,b is a moderately favourable target for transmission spectroscopy due to its small scale height. We showed the Rayleigh-like feature of approximately 26 scale heights is not reproducible.

% flatex input end: [dicussion.tex]

\begin{acknowledgements}
        We thank P.V. Sada and F. T. O'Donovan for providing their transit observations. 
        We thank Calar Alto Observatory for allocation of director's discretionary time to this programme.
        M.V. would like to thank the project VEGA 2/0143/14.
        The STELLA facility was funded by the Science and Culture Ministry of the German State of  Brandenburg (MWFK) and the German Federal Ministry for Education and Research (BMBF).
\end{acknowledgements}

\bibliographystyle{aa}
\bibliography{biblo}

\end{document}